\documentclass[11pt,a4paper,oneside]{article}

\pdfoutput=1
\usepackage{jheppub}
\usepackage{verbatim}
\usepackage{graphicx}
\usepackage{mathrsfs}
\usepackage{hyperref}
\usepackage{appendix}
\usepackage{caption}
\usepackage{float}
\usepackage{subcaption}
\usepackage{array,multirow}
\usepackage{ulem}
\usepackage{booktabs}
\usepackage{graphicx}
\usepackage{amssymb}
\usepackage{relsize}
\usepackage{slashed}
\usepackage{amsmath}
\usepackage{setspace}
\usepackage{lscape}
\usepackage{alltt}
\usepackage{epstopdf}

\usepackage{adjustbox}

\usepackage[dvipsnames]{xcolor}

\title{\bf EFT Diagrammatica  II: Tracing the UV origin of bosonic D6 CPV and D8 SMEFT operators}

\author{\bf Wrishik Naskar,}

\author{\bf Suraj Prakash,}

\author{\bf and Shakeel Ur Rahaman}

\emailAdd{wrishiknaskar@gmail.com, surajprk@iitk.ac.in, shakel@iitk.ac.in}

\affiliation{\sf Indian Institute of Technology Kanpur, Kalyanpur, Kanpur 208016, Uttar Pradesh, INDIA}

\begin{document}

\abstract{In recent times,  SMEFT, along with a superlative repertoire of theoretical and computational tools, has emerged as an efficacious platform to test the viability of proposed BSM scenarios. With symmetry as the backbone, higher mass dimensional ($\geq 5$) SMEFT operators constitute the \textit{lingua franca} for studying and comparing the direct or indirect effects of UV models on low energy observables. The steady increase in the accessible energy scales for contemporary particle collision experiments prompts us to inspect effective operators beyond the leading order and investigate their measurable impact as well as their connections with the appropriate BSM proposals. We take the next step in delineating the possible UV roots of SMEFT operators by extending our diagrammatic approach, previously employed for CP, baryon, and lepton number conserving dimension-6 operators, to the complete set of purely bosonic SMEFT operators up to mass dimension-8. We catalogue a diverse array of Feynman diagrams elucidating how the operators encapsulate heavy field propagators while abiding by a notion of minimalism. } 
	
\maketitle

\section{Introduction}\label{sec:intro}

The Standard Model (SM) of Particle Physics has relished a lot of success owing to a multitude of very precise predictions about the features of the subatomic world and their excellent agreement with experiments. The discovery of the Higgs boson \cite{ATLAS:2012yve,CMS:2012qbp} elevated the SM from merely a model to a bona fide theory of fundamental particles. Despite all these triumphs, SM fails to account for not only the dark sector but also several aspects of the visible universe. The most glaring issues have been the observed non-zero masses of neutrinos, the matter-antimatter asymmetry, along with the inexplicable dark matter and dark energy that constitute ninety-five percent of the entire universe.

During the last few decades, several ingenious models have been proposed that have sought to ameliorate our lack of understanding of the subatomic world. Early proposals that garnered a lot of popularity were radical ideas such as Grand Unified Theories (GUTs) and Supersymmetry or even an amalgam of the two. Each of these introduces multiple new degrees of freedom beyond the SM ones, and their symmetry groups contain the SM internal symmetry $SU(3)_C \times SU(2)_L \times U(1)_Y$ as a subset. Contemporary efforts have focused more on minimal extensions of the SM, where the internal symmetry is left unaltered, while the number of degrees of freedom is increased by adding one or two fields. A common underlying feature of all such beyond Standard Model (BSM) proposals is the presence of multiple energy scales within them, characterized by the hierarchy among particle masses. The lack of direct experimental detection of BSM resonances necessitates the use of a framework that can not only translate the interactions of the BSM fields in terms of SM ones but also enables us to conduct comparisons between different BSM scenarios against a common backdrop.  

Effective Field Theory (EFT) \cite{Weinberg:1980wa, Georgi:1994qn} provides us with the necessary set of tools for studying phenomena that encompass different energy scales. Therefore, it is the most suitable framework for addressing the contemporary problems in particle physics. EFT based theoretical and computational tools can enable us to conduct indirect analyses and adjudge the veracity of various new physics proposals, even in the absence of a comprehensive understanding of their Ultra-Violet (UV) origin. In the context of differentiating between BSM scenarios, the Standard Model Effective Field Theory (SMEFT) is the required common backdrop; for a detailed review, see \cite{Brivio:2017vri}. SMEFT incorporates higher mass ($>4$) dimension operators \cite{BUCHMULLER1986621,Grzadkowski:2010es,Lehman:2014jma,Murphy:2020rsh,Li:2020gnx,Li:2020xlh,Liao:2020jmn} and thus accommodates corrections to the SM parameters and measurables while also providing novel predictions such as flavour violation \cite{Buchalla:1995vs,Calibbi:2017uvl}, non-zero baryon and lepton numbers \cite{Weinberg:1979sa,Hambye:2017qix,Kobach:2016ami}, and CP-violation \cite{Branco:1986my,Yang:1997iv} among others.  Specific model-dependent analyses have been conducted within the \textit{top-down} EFT formalism \cite{Henning:2014wua,Haisch:2020ahr,Jiang:2018pbd,Gherardi:2020det,deBlas:2017xtg,deBlas:2014mba,Bilenky:1993bt,Dawson:2017vgm,Dedes:2021abc}. Automated tools such as \texttt{CoDEx} \cite{Bakshi:2018ics}, \texttt{MatchingTools} \cite{Criado:2017khh}, \texttt{STrEAM} \cite{Cohen:2020qvb}, \texttt{Matchmakereft} \cite{Carmona:2021xtq}, \texttt{SuperTracer} \cite{Fuentes-Martin:2020udw}  have made it convenient to conduct such analyses for a variety of BSM models. At the same time, the \textit{bottom-up} formalism \cite{Ellis:2018gqa,Dawson:2019clf,Dawson:2020oco,Brivio:2019ius,Li:2022abx,Gargalionis:2020xvt} facilitates model-independent studies that ultimately allow us to enforce constraints on the SMEFT free parameters, i.e., the Wilson coefficients corresponding to the effective operators.

In \cite{DasBakshi:2021xbl}, we had highlighted a novel approach towards addressing the ``inverse problem", i.e., pinpointing the valid BSM proposals in the event of the observation of anomalies or disagreement between SM predictions and experimental results. Employing simple symmetry based arguments, we unfolded CP, baryon and lepton number conserving SMEFT operators of mass dimension-6 to tree- and one-loop-level Feynman diagrams revealing heavy field propagators. The direct as well as indirect relations between SMEFT operators of mass dimension-6 and precision observables has been well-documented \cite{Alonso:2013hga,Berthier:2015oma,Anisha:2020ggj,Anisha:2021hgc,Bakshi:2020eyg}. By cataloguing possible heavy field quantum numbers corresponding to individual operators, we established direct links between BSM proposals and low-energy observables, thus providing concrete motivations for specific model-dependent analyses. 

Recently, the SMEFT operators of mass dimension-8 have been garnering a lot of attention within the high energy physics community. In the coming years, as we usher into the era of higher luminosity at the LHC, operators of mass dimension-6 alone may not be sufficient to reconcile theoretical calculations with the experimental findings, and inclusion of the next order i.e., $1/\Lambda^4$ suppressed operators will become highly significant. Within scattering amplitudes, the dimension-8 (D8)  operators can intermix with the renormalizable SM interactions to generate interference terms and offer corrections of $\mathcal{O}(1/\Lambda^4)$ to the cross-section of the process under study. This is the same order of correction that the self-mixing of dimension-6 (D6) operators produces. So, if there is no \textit{a priori} constraint preventing the inclusion of $\mathcal{O}(1/\Lambda^4)$ corrections, the inclusion of the D8  contributions becomes customary  \cite{Dawson:2021xei,Corbett:2021eux,Hays:2018zze,Dawson:2020oco,Grinstein:1991cd}. Additionally, the D8 operators present certain novel features which are absent in both the SM as well as the D6 operators. Most notably, vector boson interactions such as the neutral triple, e.g., $ ZZ\gamma, \,\,  Z\gamma\gamma $ and quartic gauge boson couplings e.g., $ ZZZZ, \,\,ZZ\gamma\gamma,\,\,  \gamma\gamma\gamma\gamma $ \cite{Eboli:2016kko,Murphy:2020rsh,Ellis:2019zex,Ellis:2020ljj} which could serve as evidences for new physics.

Recognising the growing interest in the sub-leading order of SMEFT, we have extended our approach and studied the links connecting CP-conserving as well CP-violating purely bosonic operator classes of mass dimension-8 to candidate UV theories.  For the sake of completeness, we have also included CP-violating D6 operators in our discussion. We have adhered to the notion of minimality outlined in \cite{DasBakshi:2021xbl}, which can be reiterated as follows: 
\begin{itemize}
	\item We have taken into account only those SM extensions where the internal symmetry remains the same as that of the SM; therefore none of our diagrams involves heavy vector boson propagators.
	
	\item While unfolding the operators into diagrams, preference has been given to those diagrams that are less varied with respect to the types of vertices as well as with respect to the variety of heavy propagators within them. This is done with the aim of ascertaining the allowed heavy field quantum numbers as closely as we can for a given operator class. We have relaxed this criteria of minimality for cases where operator unfolding can only be accomplished with more than one heavy propagator and (or) the external states necessitate the inclusion of a wider variety of vertices. 
	
	\item For a given operator class, the progression from tree-level to one-loop diagrams and then to two-loop diagrams becomes necessary when the lower order diagram cannot provide the links between the particular operator and certain classes of BSM models. For instance, for operators with $H$ and $H^{\dagger}$ as the external states, only a finite number of heavy scalars appear through tree-level diagrams, but a wider variety are accessible if we take  one-loop diagrams into account. Similarly, unless SM fermions are present as the external states in an operator, heavy fermions only emerge through one-loop diagrams.
\end{itemize}
It must be emphasized that the notion of minimality differs based on the operator class and the family of BSM extensions being discussed. For example, while unfolding operators composed only of the SM scalar and its derivative to reveal heavy scalar propagators, the most minimal scenario corresponds to a tree-level diagram. On the other hand, for CP-violating operators constituted of SM field strength tensors, the most minimal structure is a two-loop diagram with a pair of heavy fermion propagators. 

It must be noted that our discussion is focussed on operator classes instead of individual operators. While the latter would have been more exhaustive and richer in detail, we discovered during our analysis that the exact patterns of the diagrams that can be obtained from the schematic unfolding of operator classes are replicated even when we consider individual operators of those classes. Although, it must be mentioned that, by limiting ourselves to the level of operator classes, the specific contributions from heavy fields to individual SMEFT operators are not explicitly revealed. Also, if one forgoes our concept of minimality, one may obtain additional heavy field quantum numbers that give rise to operators of certain classes. Yet, the choice to restrict our focus to operator classes and the adherence to a pre-defined notion of minimality helps us to determine the BSM origin of SMEFT operators in a structured manner without having to worry about the vast multitude of operators at dimension-8. Our approach sufficiently achieves the aim of highlighting a well-defined rationale for conducting phenomenological analyses on certain SM extensions based on their links with SMEFT operators and thus with observables.

The structure of the article is as follows: we have started by outlining the building blocks of our construction, i.e., Lorentz invariant vertices describing interactions between light SM fields and possible heavy fields in section \ref{sec:vertices}. This is followed by a discussion on CP-violating D6 operators and their UV roots in section \ref{sec:cpv}. Next, we have conducted an extensive examination of the bosonic sector at mass dimension-8, in section \ref{sec:dim-8-bosonic}. We have constructed tree-level, one-loop as well as two-loop diagrams where necessary. Based on these diagrams we have catalogued heavy field quantum numbers for each sub-class of operators. At the end of section \ref{sec:dim-8-bosonic}, we have provided comparisons as well as validation of a subset of our results against recent literature describing similar connections between SMEFT operators and heavy fields. In section \ref{sec:dim6-dim8-common}, we have provided a commentary on the common UV origin of D6 and D8 operators while also emphasizing the subtle ways in which they differ from each other. Through this, we have also underlined the phenomenological significance of the operators discussed in this article.

\section{Fixing heavy field quantum numbers based on fundamental vertices}\label{sec:vertices}

The first and the most vital step of our systematic procedure of unfolding effective operators into Feynman diagrams is enumerating the building blocks of these diagrams, i.e., listing all possible vertices that would in turn constitute those diagrams. In the context of this work, the following points must be noted:

\begin{itemize}
	\item Since our focus is entirely on purely bosonic operators, the only possible external states are the SM scalar $\phi \in \{H,\, H^{\dagger}\}$ and the field strength tensors $X_{\mu\nu} \in \{G_{\mu\nu}^A,\, W_{\mu\nu}^I,\, B_{\mu\nu}\}$, as well as their dual tensors $\tilde{X}_{\mu\nu}$.
	
	\item  While, many of the operators contain covariant derivatives acting on the fields, we have not explicitly highlighted them in the diagrams, since their presence does not affect the heavy field quantum numbers. This is one notable departure from the conventions established in \cite{DasBakshi:2021xbl}, where we had elucidated the symbolic contraction of Lorentz indices in each diagram.
	
	\item The vertices as well as the diagrams have been constructed so as to directly match the external states of the operators, these should not be confused with low-energy process diagrams, this is why we have highlighted field strength tensors rather than the vector bosons and scalar or fermion multiplets rather individual fields in the diagrams that follow. 
\end{itemize}

\begin{table}[!htb]
	\centering
	\renewcommand{\arraystretch}{2.2}
	{\small\begin{tabular}{||c|c||c|c||}
			\hline
			\textsf{Symbol}&
			\textsf{Represents}&
			\textsf{Symbol}&
			\textsf{Represents}\\
			\hline
			
			\includegraphics[height=0.4cm, width=2cm]{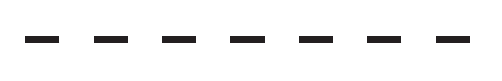}&
			Light (SM) scalar&
			\includegraphics[height=0.4cm, width=2cm]{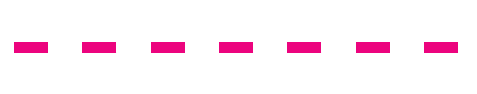}&
			Heavy scalar\\
						
			\includegraphics[height=0.4cm, width=2cm]{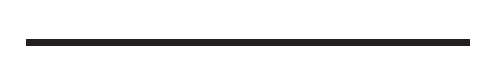}&
			Light (SM) fermion&
			\includegraphics[height=0.4cm, width=2cm]{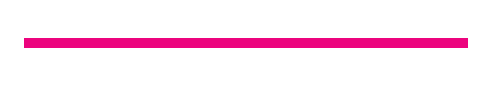}&
			Heavy fermion\\

			\includegraphics[height=0.4cm, width=2cm]{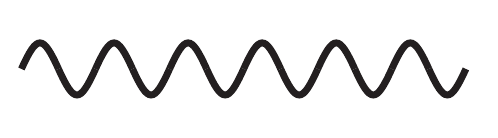}&
			Field strength tensor&
			\includegraphics[height=0.2cm, width=2cm]{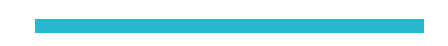}&
			2nd Heavy fermion\\
			
			\hline
			
	\end{tabular}}
	\caption{\sf Representations for tree-level light and heavy propagators of various spins.}
	\label{table:symbol-legend}
\end{table}

\begin{figure}[h]
	\centering
	\renewcommand{\thesubfigure}{\roman{subfigure}}
	\hspace*{0.6cm}\begin{subfigure}[h]{4cm}
		\centering
		\includegraphics[scale=0.4]{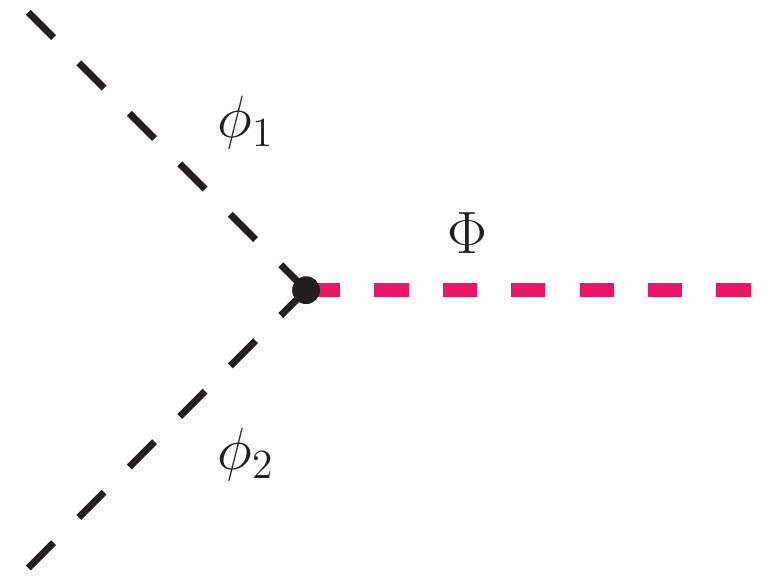}
		\caption{\hypertarget{vertex-1}{$V{}_1$}}\label{vertex:V1}
	\end{subfigure}
	\begin{subfigure}[h]{4cm}
		\centering
		\includegraphics[scale=0.4]{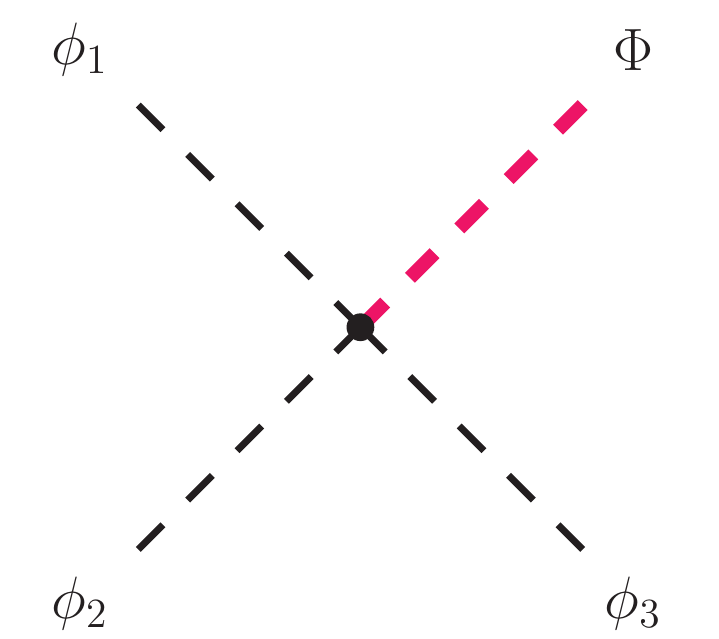}
		\caption{\hypertarget{vertex-2}{$V{}_2$}}\label{vertex:V2}
	\end{subfigure}
	\begin{subfigure}[h]{4cm}
		\centering
		\includegraphics[scale=0.4]{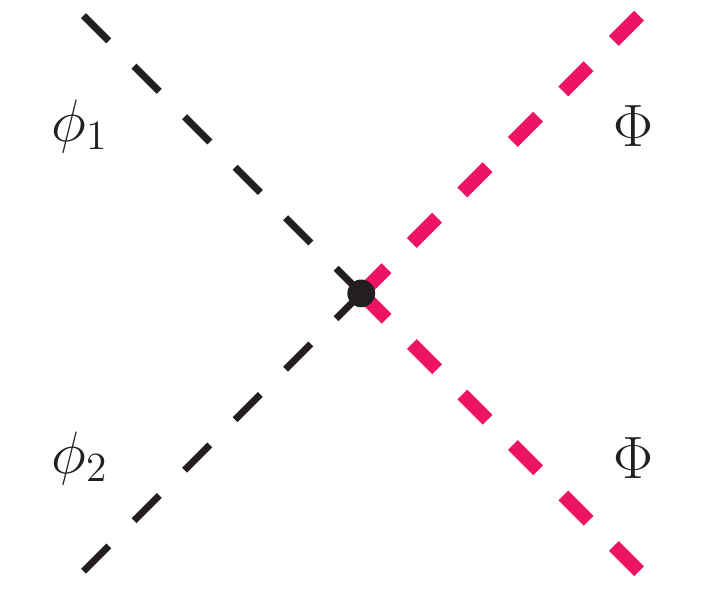}
		\caption{\hypertarget{vertex-3}{$V{}_3$}}\label{vertex:V3}
	\end{subfigure}
	\newline\hspace*{-0.6cm}
	\begin{subfigure}[h]{3.7cm}
		\centering
		\includegraphics[scale=0.4]{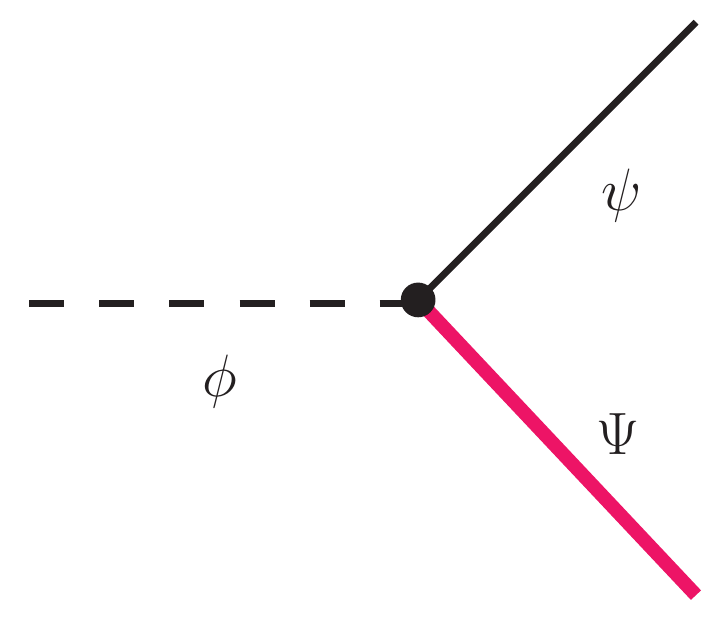}
		\caption{\hypertarget{vertex-4}{$V{}_4$}}\label{vertex:V4}
	\end{subfigure}
	\begin{subfigure}[h]{3.7cm}
		\centering
		\includegraphics[scale=0.4]{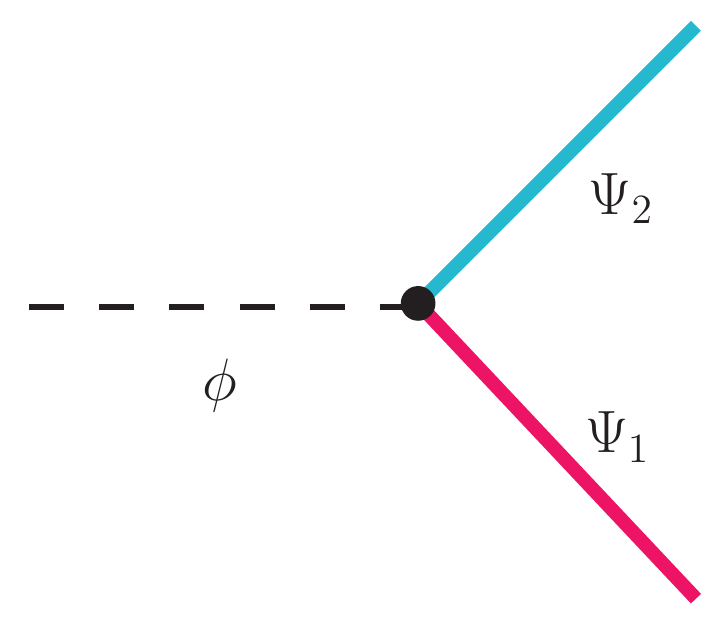}
		\caption{\hypertarget{vertex-5}{$V{}_5$}}\label{vertex:V5}
	\end{subfigure}
	\begin{subfigure}[h]{3.7cm}
		\centering
		\includegraphics[scale=0.4]{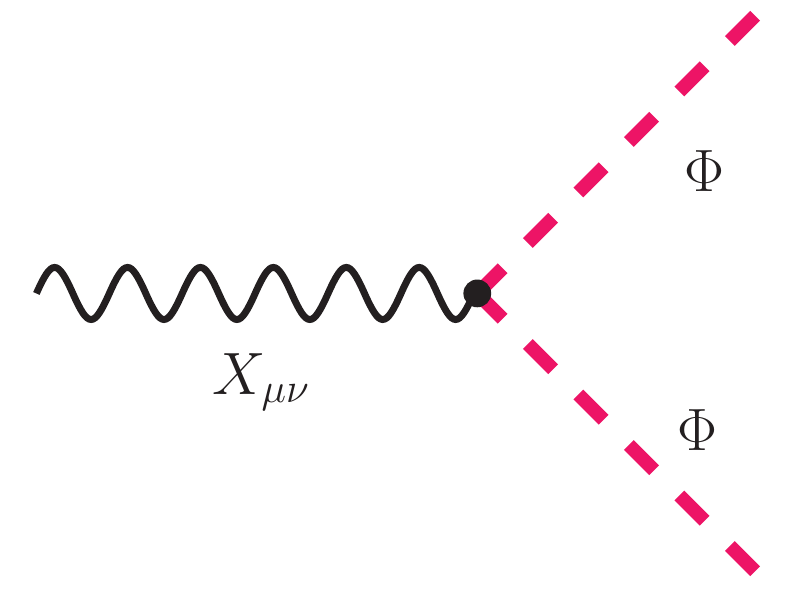}
		\caption{\hypertarget{vertex-6}{$V{}_6$}}\label{vertex:V6}
	\end{subfigure}
	\begin{subfigure}[h]{3.7cm}
		\centering
		\includegraphics[scale=0.4]{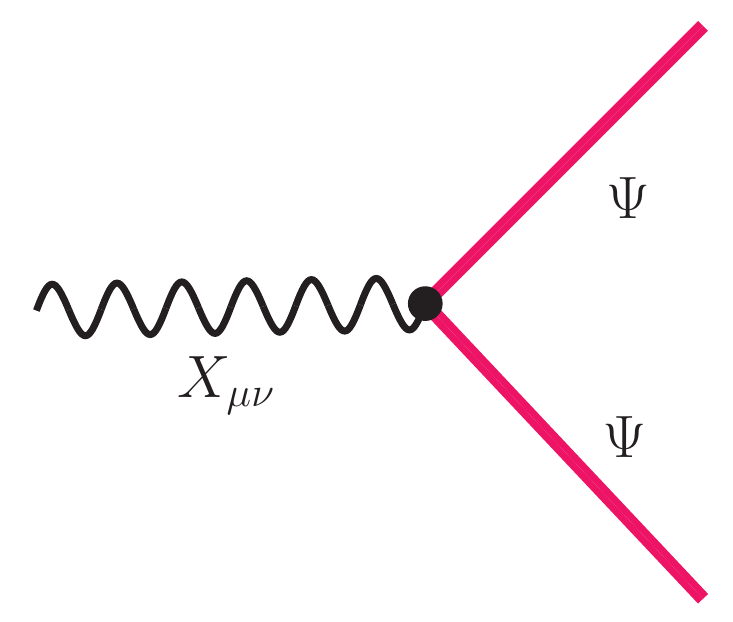}
		\caption{\hypertarget{vertex-7}{$V{}_7$}}\label{vertex:V7}
	\end{subfigure}
	\caption{\sf Vertices describing interactions between light (SM) and heavy fields. The black lines denote SM fields, whereas the pink and blue lines denote heavy fields. In $V_5$, two unique colors have been used to explicitly highlight the presence of two heavy fields.}
	\label{fig:vertices}
\end{figure}

\noindent The vertices relevant to our discussion have been shown in Fig.~\ref{fig:vertices} (additional vertices constituted of the same propagators but not relevant to the diagrams described in this article have been depicted in appendix \ref{sec:extra-vertices}) and the various internal and external lines used within these have been described in Table~\ref{table:symbol-legend}. The heavy field quantum numbers in each case can be fixed based on symmetry arguments and by using the $SU(3)_C \times SU(2)_L \times U(1)_Y$ quantum numbers of the SM fields, see Table~\ref{table:SM-fields}. We have provided detailed descriptions for each case below:

\begin{enumerate}
	\item V${}_1$: A trilinear scalar vertex with one heavy and two light fields. In this case, we have two\footnote{The third possibility $\phi_1 = \phi_2 = H^{\dagger}$ simply gives the conjugate of the result of case (b), therefore we have chosen not to enumerate it. } unique possibilities:
	
	\begin{enumerate}
		\item $\phi_1 = H$, $\phi_2 = H^{\dagger}$: Since $H$ transforms as $(1,2,\frac{1}{2})$, this implies that the only possible heavy scalar quantum numbers are {\boldmath$\Phi \in \{ (1,1,0),\,\,(1,3,0)\}$}.
		
		\item $\phi_1 =  \phi_2 = H$: This implies {\boldmath$\Phi \in \{(1,3,1)\}$}\footnote{The exclusion of $\Phi \in \{(1,1,1)\}$ can be explained as follows: the interaction between 2 $H$'s and one such $\Phi$ can be described by the term $\kappa\,\epsilon_{ij}\,H^{i}H^{j}\Phi^*$ where $\kappa$ is the coupling constant, and $\epsilon_{ij}$ is the completely antisymmetric rank-2 tensor which is required to construct an $SU(2)$ singlet from the product of two $SU(2)$ doublets. The product $H^i \, H^j$ is symmetric with respect to particle exchange and thus is symmetric in the indices $i, j$ whereas $\epsilon_{ij}$ is antisymmetric in the same two indices. As a result of this, the overall term being a product of a symmetric and an antisymmetric piece vanishes.}.
	\end{enumerate}

	\item V${}_2$: A quartic scalar vertex with one heavy and three light fields. Again, we have two distinct sub-cases:
	
	\begin{enumerate}
		\item $\phi_1 = \phi_2 = H$, $\phi_3 = H^{\dagger}$: This implies {\boldmath$\Phi \in \{ (1,2,\frac{1}{2}),\,\,(1,4,\frac{1}{2})\}$}.
		
		\item $\phi_1 =  \phi_2 = \phi_3 = H$: This implies {\boldmath$\Phi \in \{ (1,2,\frac{3}{2}),\,\,(1,4,\frac{3}{2})\}$}.
	\end{enumerate}
	
	\item V${}_3$: A quartic scalar vertex with two heavy and two light fields. With $\phi_1 = H$ and $\phi_2 = H^{\dagger}$, this vertex is ubiquitous across all models containing a second scalar apart from the SM Higgs. In this case, the heavy field quantum number cannot be determined uniquely and it can have arbitrary quantum numbers under $SU(3)_C \times SU(2)_L \times U(1)_Y$, i.e., {\boldmath$\Phi \in \{ (R_C,R_L,Y)\}$}. Here, $R_C$, $R_L$ denote valid representations under the $SU(3)_C$ and $SU(2)_L$ groups respectively and $Y$ refers to the $U(1)_Y$ hypercharge.
	
	A second more elaborate case corresponds to when $\phi_1 =  \phi_2 = H$. In that case we require two heavy scalars, whose quantum numbers must be such so that all four scalars form an overall singlet. We have not delved any deeper into such cases in the remainder of this work since this case departs strongly from our notion of minimality.
	
	\item  V${}_4$: A Yukawa-like vertex with a light scalar, a light fermion and a heavy fermion. Owing to the rich fermion sector of the SM, there are multiple possibilities and these have been catalogued in Table~\ref{table:V4-quantum-numbers}. 
	
	\begin{table}[h]
		\centering
		\renewcommand{\arraystretch}{2.0}
		\begin{tabular}{||c|c|c||c|c|c||}
			\hline
			\hline
			$\phi$&
			$\psi$&
			$\Psi$ \textsf{(quantum numbers)}&
			$\phi$&
			$\psi$&
			$\Psi$ \textsf{(quantum numbers)}\\
			\hline
			\hline
			
			$H$&
			$q_L$&
			$\{(\bar{3}, 1, -\frac{2}{3}),\,(\bar{3}, 3, -\frac{2}{3})\}$&
			$H^{\dagger}$&
			$q_L$&
			$\{(\bar{3}, 1, \frac{1}{3}),\,(\bar{3}, 3, \frac{1}{3})\}$\\
			
			$H$&
			$l_L$&
			$\{(1, 1, 0),\,(1, 3, 0)\}$&
			$H^{\dagger}$&
			$l_L$&
			$\{(1, 1, 1),\,(1, 3, 1)\}$\\
			
			$H$&
			$u_R$&
			$\{(\bar{3}, 2, -\frac{7}{6})\}$&
			$H^{\dagger}$&
			$u_R$&
			$\{(\bar{3}, 2, -\frac{1}{6})\}$\\
			
			$H$&
			$d_R$&
			$\{(\bar{3}, 2, -\frac{1}{6})\}$&
			$H^{\dagger}$&
			$d_R$&
			$\{(\bar{3}, 2, \frac{5}{6})\}$\\
			
			$H$&
			$e_R$&
			$\{(1, 2, \frac{1}{2})\}$&
			$H^{\dagger}$&
			$e_R$&
			$\{(1, 2, \frac{3}{2})\}$\\
			
			\hline
		\end{tabular}
	\caption{\sf Lists of heavy field quantum numbers for different choices of the light scalar and light fermion in the vertex $V_4$.}
	\label{table:V4-quantum-numbers}
	\end{table}
	
	\item V${}_5$:   A Yukawa-like vertex with a light scalar and two heavy fermions\footnote{While we have refrained from discussing scenarios with two distinct heavy scalars at the same vertex, we cannot exclude the similar case involving fermions. This is owing to the fact that CP-violating bosonic SMEFT operators cannot be generated by a single heavy fermion. We discuss this in more detail in section~\ref{sec:cpv}.}. Once again, the quantum numbers of $\Psi_1$ and $\Psi_2$, i.e., $(R_{C_1},R_{L_1},Y_1)$ and $(R_{C_2},R_{L_2},Y_2)$ cannot be fixed exactly but we can impose the following constraints on them:
	\begin{eqnarray}\label{eq:vertex-5-rel}
		R_{C_1} \otimes R_{C_2} = 1, \hspace{1cm} R_{L_1} \otimes R_{L_2} = 2,  \hspace{1cm} Y_1 = Y_2  \pm \frac{1}{2}.
	\end{eqnarray}
	In the last relation, $+$ or $-$ appears depending on whether $\phi = H^{\dagger}$ or $H$ at the vertex.
	
	\item V${}_6$ and V${}_7$: Here, one of the SM field strength tensors ($B_{\mu\nu}, \, W^I_{\mu\nu}, \, G^A_{\mu\nu}$) appear as the lighter field. In this case, we again have the freedom to assign arbitrary quantum numbers to the heavy field (scalar as well as fermion), except the constraint of non-triviality imposed on one of the three quantum numbers depending on $X_{\mu\nu}$:
	\begin{eqnarray}\label{eq:vertex-6-rel}
		X_{\mu\nu} \equiv B_{\mu\nu} \Rightarrow Y \neq 0, \hspace{0.3cm} X_{\mu\nu} \equiv W^I_{\mu\nu} \Rightarrow R_L \neq 1, \hspace{0.3cm} X_{\mu\nu} \equiv G^A_{\mu\nu} \Rightarrow R_C \neq 1.
	\end{eqnarray}
	In each case, the other quantum numbers can assume arbitrary of values. 
\end{enumerate}

\noindent These vertices form the rudiments of the Feynman diagrams that appear in the next sections of this article. The forthcoming discussion revolves around tree-level, one-loop diagrams, as well as two-loop diagrams (where necessary). The following points must be emphasized regarding the appearance of different diagrams across the various operator classes:

\begin{itemize}
	\item Tree-level diagrams containing a heavy scalar propagator appear in a small number of cases and these pinpoint the heavy field quantum numbers exactly. Since the external states consist of $\phi$'s and (or) $X_{\mu\nu}$'s, we do not encounter tree-level diagrams with a heavy fermion propagator. 
	
	\item One-loop diagrams composed entirely of a single heavy scalar are ubiquitous across the (CP-conserving dimension-8) operator classes considered in this work. These encapsulate a wide variety of heavy fields.
	
	\item One-loop diagrams, composed entirely of a single heavy fermion, appear when the external state consists only of $X_{\mu\nu}$'s. These also cover a wide variety of heavy fields with the quantum numbers constrained by the specific $X_{\mu\nu}$ present at the vertices.
	
	\item One-loop diagrams exhibiting light-heavy mixing between an SM fermion and a heavy fermion are vital in scenarios where the external states contain SM scalars. In these cases, the heavy field-quantum numbers can be inferred precisely. In cases such as for the $X^4$ operator class at dimension-8, where all heavy fermion possibilities are encompassed by the  one-loop diagram consisting of a single heavy fermion, light-heavy mixing is not necessary to account for additional fermions. 
	
	\item One-loop diagrams exhibiting mixing between two heavy fermions are necessary to trace the UV origin of the CP-violating subset of operator classes containing $X_{\mu\nu}$'s. These are also essential to explain the embedding of general heavy fermions with arbitrary quantum numbers within operators made up of only the SM scalar, its conjugate and their derivatives because in such cases one-loop diagrams made up of a unique fermion do not exist.
	
	\item Two-loop diagrams with two heavy fermion and one light scalar propagator are unavoidable if we attempt to explain the UV origin of CP-violation through operators of the $X^3$ and $X^4$ classes.      
\end{itemize}

\section{CP-violating D6 operators}\label{sec:cpv}

CP-violation is a critical component of the matter-antimatter asymmetry puzzle \cite{Sakharov:1967dj}. Within the SM, the presence of a phase in the CKM matrix \cite{Kobayashi:1973fv} alone is not sufficient to explain baryogenesis \cite{Cohen:1993nk,Grzadkowski:1993gh,deVries:2017ncy}, additional sources of CP-violation can be described through the inclusion of the CP-violating effective operators \cite{Yang:1997iv}. This motivates the search for CP-violation at current and future particle collider programs \cite{Aad:2020cpH,Sirunyan:2020cpv}. Any tangible measurement will essentially point towards BSM sources. 

\subsection{Signature of CP-violation in SMEFT}

Extensive studies have been conducted on CP-violation in the context of Higgs physics \cite{Brehmer:2018prd,Bernlochner:2019prb,Englert:2019prd}. Within new physics proposals, CP-violation manifests through interactions of the form: {\boldmath$(a + b\, \gamma_5)\, \overline{\Psi}_1\Psi_2\,\phi$} \cite{Bakshi:2021,Bakshi:2021prd,Abe:2017sam}. Processes described by fermion loops containing an overall odd number of $\gamma_5$ at the vertices can carry the signature of CP-violation. The $\gamma_5$ matrix generates rank-4 Levi-Civita tensors based on the following relation:
\begin{equation}
	2\,i\,\sigma_{\rho\delta}\,\gamma_5 = \epsilon_{\mu\nu\rho\delta}\,\sigma^{\mu\nu}.
\end{equation}
Therefore, for the SMEFT operators, the most overt signature of CP-violation is the presence of an odd number of rank-4 Levi-Civita tensors $\epsilon^{\mu\nu\rho\sigma}$. Thus, operators containing the duals of field strength tensors:
\begin{eqnarray}
	\tilde{X}_{\mu\nu} = \frac{1}{2} \, \epsilon_{\mu\nu\rho\sigma} \, X^{\rho\sigma},
\end{eqnarray}

\noindent encapsulate possible sources of CP-violation. Dimension-6 SMEFT operators that hint towards CP-violation have been listed below, categorized by class:
  \begin{eqnarray}\label{eq:dim-6-cpv-ops}
  	\phi^2X^2 \, &\rightarrow& \hspace{0.15cm}
  	\mathcal{Q}_{H\tilde{G}}:\, (H^\dagger\, H)\,(\tilde{G}^A_{\mu\nu}\,G^{A\mu\nu}), \hspace{1.3cm}
  	\mathcal{Q}_{H\tilde{W}}:\, (H^\dagger\,H)\,(\tilde{W}^I_{\mu\nu}\,W^{I\mu\nu}), \nonumber\\
  	&&	 \hspace{0.15cm} \mathcal{Q}_{H\tilde{B}}:\, (H^\dagger\, H)\,(\tilde{B}_{\mu\nu}\,B^{\mu\nu}), \hspace{1.5cm}
  	\mathcal{Q}_{H\tilde{W}B}:\, (H^\dagger\,\tau^I\, H)\,(\tilde{W}^I_{\mu\nu}\,B^{\mu\nu}),\nonumber\\
  	X^3\, &\rightarrow& \hspace{0.15cm} \mathcal{Q}_{\tilde{G}}:\, f^{ABC}\tilde{G}^{A\mu}_{\nu}\,G^{B\nu}_\rho \,G^{C\rho}_\mu, \hspace{1.55cm} \mathcal{Q}_{\tilde{W}}:\, \varepsilon^{IJK}\tilde{W}^{I\mu}_{\nu}\,W^{J\nu}_\rho\, W^{K\rho}_\mu. 
  \end{eqnarray}

It has been established in recent works \cite{Bakshi:2021,Bakshi:2021prd} that while a single heavy fermion in the loop can lead to the CP conserving counterparts of the operators listed in Eq.~\eqref{eq:dim-6-cpv-ops}, the CP-violating ones necessitate the inclusion of a second heavy fermion. Also, the degree of non-triviality increases when we consider CP-violating operators of the $X^3$ class, where in order to accommodate the $\gamma_5$ matrix, we require two-loop diagrams with a light scalar propagator in addition to the heavy fermions.

\subsection{Unfolding the $\phi^2X^2$ and $X^3$ operator classes}

\begin{figure}[!htb]
	\centering
	\renewcommand{\thesubfigure}{\roman{subfigure}}
	\hspace*{-0.4cm}\begin{subfigure}[h]{3.7cm}
		\centering
		\includegraphics[scale=0.4]{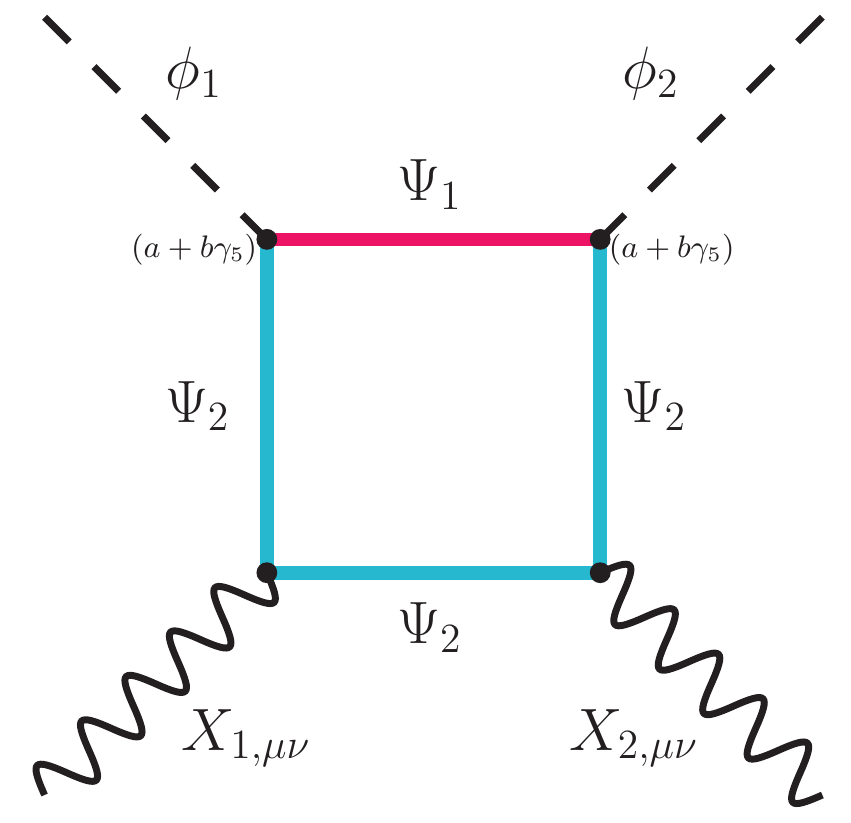}
		\caption{}\label{fig:phi2x2-1}
	\end{subfigure}
	\begin{subfigure}[h]{3.7cm}
		\centering
		\includegraphics[scale=0.4]{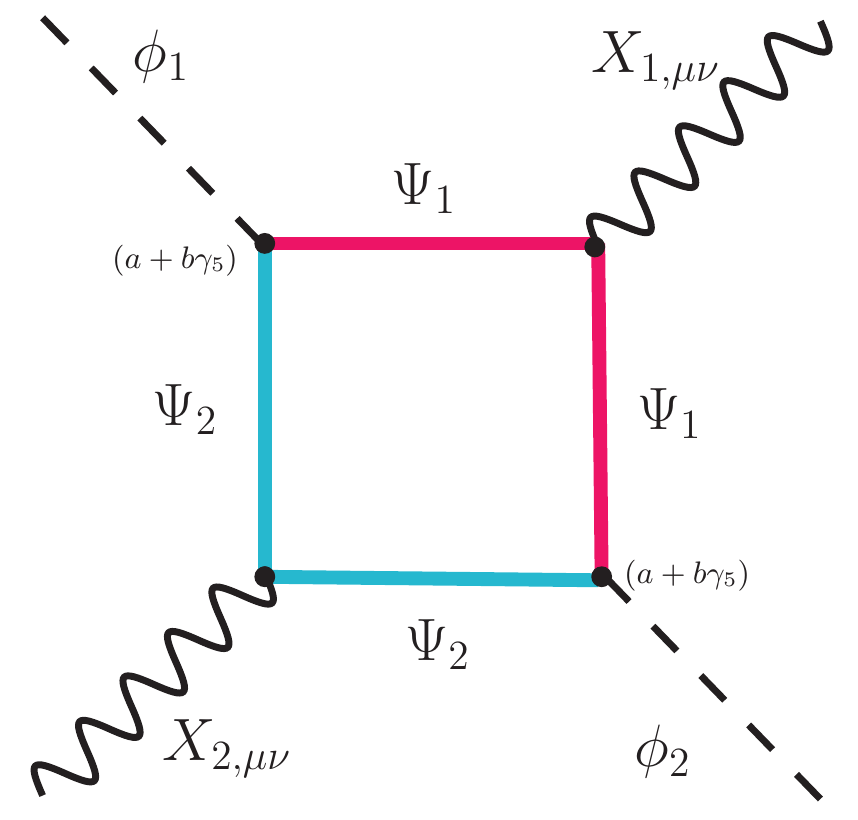}
		\caption{}\label{fig:phi2x2-2}
	\end{subfigure}
	\begin{subfigure}[h]{3.7cm}
		\centering
		\includegraphics[scale=0.4]{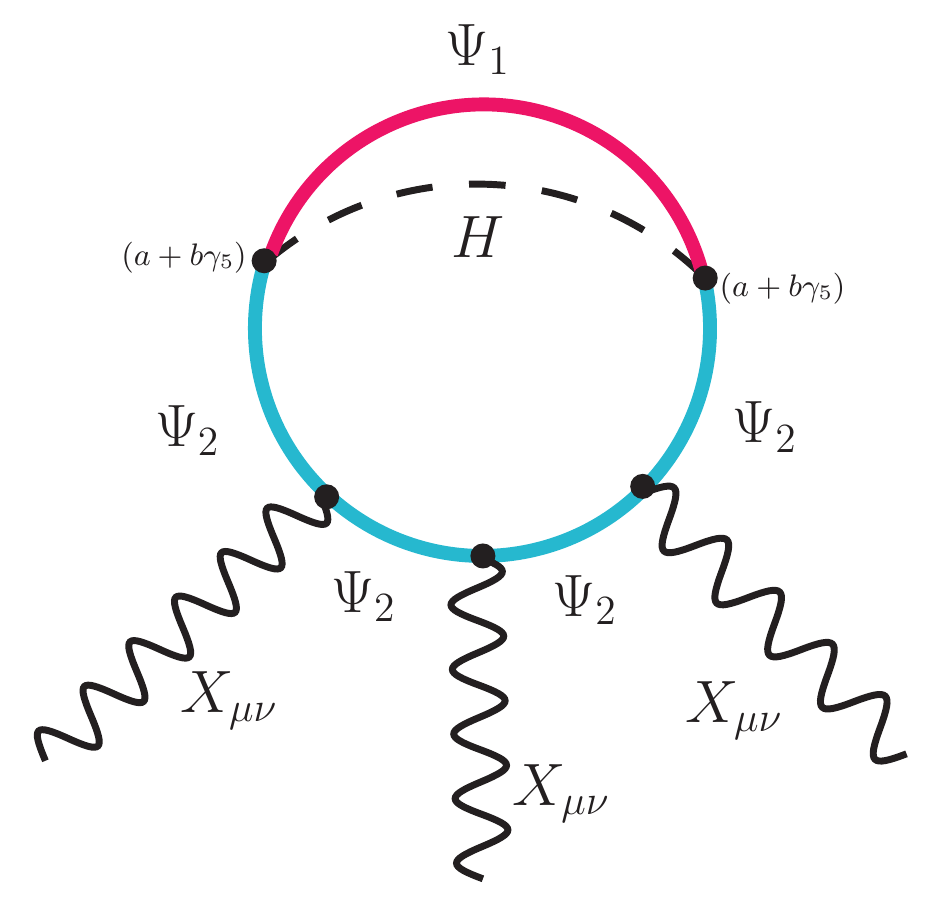}
		\caption{}\label{fig:x3-1}
	\end{subfigure}
	\begin{subfigure}[h]{3.7cm}
		\centering
		\includegraphics[scale=0.37]{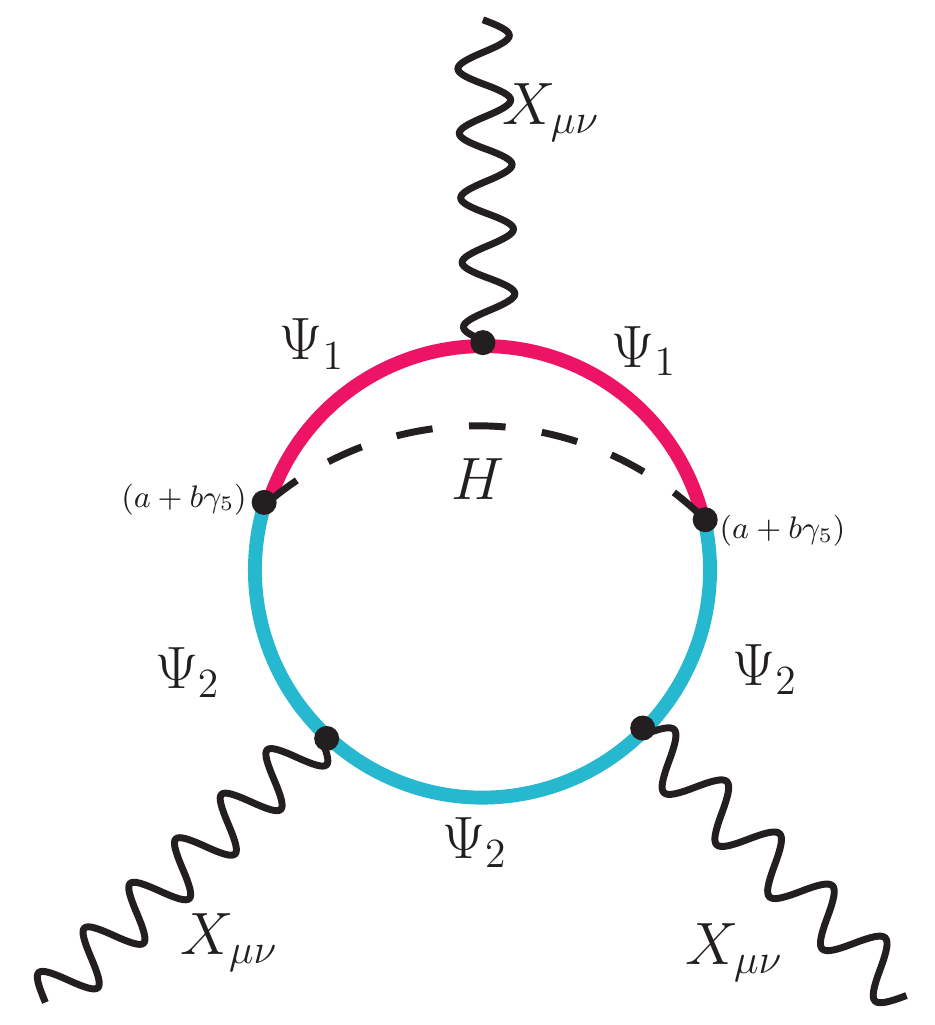}
		\caption{}\label{fig:x3-2}
	\end{subfigure}
	\caption{\sf Unfolding SMEFT operators of mass dimension-6 belonging to (i), (ii) the $\phi^2X^2$ class into one-loop and (iii), (iv) the $X^3$ class into two-loop  Feynman diagrams involving heavy fermion propagators. In order to be concise in our discussion, we have only considered a general Yukawa vertex with the coupling parametrised as $a + b\,\gamma_5$.}
	\label{fig:phi2x2-x3-unfolding}
\end{figure}

The schematic manner in which the operators of the $\phi^2X^2$ ($\mathcal{Q}_{H\tilde{G}}$, $\mathcal{Q}_{H\tilde{W}}$, $\mathcal{Q}_{H\tilde{B}}$ and $\mathcal{Q}_{H\tilde{W}B}$) and $X^3$ ($\mathcal{Q}_{\tilde{G}}$, $\mathcal{Q}_{\tilde{W}}$) classes can be unfolded into one- and two-loop diagrams respectively, involving two heavy fermions ($\Psi_1$ and $\Psi_2$), is shown in Fig.~\ref{fig:phi2x2-x3-unfolding}. The quantum numbers of $\Psi_1$ and $\Psi_2$ can be determined by first identifying the vertices as either \hyperlink{vertex-5}{$V{}_5$} or \hyperlink{vertex-7}{$V{}_7$} and then simultaneously satisfying the relevant relations -  Eq.~\eqref{eq:vertex-5-rel} or \eqref{eq:vertex-6-rel} at each vertex. These constraints have been reiterated below:
\begin{eqnarray}\label{eq:cpv-constraint}
	R_{C_1} \otimes R_{C_2} &=& 1, \hspace{1cm}   X_{\mu\nu} \equiv G^A_{\mu\nu} \Rightarrow R_C \neq 1, \nonumber\\
	R_{L_1} \otimes R_{L_2} &=& 2,  \hspace{1cm} X_{\mu\nu} \equiv W^I_{\mu\nu} \Rightarrow R_L \neq 1,   \nonumber\\
	Y_1 = Y_2  &\pm& \frac{1}{2},  \hspace{1cm} X_{\mu\nu} \equiv B_{\mu\nu} \Rightarrow Y \neq 0. 
\end{eqnarray}

The requirement of two heavy fermions can be understood based on the presence of vertices of the \hyperlink{vertex-5}{$V{}_5$} category within each of these diagrams. On top of that, these fermions must be vector-like because the difference between the couplings of the left- and right-chiral parts indicates the violation of CP-symmetry. This is why the parametrisation of the Yukawa coupling constant as $(a + b\,\gamma_5)$ with $a \neq b \neq 0$ is necessary. The projection operators $ \mathbb{P_L},\, \mathbb{P_R} $ or $\gamma_5$ project out this exact disparity between the left- and right-chiral sectors. This signature of CP-violation consequently appears within the Wilson coefficients of the aforementioned SMEFT operators.

\subsection{Validation of the results}
As described above, our results correspond to SM extensions containing vector-like fermions (VLFs).  To substantiate our results, we have inspected recent works that shed light on the physics of UV models containing VLFs. These analyses utilize the \textit{top-down} procedure of integrating out the VLFs to obtain the CP-violating D6 SMEFT operators.

\begin{table}[h]
	\centering
	\renewcommand{\arraystretch}{2.5}
	\begin{tabular}{||c|c|c||}
			\hline
			\hline
			\textsf{Operator}&
			\textsf{Quantum numbers of VLF pairs}&
			\textsf{References}\\
			\hline
			\hline
			
			$\mathcal{Q}_{H\tilde{G}}$, $\mathcal{Q}_{\tilde{G}}$&
			\{$(R_C,2,Y),(R_C,1,Y\pm1/2)$\}&
			\cite{Huo:2015exa,Angelescu:2020yzf}\\
			
			$\mathcal{Q}_{H\tilde{W}}$, $\mathcal{Q}_{H\tilde{B}}$, $\mathcal{Q}_{HW\tilde{B}}$, $\mathcal{Q}_{\tilde{W}}$&
			$\{(1,2,Y),(1,1,Y\pm1/2)\},\,\, \{(1,3,1),(1,2,1/2)\}$&
			\cite{Angelescu:2020yzf,Bakshi:2021prd,Bakshi:2021}\\
			
			\hline
	\end{tabular}
	\caption{\sf Heavy field representations that have been found to yield CP-violating SMEFT operators after being integrated out along with the works reporting these connections.}
	\label{table:cpv-validation}
\end{table}

Quantum numbers of the heavy fields that beget individual (CP-violating) operators of the $\phi^2X^2$ and $X^3$ classes, along with references to the works that have studied them, have been mentioned in Table~\ref{table:cpv-validation}. A quick inspection of the contents of Table~\ref{table:cpv-validation} reveals that the mentioned heavy field quantum numbers indeed satisfy the constraints outlined in Eq.~\eqref{eq:cpv-constraint} and they thus form a subset of our results. 

\section{The bosonic sector at D8}\label{sec:dim-8-bosonic}

The bosonic sector of SMEFT at mass dimension-8 is constituted by operators containing the SM scalar, its conjugate, their covariant derivative, the field strength tensors corresponding to the gauge groups, and their dual tensors. In our study, we have subdivided these operators into three broad categories:

\begin{enumerate}
	\item Those with only $\phi \in \{H, H^{\dagger}\}$ and their covariant derivatives as external states.
	
	\item Those with $X_{\mu\nu} \in \{G_{\mu\nu}^A,\, W_{\mu\nu}^I,\, B_{\mu\nu}\}$ and their dual tensors as external states.
	
	\item Those containing a mix of scalars, their covariant derivatives, field strength tensors and their duals.
\end{enumerate}

Each category has further been subdivided based on the number of derivatives. We have drawn tree-level (\textit{where applicable}), one-loop and two-loop (\textit{where necessary}) diagrams for the various cases, highlighting the heavy field propagators within them. After identifying the vertices appearing in those diagrams, we have shed light on the permitted quantum numbers for the heavy field(s). On account of the fact, that the covariant derivatives do not influence the heavy field quantum numbers, we have not provided any symbolic representation for them in our diagrams. Also, as opposed to listing diagrams exhaustively, taking into account every permutation of the external legs, we have opted to be concise in our presentation by only providing schematic diagrams that encapsulate the the information related to multiple operators.

\subsection{External states: Only $\phi$}

This category consists of operators belonging to the $\phi^8$, $\phi^6\mathcal{D}^2$ and $\phi^4\mathcal{D}^4$ classes. The complete list of independent operators has been catalogued in Table~\ref{tab:smeft8class_1_2_3}. We have unfolded these operators, to reveal heavy propagators, in a systematic way starting with tree-level diagrams then proceeding towards one-loop diagrams of different varieties (based on the number of heavy propagators within the loops). 

\subsubsection*{\centering$ \boxed{\,\,\,\,\,\text{\texttt{Tree-level}}\,\,\,\,\,}$}
	
	\begin{figure}[h]
		\centering
		\renewcommand{\thesubfigure}{\roman{subfigure}}
		\hspace{-1cm}
		\begin{subfigure}[t]{4.5cm}
			\centering
			\includegraphics[scale=0.35]{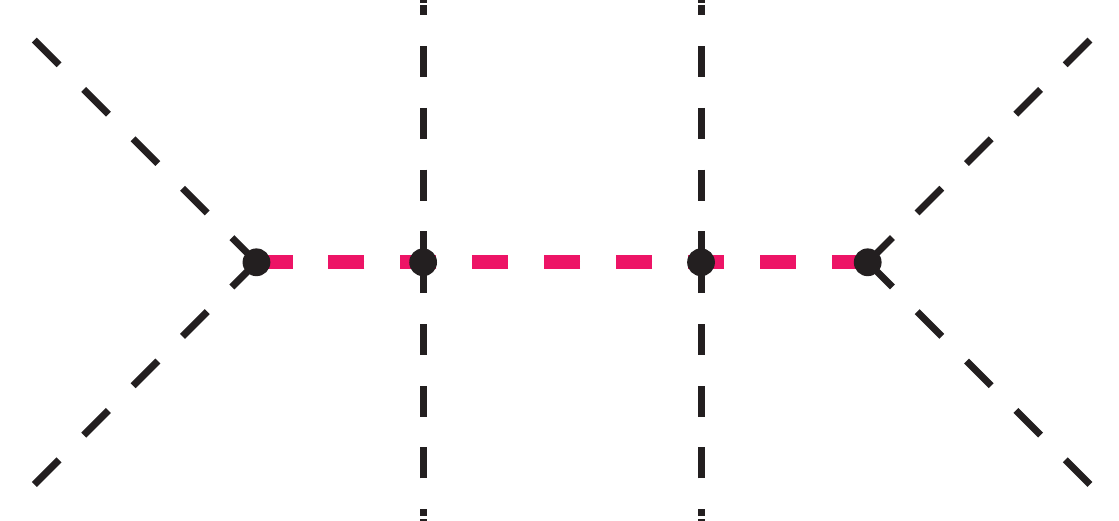}
			\caption{$\phi^8$}
			\label{subfig:phi8-1}
		\end{subfigure}
		\begin{subfigure}[t]{4.5cm}
			\centering
			\includegraphics[height=2.2cm,width=3.5cm]{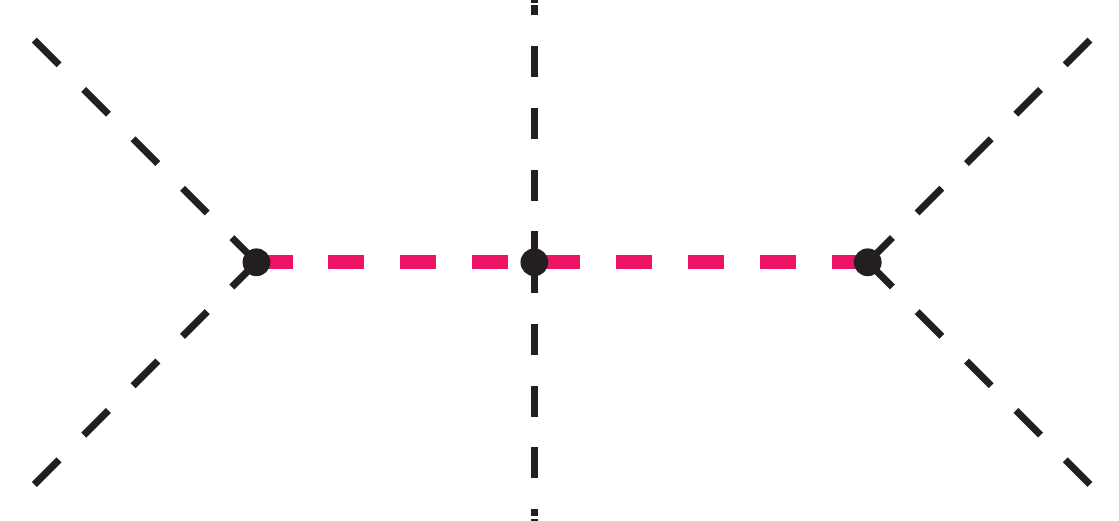}
			\caption{$\phi^6\mathcal{D}^2$}
			\label{subfig:phi6d2-2}
		\end{subfigure}
		\begin{subfigure}[t]{4cm}
			\centering
			\includegraphics[scale=0.3]{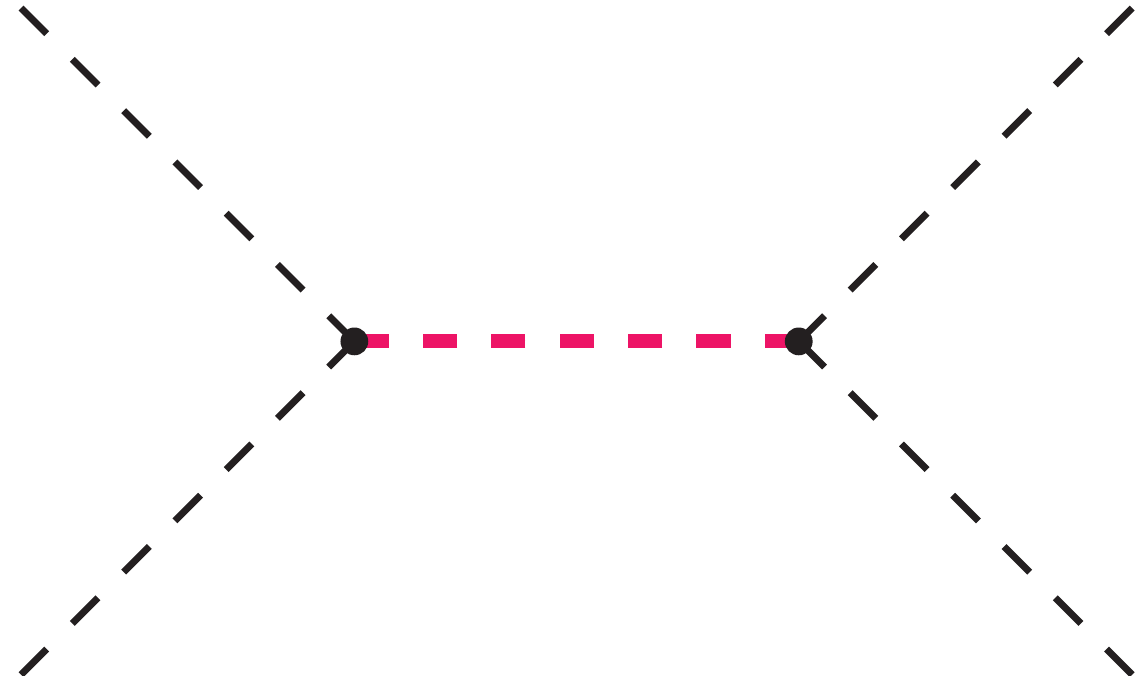}
			\caption{$\phi^4\mathcal{D}^4$}
			\label{subfig:phi4d4-1}
		\end{subfigure}
		\hspace*{-1cm}
		\begin{subfigure}[t]{4.5cm}
			\centering
			\includegraphics[scale=0.35]{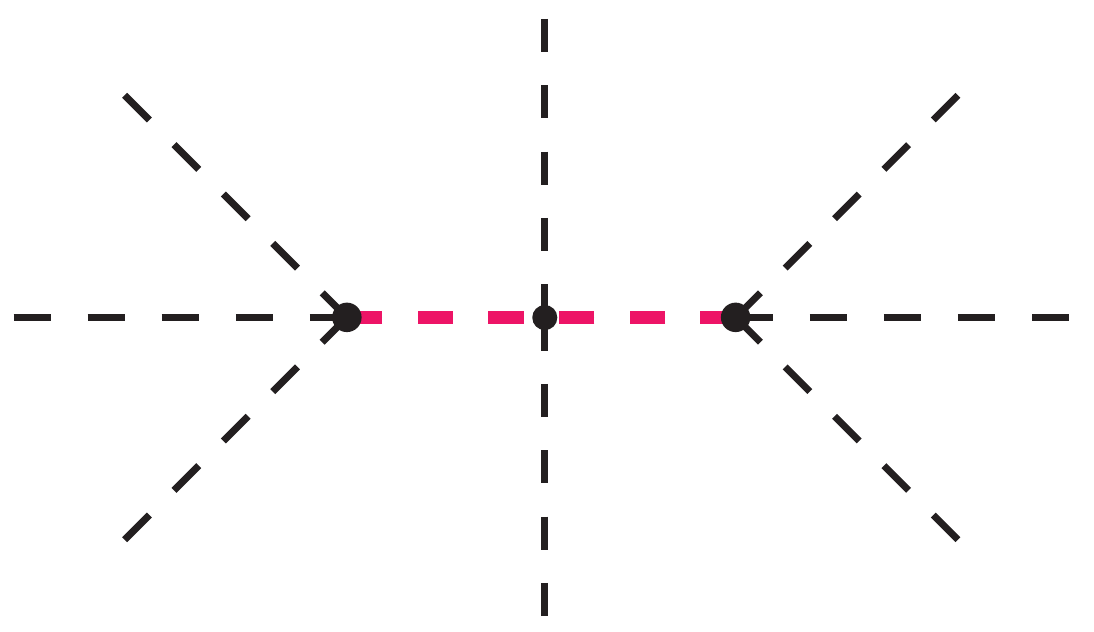}
			\caption{$\phi^8$}
			\label{subfig:phi8-2}
		\end{subfigure}
		\begin{subfigure}[t]{4.5cm}
			\centering
			\includegraphics[height=2.2cm,width=3.5cm]{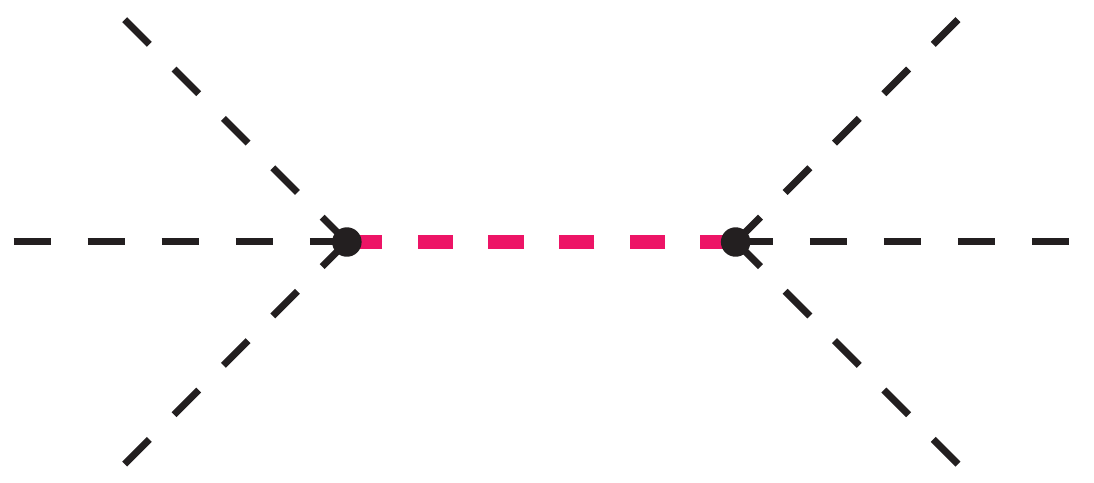}
			\caption{$\phi^6\mathcal{D}^2$}
			\label{subfig:phi6d2-1}
		\end{subfigure}
		\caption{\sf Schematic unfolding of $\phi^8$, $\phi^6\mathcal{D}^2$ and $\phi^4\mathcal{D}^4$ classes of D8 SMEFT operators into tree-level diagrams.}
		\label{fig:phi-only-tree}
	\end{figure}

\noindent Tree-level schematic diagrams corresponding to the $\phi^8$, $\phi^6\mathcal{D}^2$ and $\phi^4\mathcal{D}^4$ have been elucidated in Fig.\,\ref{fig:phi-only-tree}. The set of heavy field quantum numbers can be obtained at the level of the operator class itself by examining the vertices within the diagrams as described below:

\begin{enumerate}
	\item Figs.\,\ref{subfig:phi8-1}, \ref{subfig:phi6d2-2}, \ref{subfig:phi4d4-1} are composed of the vertices: \{\hyperlink{vertex-1}{$V{}_1$}, \hyperlink{vertex-3}{$V{}_3$}\}, \{\hyperlink{vertex-1}{$V{}_1$}, \hyperlink{vertex-3}{$V{}_3$}\} and \{\hyperlink{vertex-3}{$V{}_3$}\} respectively. Based on the discussion in section \ref{sec:vertices}, we know that \hyperlink{vertex-3}{$V{}_3$} is ubiquitous in all scalar extensions of the SM and permits assigning arbitrary quantum numbers to the heavy field. On the other hand \hyperlink{vertex-1}{$V{}_1$} only permits a finite number of cases, i.e.,
	\begin{eqnarray}\label{eq:dim8-tree1}
		\boldmath \Phi\in\{(1,3,0),(1,1,0),(1,3,1)\}.
	\end{eqnarray} 

	More concrete connections between specific operators of a given class and distinct heavy quantum numbers can be inferred based on the arrangement of $H$ and $H^{\dagger}$ at the vertices, which itself is necessitated by the proper contraction of Lorentz indices associated with the derivatives accompanying these fields in the operator in the case of $\phi^6\mathcal{D}^2$ and $\phi^4\mathcal{D}^4$ operators.
	
	\item Figs.\,\ref{subfig:phi8-2}, \ref{subfig:phi6d2-1} are  are composed of the vertices \{\hyperlink{vertex-2}{$V{}_2$}, \hyperlink{vertex-3}{$V{}_3$}\} and \{\hyperlink{vertex-2}{$V{}_2$}\} respectively. Once again, while \hyperlink{vertex-3}{$V{}_3$} permits arbitrary quantum numbers for the heavy field, \hyperlink{vertex-2}{$V{}_2$} leads to only a finite number of cases, i.e.,
	\begin{eqnarray}\label{eq:dim8-tree2}
		\boldmath \Phi\in\{(1,4,\frac{3}{2}),(1,2,\frac{3}{2}),(1,4,\frac{1}{2}),(1,2,\frac{1}{2})\}.
	\end{eqnarray}
\end{enumerate}
\hrule

\subsubsection*{\centering$ \boxed{\,\,\,\,\,\text{\texttt{Heavy-loop}}\,\,\,\,\,}$}

\begin{figure}[h]
		\centering
		\renewcommand{\thesubfigure}{\roman{subfigure}}
		\begin{subfigure}[t]{4.5cm}
			\centering
			\includegraphics[scale=0.37]{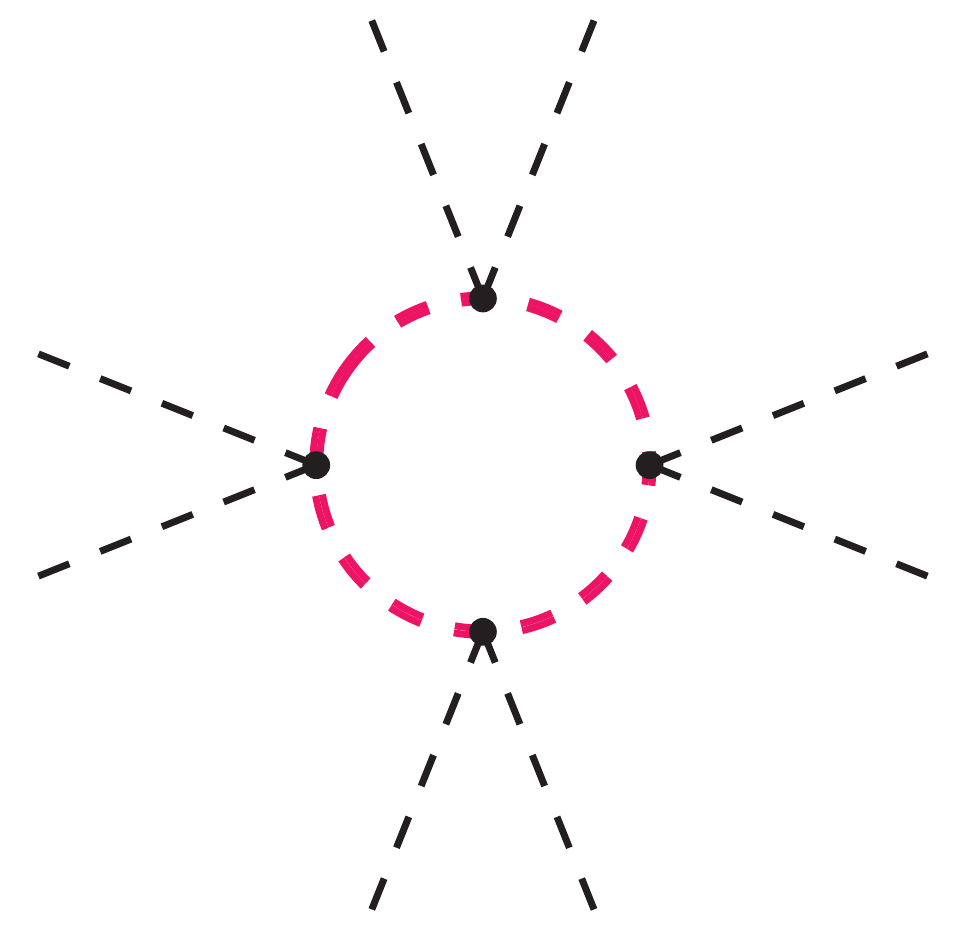}
			\caption{$\phi^8$}
			\label{subfig:phi8-3}
		\end{subfigure}
		\begin{subfigure}[t]{4.5cm}
			\centering
			\includegraphics[scale=0.3]{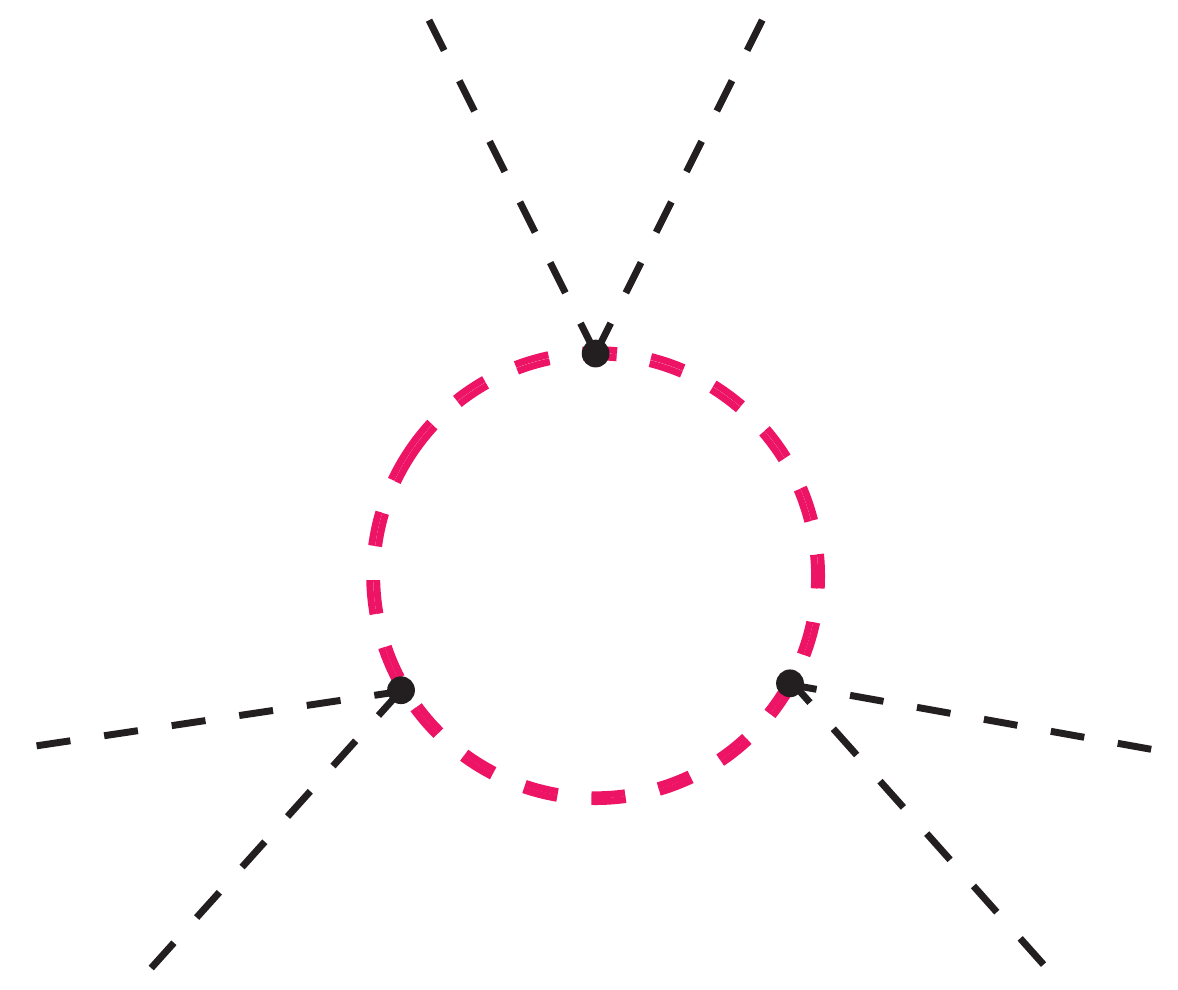}
			\caption{$\phi^6\mathcal{D}^2$}
			\label{subfig:phi6d2-3}
		\end{subfigure}
		\begin{subfigure}[t]{4cm}
			\centering
			\includegraphics[scale=0.3]{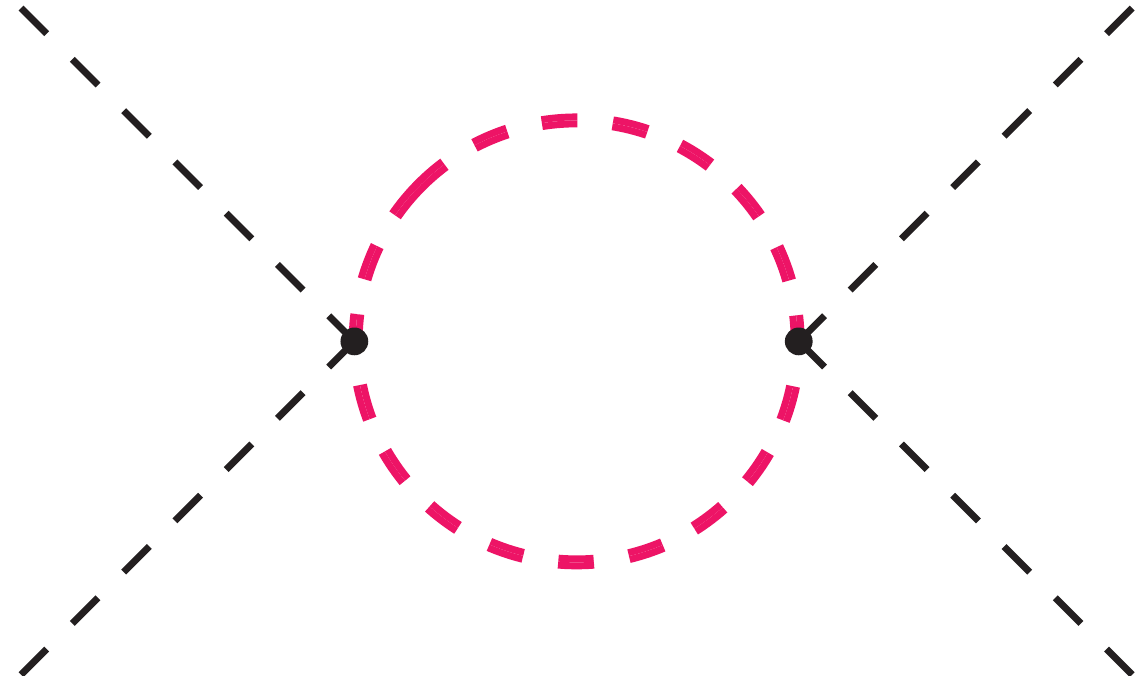}
			\caption{$\phi^4\mathcal{D}^4$}
			\label{subfig:phi4d4-2}
		\end{subfigure}
		\caption{\sf One-loop schematic diagrams revealing heavy scalar propagators enveloped within (i) $ \phi^8 $, (ii) $ \phi^6\mathcal{D}^2 $, and (iii) $ \phi^4\mathcal{D}^4 $ classes of SMEFT operators.}
		\label{fig:phi-only-heavy-loop}
\end{figure}
Fig.\,\ref{fig:phi-only-heavy-loop} contains one-loop schematic diagrams corresponding to the $\phi^8$, $\phi^6\mathcal{D}^2$ and $\phi^4\mathcal{D}^4$ operator classes, where the entire loop is composed of a distinct heavy scalar. The constituent vertex for each diagram is \hyperlink{vertex-3}{$V{}_3$}. The quartic interaction appearing at each vertex of these diagrams encompasses all possible heavy scalars. Therefore, the quantum numbers, in this case, remains arbitrary, i.e., $\Phi\in(R_C,R_L,Y).$
\\
\hrule

\subsubsection*{\centering$ \boxed{\,\,\,\,\,\text{\texttt{Light-heavy mixing}}\,\,\,\,\,}$}

\begin{figure}[!htb]
		\centering
		\renewcommand{\thesubfigure}{\roman{subfigure}}
		\hspace*{-0.8cm}
		\begin{subfigure}[t]{4.5cm}
			\centering
			\includegraphics[scale=0.3]{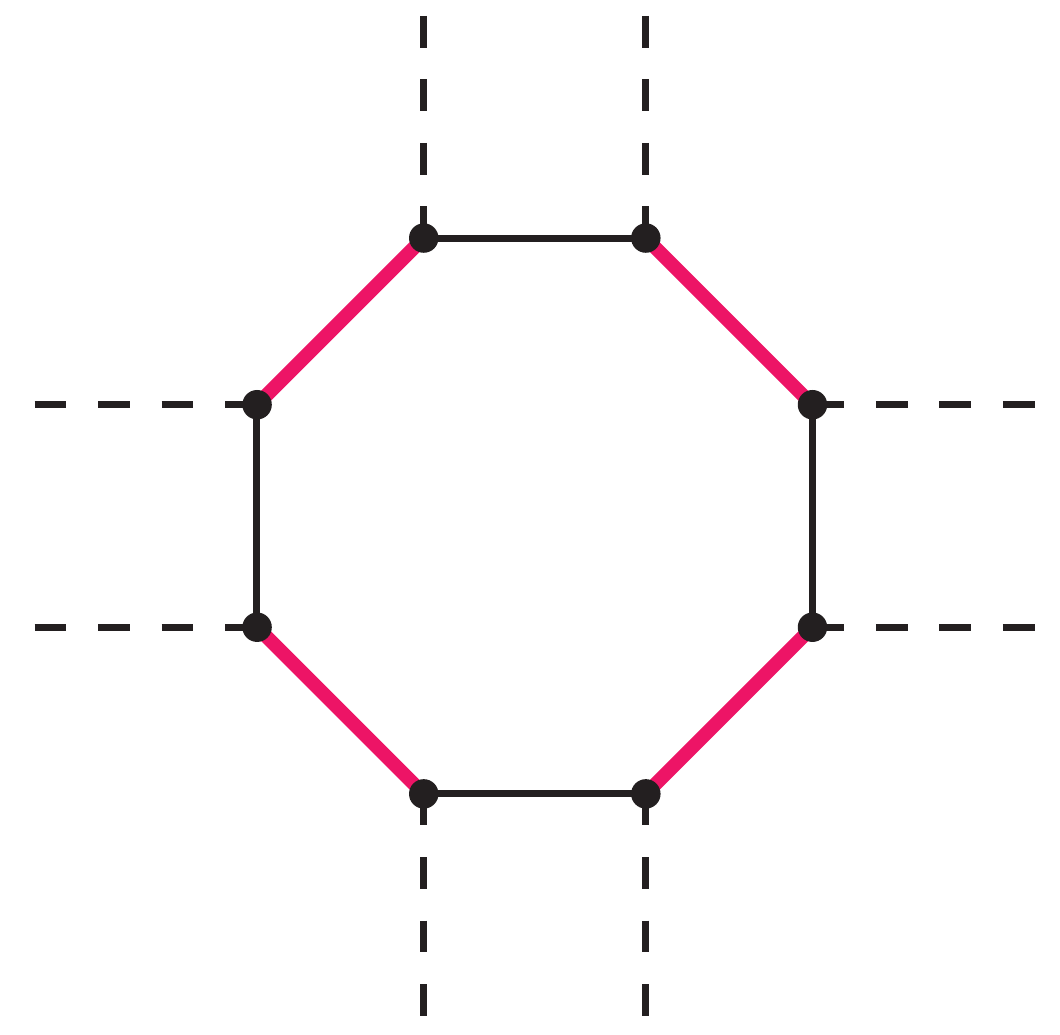}
			\caption{$\phi^8$}
			\label{subfig:phi8-4}
		\end{subfigure}
		\begin{subfigure}[t]{4.5cm}
			\centering
			\includegraphics[scale=0.44]{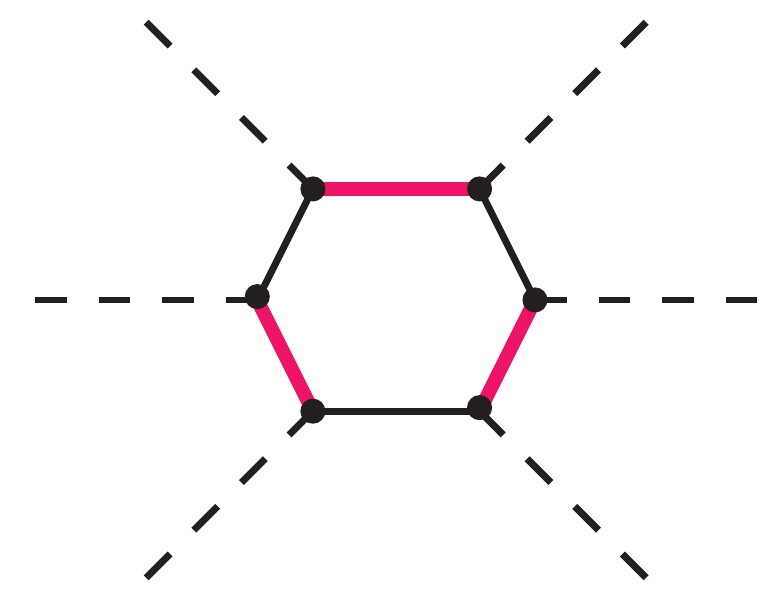}
			\caption{$\phi^6\mathcal{D}^2$}
			\label{subfig:phi6d2-4}
		\end{subfigure}
		\begin{subfigure}[t]{3.7cm}
			\centering
			\includegraphics[scale=0.3]{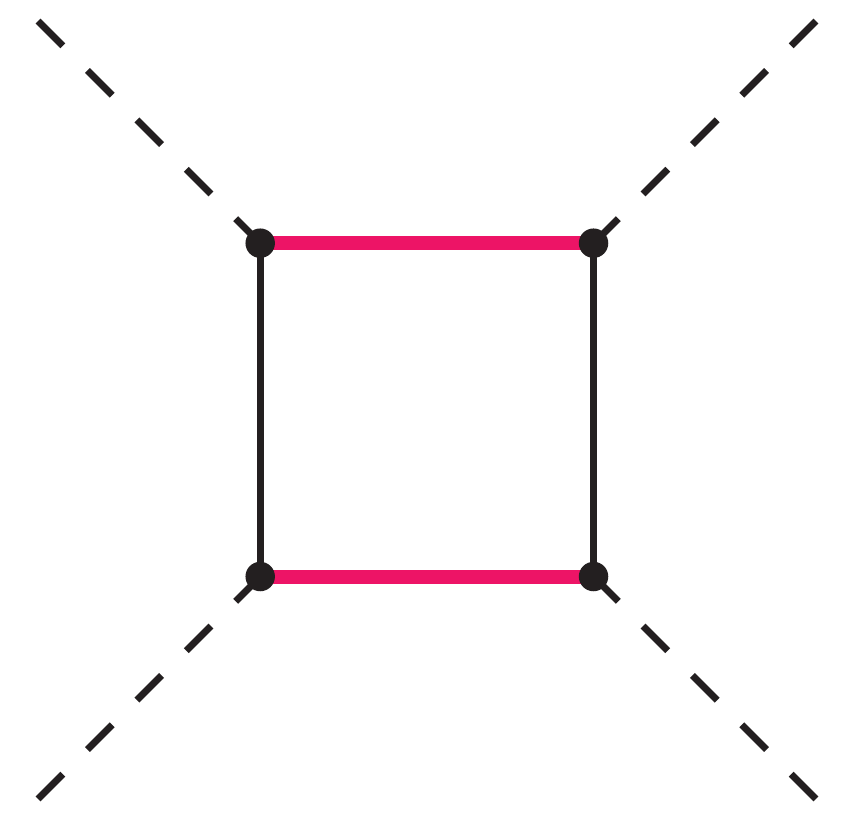}
			\caption{$\phi^4\mathcal{D}^4$}
			\label{subfig:phi4d4-3}
		\end{subfigure}
		\caption{\sf Schematic diagrams revealing light-heavy fermion mixing incorporated within (i) $ \phi^8 $, (ii) $ \phi^6\mathcal{D}^2 $, and (iii) $ \phi^4\mathcal{D}^4 $ classes of SMEFT operators.}
		\label{fig:phi-only-light-heavy}
\end{figure}
Fig.\,\ref{fig:phi-only-light-heavy} contains one-loop schematic diagrams corresponding to the $\phi^8$, $\phi^6\mathcal{D}^2$ and $\phi^4\mathcal{D}^4$ operator classes, that exhibit mixing between a light (SM) and a heavy fermion. The constituent vertex for each diagram is \hyperlink{vertex-4}{$V{}_4$}. The heavy field quantum numbers can be uniquely determined based on the choice of the light (SM) fermion as illustrated in Table~\ref{table:V4-quantum-numbers}.
\\
\hrule
\subsubsection*{\centering$ \boxed{\,\,\,\,\,\text{\texttt{Heavy-heavy mixing}}\,\,\,\,\,}$}

\begin{figure}[!htb]
	\centering
	\renewcommand{\thesubfigure}{\roman{subfigure}}
	\hspace*{-0.8cm}
	\begin{subfigure}[t]{4.5cm}
		\centering
		\includegraphics[scale=0.3]{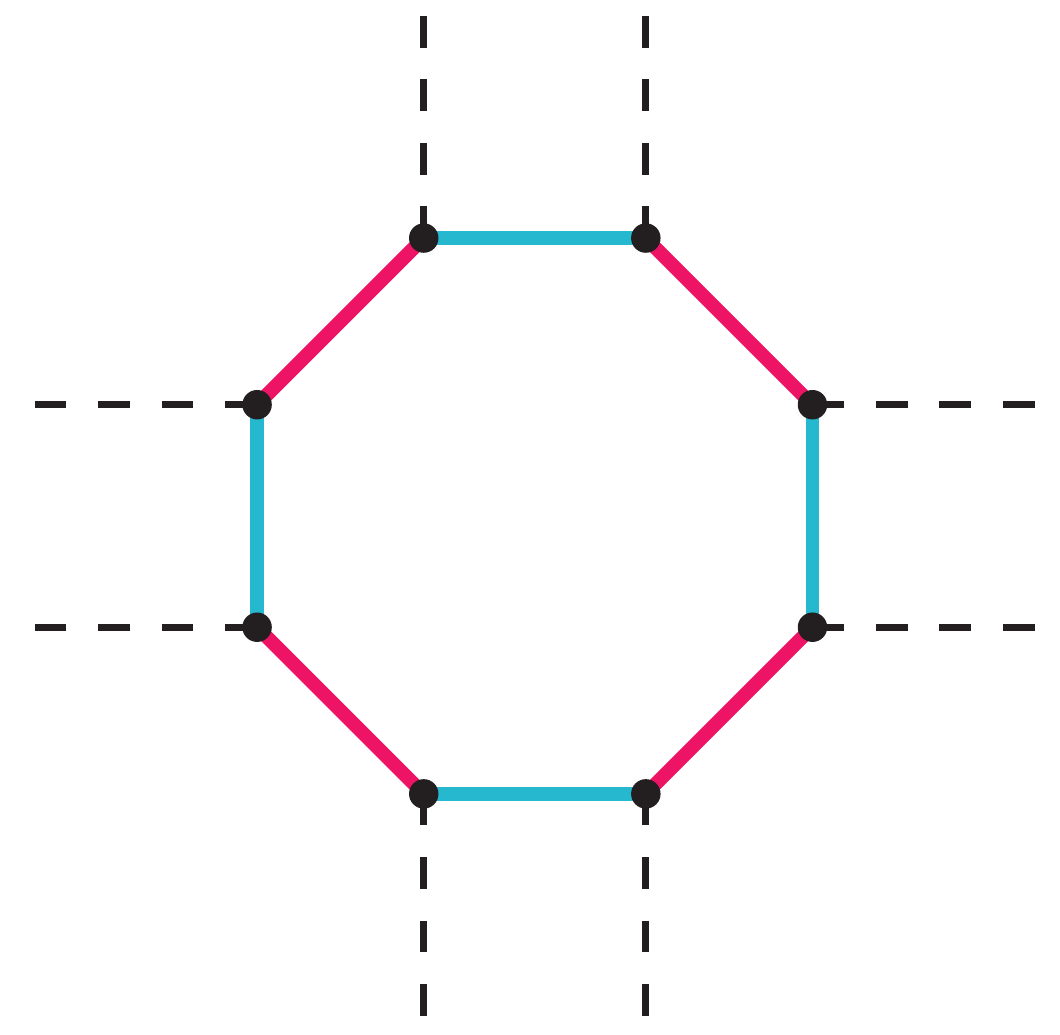}
		\caption{$\phi^8$}
		\label{subfig:phi8-5}
	\end{subfigure}
	\begin{subfigure}[t]{4.5cm}
		\centering
		\includegraphics[scale=0.44]{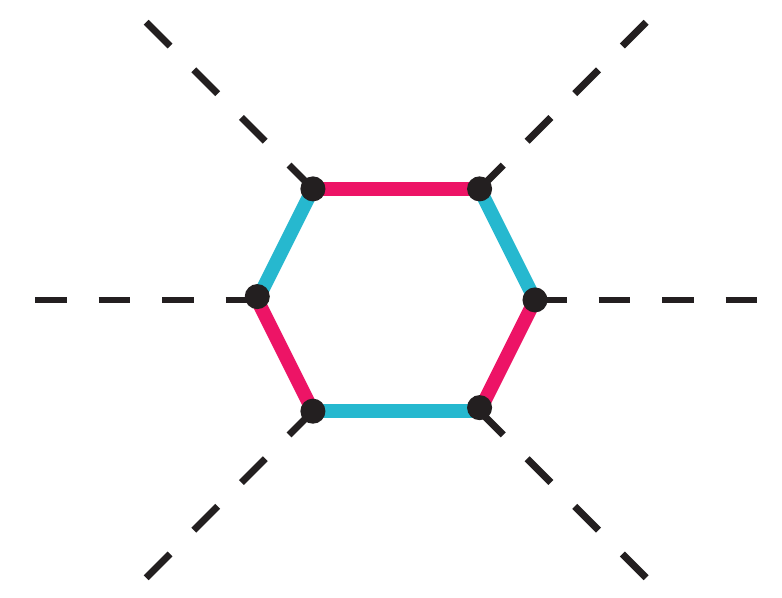}
		\caption{$\phi^6\mathcal{D}^2$}
		\label{subfig:phi6d2-5}
	\end{subfigure}
	\begin{subfigure}[t]{3.7cm}
		\centering
		\includegraphics[scale=0.3]{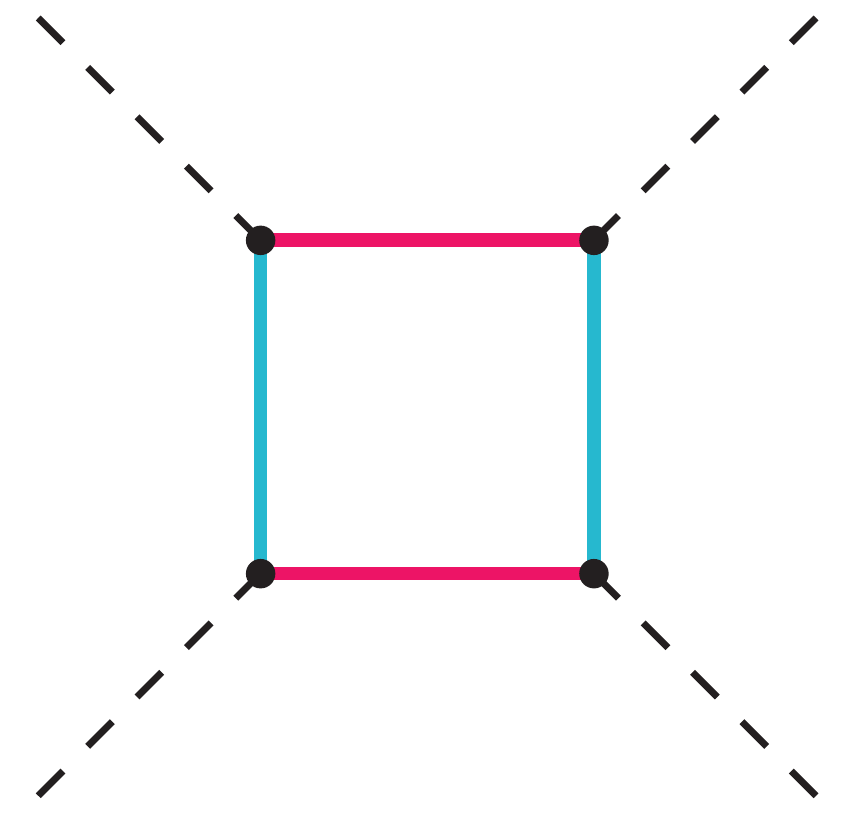}
		\caption{$\phi^4\mathcal{D}^4$}
		\label{subfig:phi4d4-4}
	\end{subfigure}
	\caption{\sf Schematic diagrams revealing heavy-heavy fermion mixing incorporated within (i) $ \phi^8 $, (ii) $ \phi^6\mathcal{D}^2 $, and (iii) $ \phi^4\mathcal{D}^4 $ classes of SMEFT operators.}
	\label{fig:phi-only-heavy-heavy}
\end{figure}

Fig.\,\ref{fig:phi-only-heavy-heavy} contains one-loop schematic diagrams corresponding to the $\phi^8$, $\phi^6\mathcal{D}^2$ and $\phi^4\mathcal{D}^4$ operator classes, that exhibit mixing between two heavy fermions. The constituent vertex for each diagram is \hyperlink{vertex-5}{$V{}_5$}. This is a broader generalization of the case depicted in Fig.\,\ref{fig:phi-only-light-heavy} and accomodates a wide variety of heavy fermion $\Psi_{1,2}$ extensions of the SM whose quantum numbers $(R_{C_{1,2}},R_{L_{1,2}},Y_{1,2})$ must satisfy Eq.~\eqref{eq:vertex-5-rel}. There have been ample surveys and discussions around extensions of the SM incorporating such heavy vector-like fermion (VLF) pairs \cite{Ellis:2014dza,Ishiwata:2013gma,Ishiwata:2015cga,Branco:1986my}. The embedding of VLF pairs within D8 SMEFT operators hints at the significance of these operators in the study of associated phenomenology.

\subsection{External states: Only $X_{\mu\nu}$}

The complete list of independent operators of the $X^4$ class have been catalogued in Table~\ref{tab:smeft8class_4}. It is noteworthy that unlike its D6 counterpart, i.e., the $X^3$ class which only contains 4 operators, the $X^4$ class consists of a wide variety of operators made up of the SM field strength tensors $G^A_{\mu\nu}$, $W^I_{\mu\nu}$, $B_{\mu\nu}$ as well as their duals. This subdivision has been vividly elucidated in Table~\ref{tab:smeft8class_4}, through proper nomenclature of the operators based on their constituents. Since, we have not taken into account SM extensions with additional gauge bosons, we only come across loop-level diagrams in this case.  
\begin{figure}[!htb]
	\centering
	\renewcommand{\thesubfigure}{\roman{subfigure}}
	\hspace*{-0.5cm}\begin{subfigure}[t]{4.2cm}
		\centering
		\includegraphics[scale=0.35]{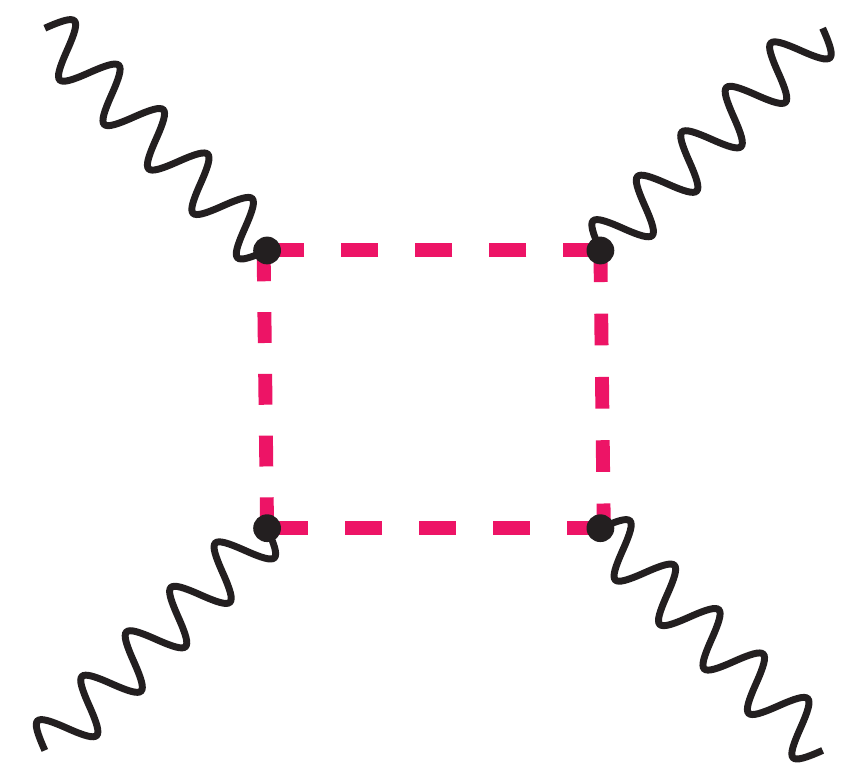}
		\caption{\textsf{Heavy loop (scalar)}}
		\label{subfig:x4-1}
	\end{subfigure}
	\begin{subfigure}[t]{4.2cm}
		\centering
		\includegraphics[scale=0.35]{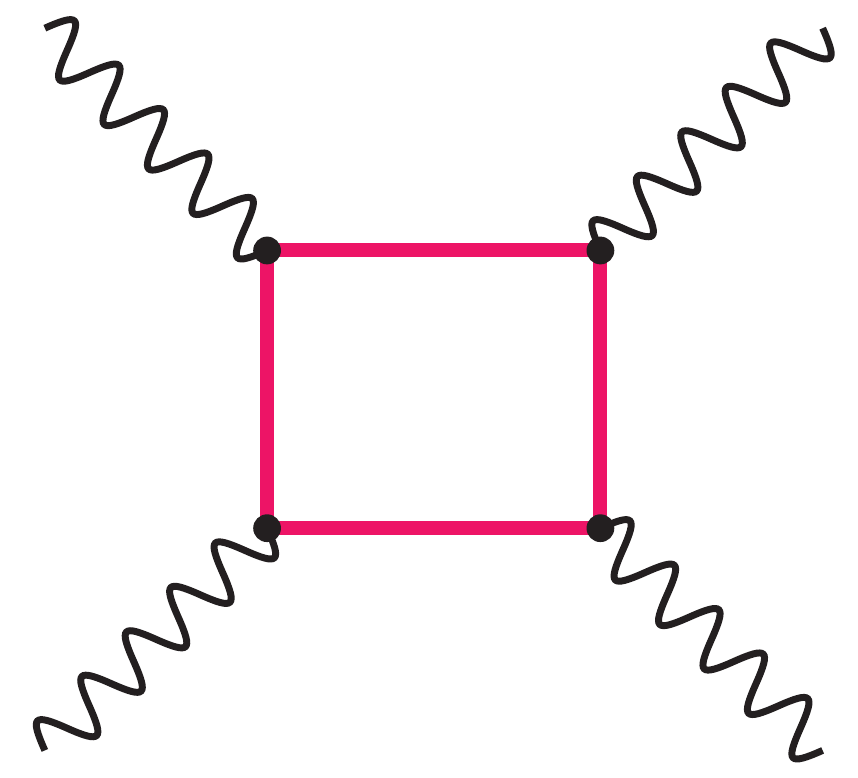}
		\caption{\textsf{Heavy loop (fermion)}}
		\label{subfig:x4-2}
	\end{subfigure}
	\begin{subfigure}[t]{4.2cm}
		\centering
		\includegraphics[scale=0.35]{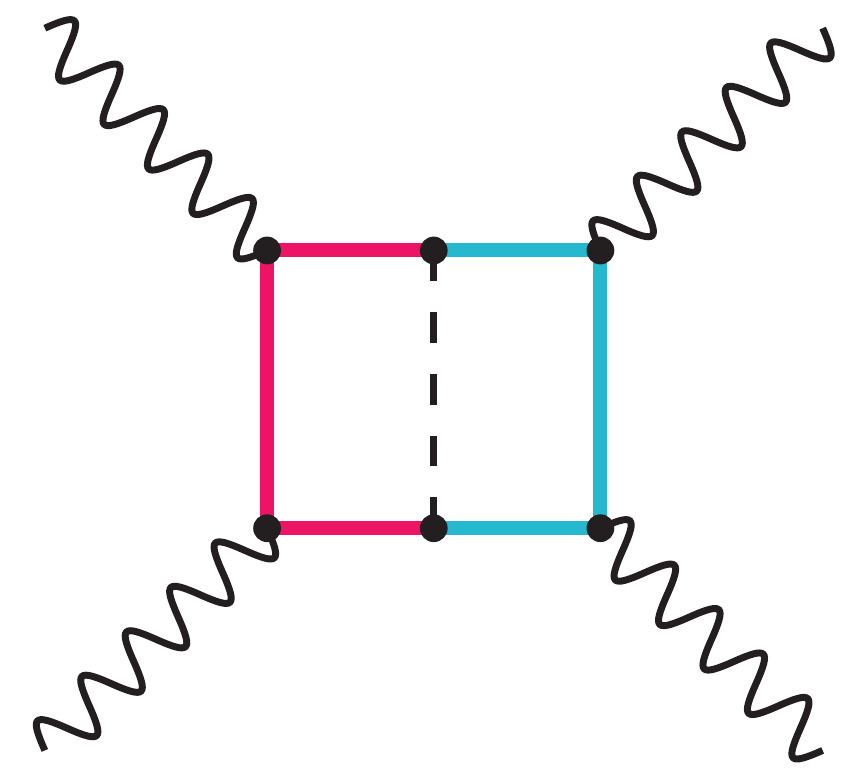}
		\caption{\textsf{Heavy-heavy mixing}}
		\label{subfig:x4-4}
	\end{subfigure}
	\caption{\sf (i),  (ii) One-loop and (iii) two-loop schematic diagrams revealing heavy field propagators enveloped within $ X^4$ class of SMEFT operators.}
	\label{fig:x4-only}
\end{figure}

Fig.\,\ref{fig:x4-only} shows one-loop as well as two-loop diagrams revealing scalar as well as fermion propagators. The inclusion of two-loop diagrams is necessitated by the operators with an odd number of dual field strength tensors ($\tilde{X}_{\mu\nu}$), which contain the signature of CP-violation. These can directly be compared with the contents of Fig.~\ref{fig:phi2x2-x3-unfolding}. The allowed heavy field quantum numbers, for each case in Fig.\,\ref{fig:x4-only}, can be ascertained as follows:

\begin{enumerate}
	\item Fig.\,\ref{subfig:x4-1}: composed of the vertex \hyperlink{vertex-6}{$V{}_6$}. As described in section~\ref{sec:vertices}, the heavy scalar quantum numbers $(R_C, R_L, Y)$ can be arbitrary, but subject to one or more of the following constraints based on the $X_{\mu\nu}$'s present within the operator:
	\begin{eqnarray}\label{eq:x4-vertex-constraint}
		X_{\mu\nu} \equiv B_{\mu\nu} \Rightarrow Y \neq 0, \hspace{0.3cm} X_{\mu\nu} \equiv W^I_{\mu\nu} \Rightarrow R_L \neq 1, \hspace{0.3cm} X_{\mu\nu} \equiv G^A_{\mu\nu} \Rightarrow R_C \neq 1.
	\end{eqnarray}

	\item Fig.\,\ref{subfig:x4-2}: composed of the vertex \hyperlink{vertex-7}{$V{}_7$}. This is similar to Fig.\,\ref{subfig:x4-1} but with the scalar loop replaced by a fermion loop. Once again, there is a freedom with respect to quantum number assignment except for a constraint in the form of Eq.~\eqref{eq:x4-vertex-constraint} depending on the various $X_{\mu\nu}$'s involved in the operator. Since this diagram covers all possible heavy fermion extensions of the SM, we have not separately considered a diagram displaying light-heavy mixing between an SM fermion and a heavy fermion, which is more restrictive to the choice of the heavy fermion.
	
	\item Fig.\,\ref{subfig:x4-4}: composed of the vertices - \{\hyperlink{vertex-5}{$V{}_5$}, \hyperlink{vertex-7}{$V{}_7$}\}. This is a two-loop diagram exhibiting mixing between two heavy fermions and the SM scalar. Similar to Figs.\,\ref{fig:x3-1} and \ref{fig:x3-2}, this diagram is necessary to accommodate SM extensions with a CP-violating signature. The heavy fermion quantum numbers must satisfy both Eqs.~\eqref{eq:vertex-5-rel} and \eqref{eq:x4-vertex-constraint} depending on the field strength tensor(s) in contact.   
\end{enumerate}

\subsection{External states: $\phi$ and $X_{\mu\nu}$}

This category consists of operators belonging to the $\phi^2X^3$, $\phi^4X^2$,  $\phi^2X^2\mathcal{D}^2$ and $\phi^4X\mathcal{D}^2$ classes. The complete list of independent operators has been catalogued in Table~\ref{tab:smeft8class_5_6_7_8}. 

\subsubsection*{\centering$ \boxed{\,\,\,\,\,\text{\texttt{Heavy-loop}}\,\,\,\,\,}$}

\begin{figure}[!htb]
	\centering
	\renewcommand{\thesubfigure}{\roman{subfigure}}
	\hspace*{-1.2cm}
	\begin{subfigure}[t]{3.7cm}
		\centering
		\includegraphics[scale=0.3]{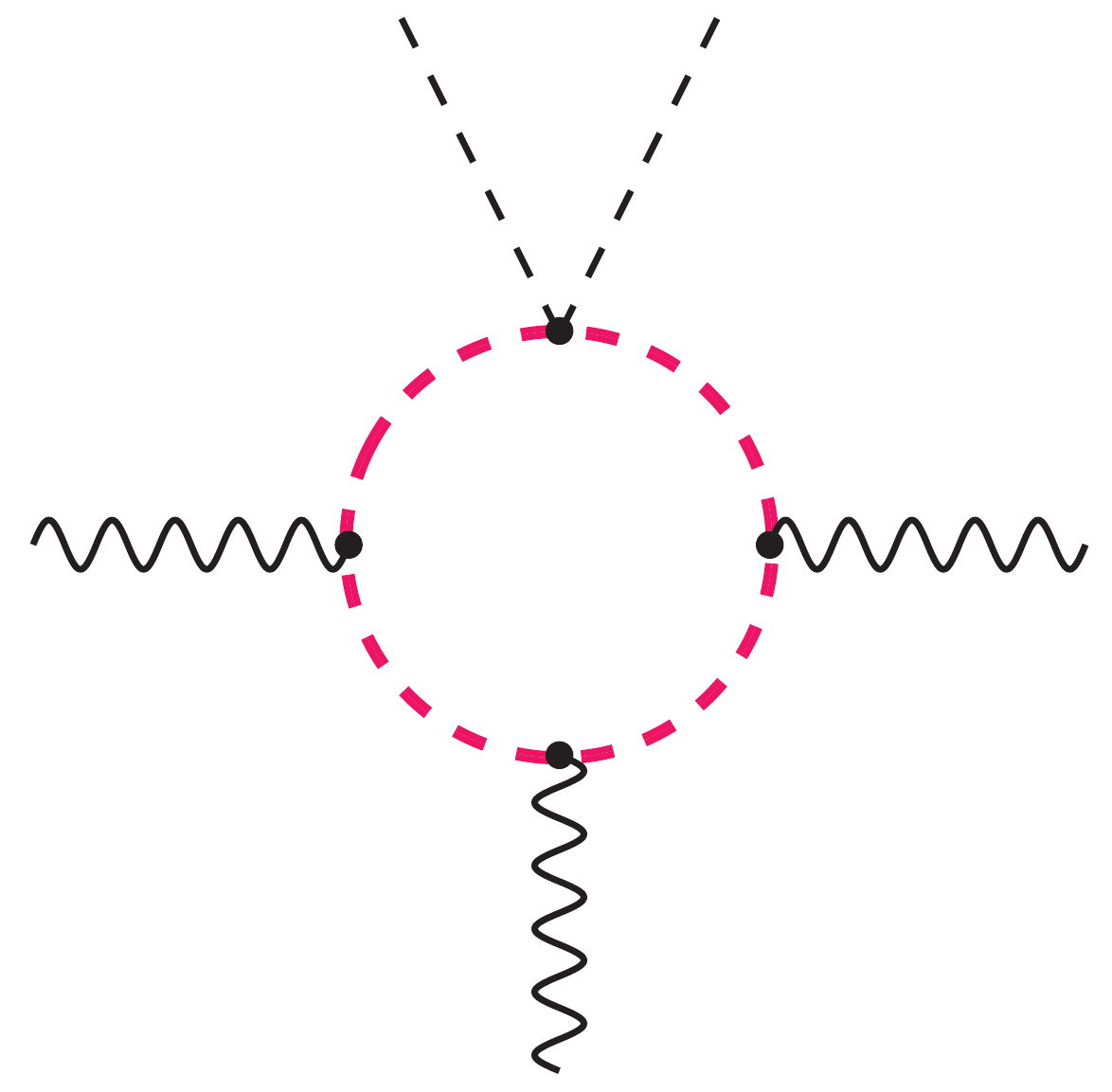}
		\caption{$\phi^2X^3$}
		\label{subfig:x3p2-1}
	\end{subfigure}
	\begin{subfigure}[t]{3.7cm}
		\centering
		\includegraphics[scale=0.3]{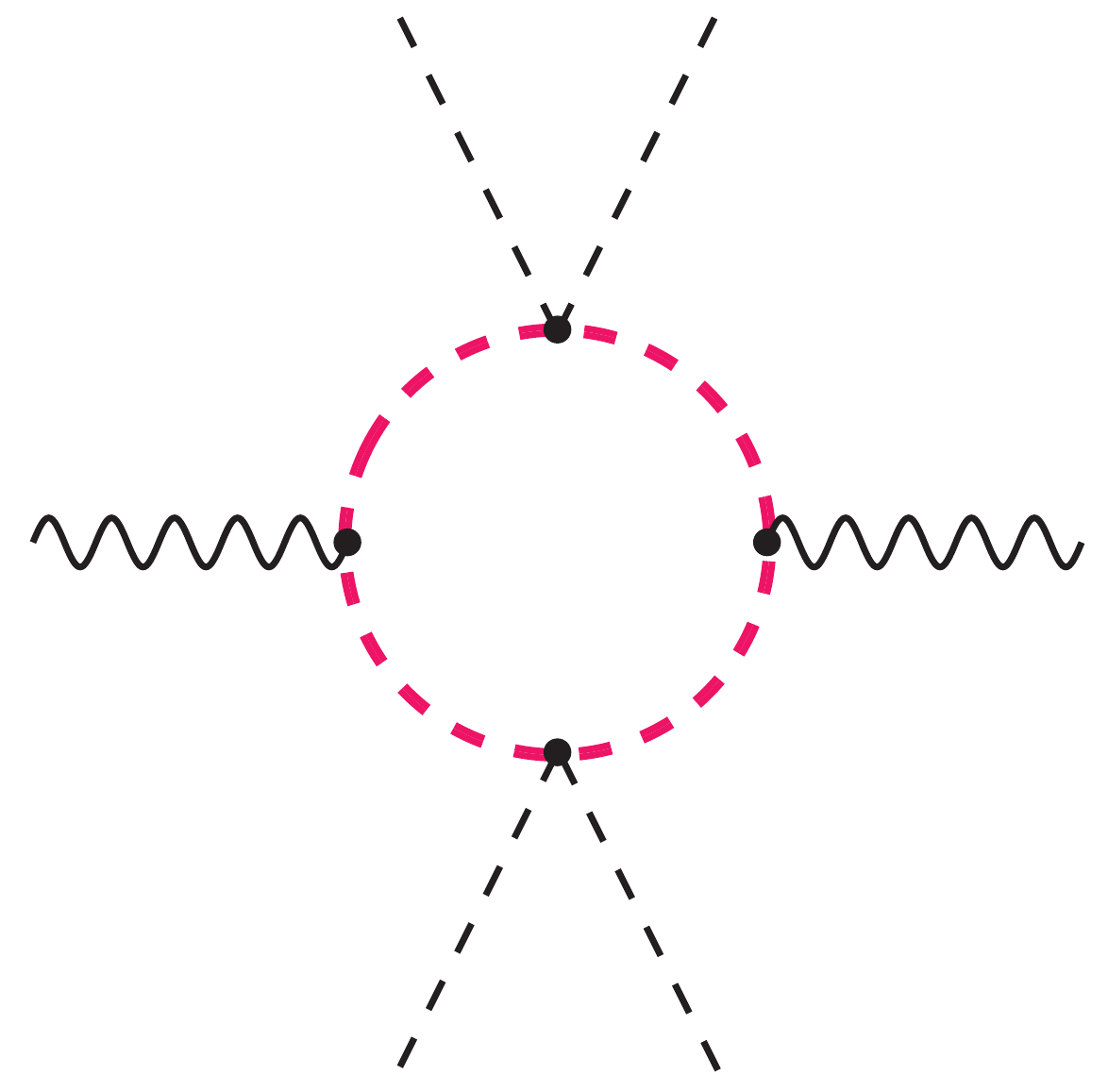}
		\caption{$\phi^4X^2$}
		\label{subfig:x2p4-1}
	\end{subfigure}
	\begin{subfigure}[t]{3.7cm}
		\centering
		\includegraphics[scale=0.4]{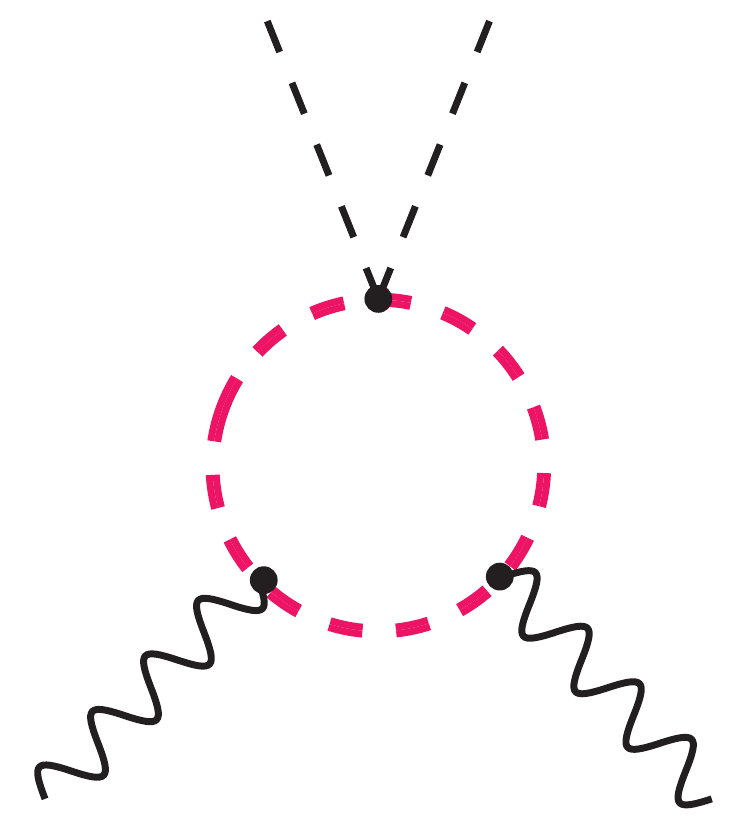}
		\caption{$\phi^2X^2\mathcal{D}^2$}
		\label{subfig:x2phi2d2-1}
	\end{subfigure}
	\begin{subfigure}[t]{3.7cm}
		\centering
		\includegraphics[scale=0.4]{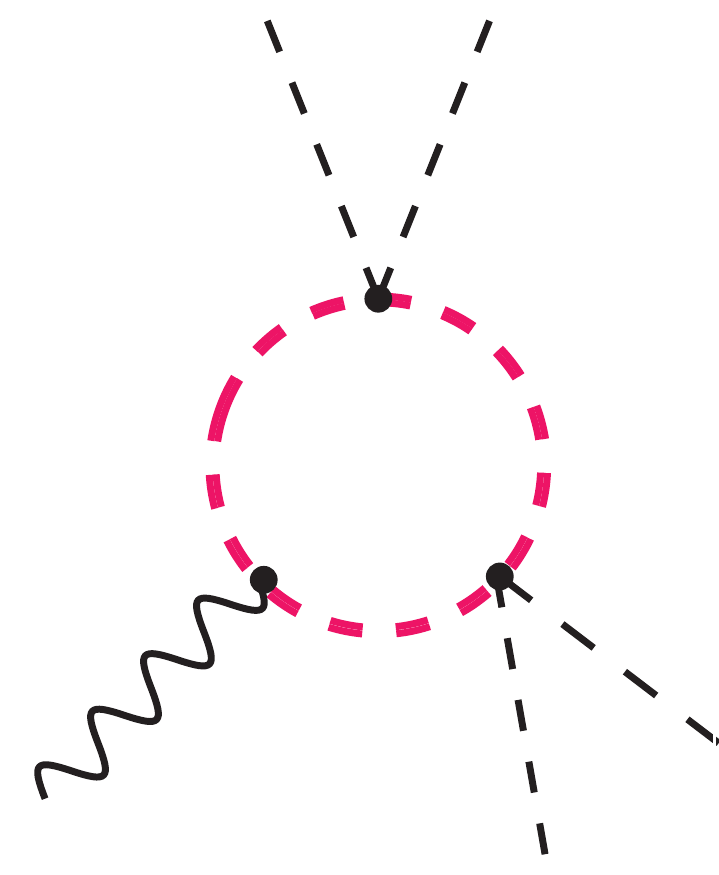}
		\caption{$\phi^4X\mathcal{D}^2$}
		\label{subfig:xphi4d2-1}
	\end{subfigure}
	\caption{\sf One-loop schematic diagrams revealing heavy field propagators enveloped within (i) $ \phi^2X^3 $, (ii) $ \phi^4X^2 $, (iii) $ \phi^2X^2\mathcal{D}^2 $, and (iv) $ \phi^4X\mathcal{D}^2 $ classes of SMEFT operators.}
	\label{fig:phi-X-heavy-loop}
\end{figure}

Fig.\,\ref{fig:phi-X-heavy-loop} contains one-loop schematic diagrams for each of the four operator classes. In each case, the entire loop is composed of a distinct heavy scalar. The constituent vertices for each diagram are \{\hyperlink{vertex-3}{$V{}_3$}, \hyperlink{vertex-6}{$V{}_6$}\}. While \hyperlink{vertex-3}{$V{}_3$} permits all heavy scalars with arbitrary quantum numbers $\Phi\in(R_C,R_L,Y)$, the presence of the field strength tensors filters out some of the possibilities per Eq.~\eqref{eq:x4-vertex-constraint}.
\\
\hrule
\subsubsection*{\centering$ \boxed{\,\,\,\,\,\text{\texttt{Light-heavy mixing}}\,\,\,\,\,}$}

\begin{figure}[!htb]
	\centering
	\renewcommand{\thesubfigure}{\roman{subfigure}}
	\hspace*{-1.6cm}
	\begin{subfigure}[t]{7.7cm}
		\centering
		\includegraphics[scale=0.37]{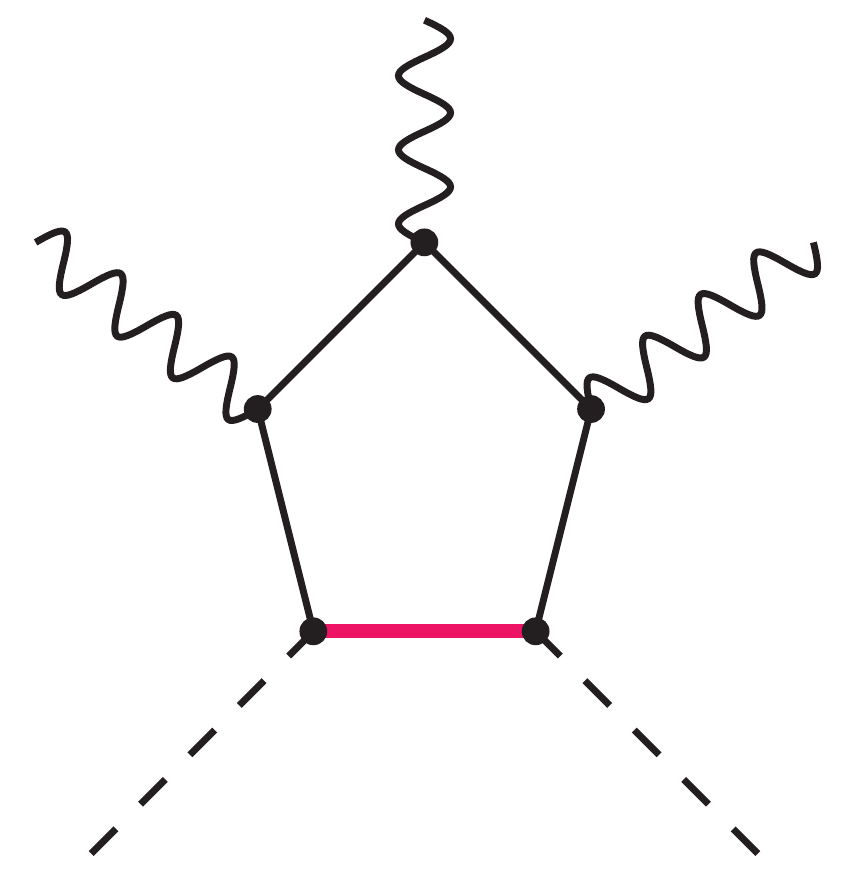}
		\includegraphics[scale=0.36]{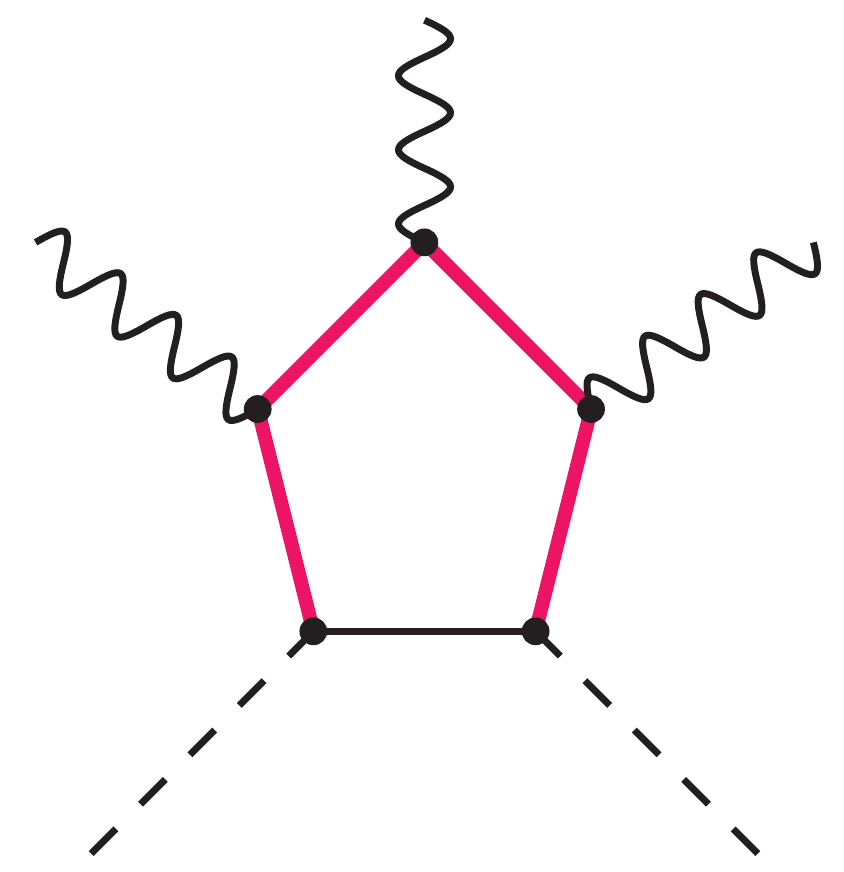}
		\caption{\textsf{$\phi^2X^3$}}
		\label{subfig:x3p2-2}
	\end{subfigure}
	\begin{subfigure}[t]{8cm}
		\centering
		\includegraphics[scale=0.27]{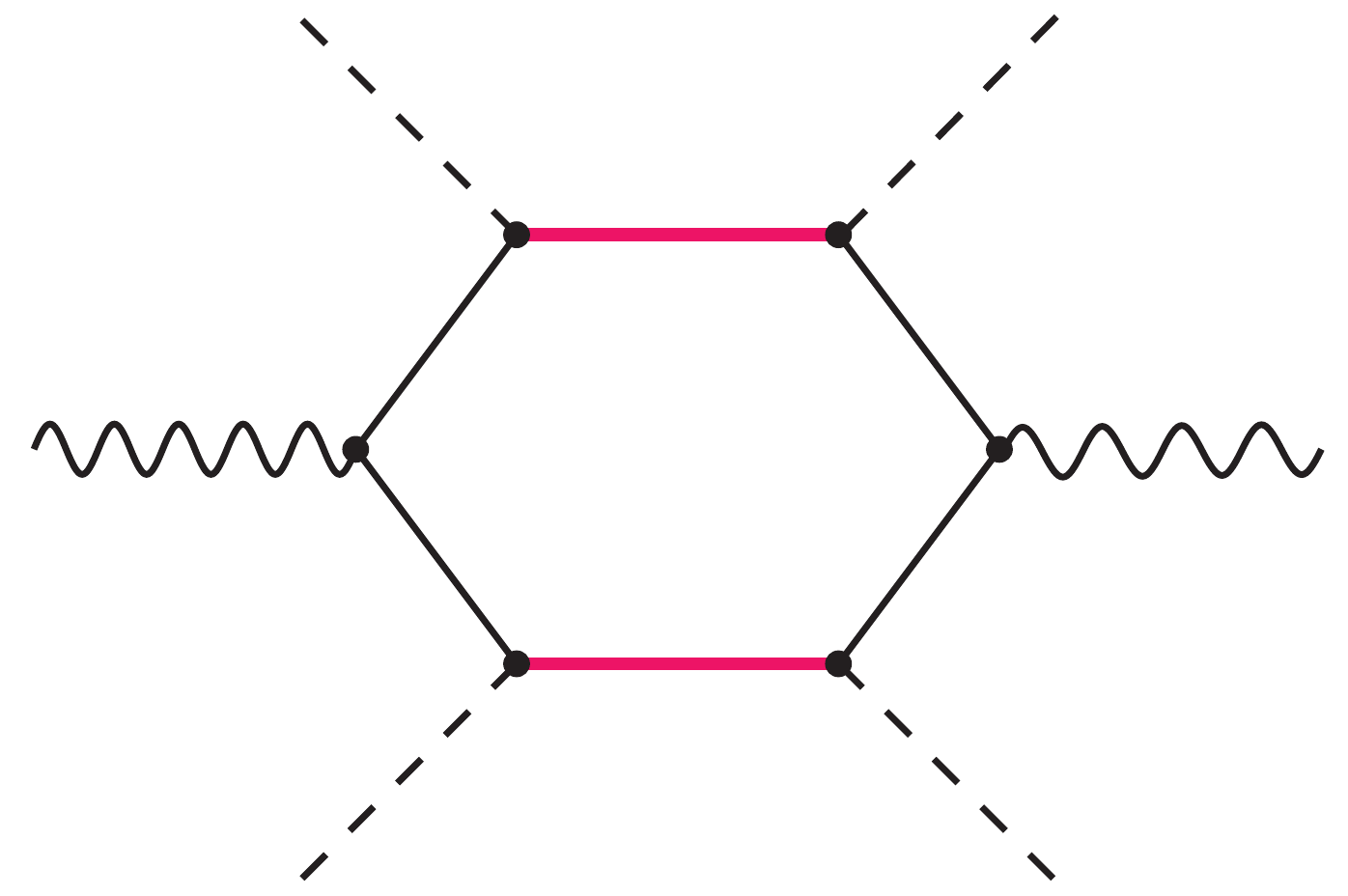}
		\includegraphics[scale=0.27]{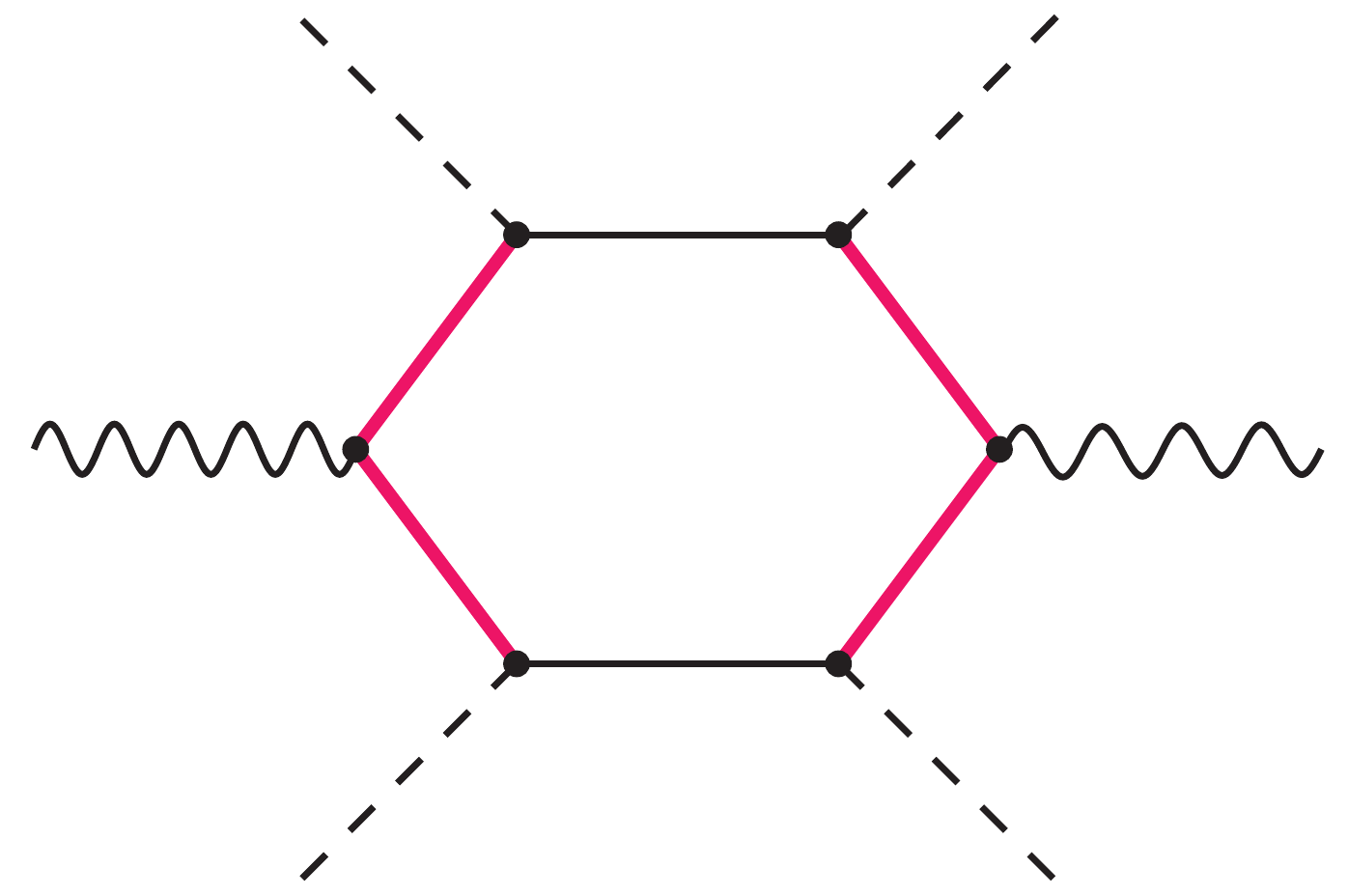}
		\caption{$\phi^4X^2$}
		\label{subfig:x2p4-2}
	\end{subfigure}
	\begin{subfigure}[t]{3.7cm}
		\centering
		\includegraphics[scale=0.35]{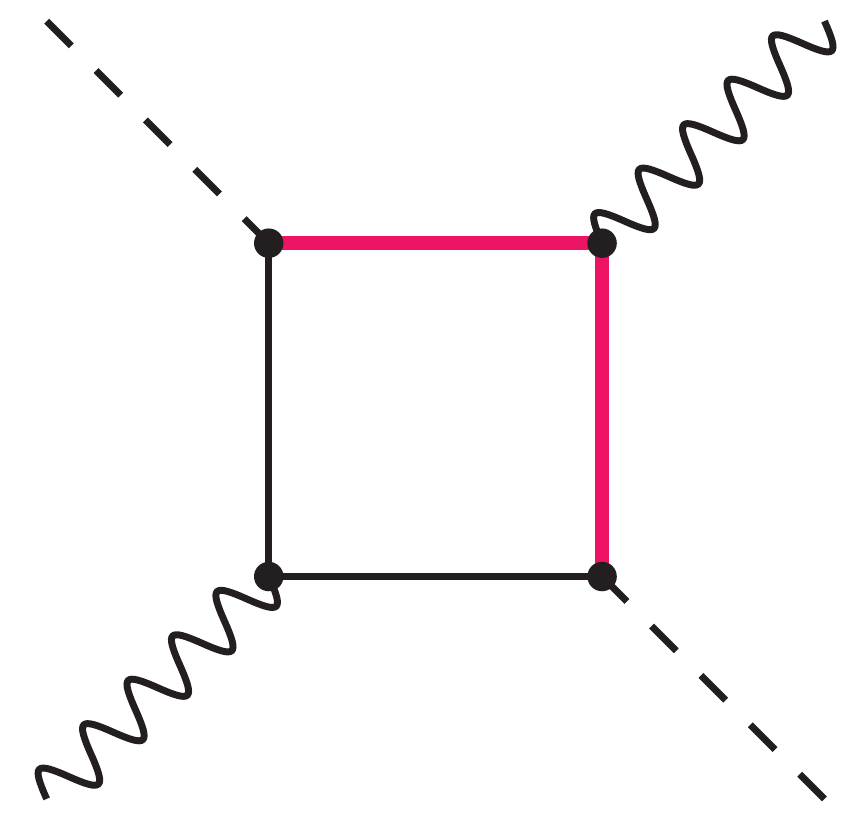}
		\label{subfig:x2phi2d2-2}
		\caption{$\phi^2X^2\mathcal{D}^2$}
	\end{subfigure}
	\begin{subfigure}[t]{7.7cm}
		\centering
		\includegraphics[scale=0.37]{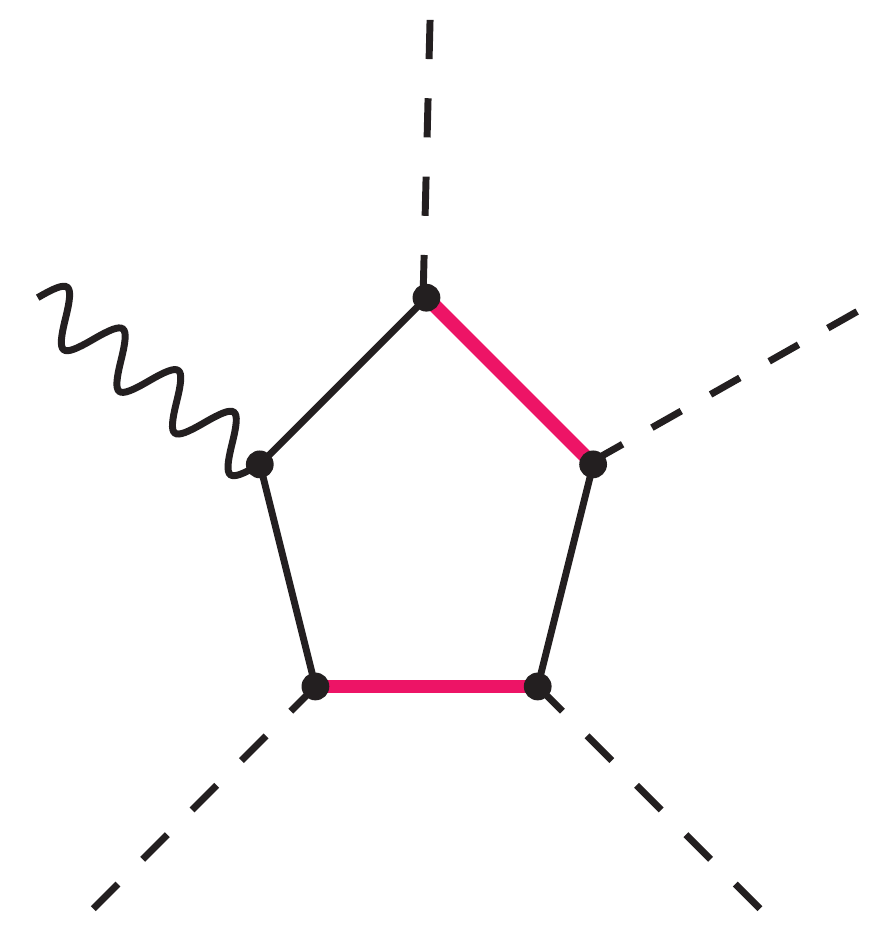}
		\includegraphics[scale=0.37]{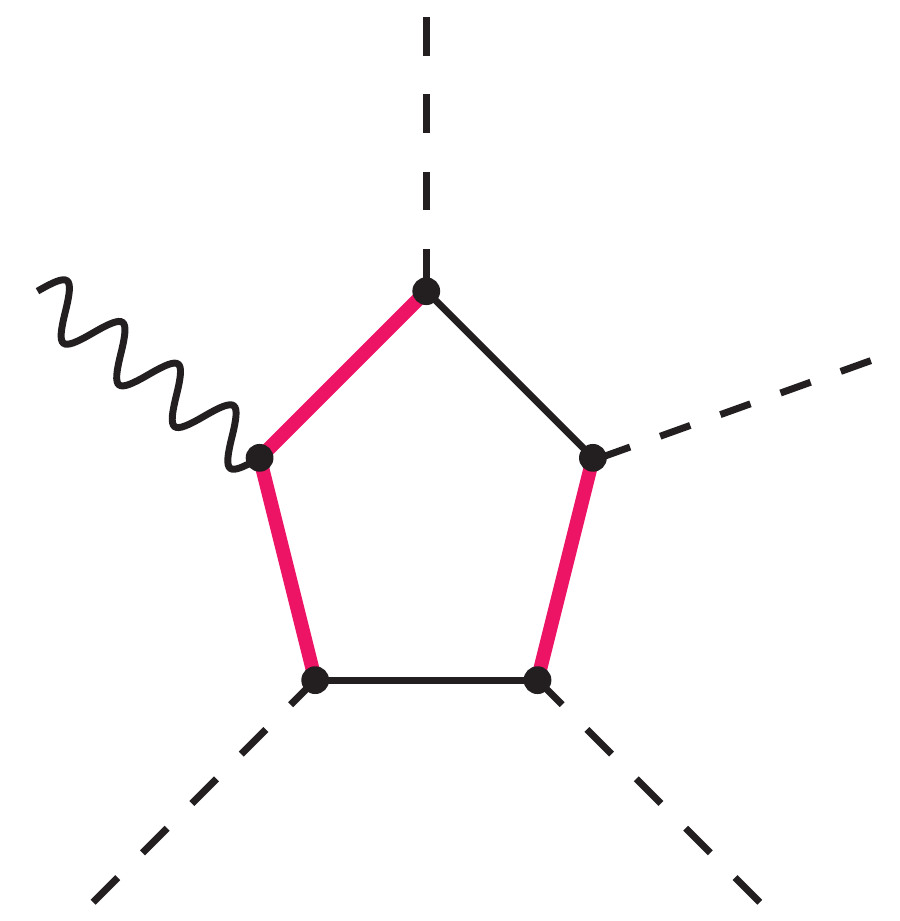}
		\caption{$\phi^4X\mathcal{D}^2$}
		\label{subfig:xphi4d2-2}
	\end{subfigure}
	\caption{\sf Schematic diagrams revealing light-heavy fermion mixing incorporated within (i) $ \phi^2X^3 $, (ii) $ \phi^4X^2 $, (iii) $ \phi^2X^2\mathcal{D}^2 $, and (iv) $ \phi^4X\mathcal{D}^2 $ classes of SMEFT operators.}
	\label{fig:phi-X-light-heavy}
\end{figure}

\noindent Fig.\,\ref{fig:phi-X-light-heavy} contains one-loop schematic diagrams corresponding to the $\phi^2X^3$, $\phi^4X^2$,  $\phi^2X^2\mathcal{D}^2$ and $\phi^4X\mathcal{D}^2$ operator classes, that exhibit mixing between a light (SM) and a heavy fermion. The constituent vertices for each case are \{\hyperlink{vertex-4}{$V{}_4$},  \hyperlink{vertex-7}{$V{}_7$}\} as well as interactions between the SM fermions and SM field strength tensors. The heavy field quantum numbers can be uniquely determined by first fixing the light fermion within the loop, which allows us to use the results of Table~\ref{table:V4-quantum-numbers}, followed by filtering them out further by imposing Eq.~\eqref{eq:x4-vertex-constraint} at the vertices where the fermions come in contact with the field strength tensors, e.g., 

\begin{itemize}
	\item If we consider the operator: {\small$Q_{W^2BH^2}^{(1)}$  $\equiv$  $\epsilon^{IJK} (H^\dag \tau^I H) B_{\mu}^{\,\nu} W_{\nu}^{J\rho} W_{\rho}^{K\mu}$} of the $\phi^2X^3$ class and examine the first sub-figure within Fig.\;\ref{subfig:x3p2-2}, we see that the light fermion is in direct contact with the $SU(2)_L$ as well as the $U(1)_Y$  field strength tensors. This filters out those SM fermions that transform trivially under $SU(2)_L$, i.e, $u_R \in (3,1,\frac{2}{3})$, $d_R \in (3,1,-\frac{1}{3})$, and $e_R \in (1,1,-1)$. Therefore, the fields that participate in light-heavy mixing in the loop are:
	
	\begin{eqnarray}
		(i)\quad \psi &= & l_L \in (1,2,-\frac{1}{2}),\,\, \Psi \in \{(1,1,0),\,(1,3,0),\,(1,1,1),\,(1,3,1)\}; \nonumber\\
		(ii)\quad \psi &=& q_L \in (3,2,\frac{1}{6}),\,\,\,\,\,\, \Psi \in \{(\bar{3},1,-\frac{2}{3}),\,(\bar{3},3,-\frac{2}{3}),\,(\bar{3},1,\frac{1}{3}),\,(\bar{3},3,\frac{1}{3})\}.
	\end{eqnarray}

	On the other hand, focussing on the second sub-figure in Fig.\;\ref{subfig:x3p2-2}, since the heavy fermion is in direct contact with the field strength tensors, there are no restrictions with respect to the selection of the SM fermion but the heavy field choices are curtailed and fields that are either $SU(2)_L$ singlets and (or) carry a zero hypercharge are excluded. The permitted combinations of light and heavy fields in the loop are:
	\begin{eqnarray}
		(i)\quad \psi &= & l_L \in (1,2,-\frac{1}{2}),\hspace{0.5cm} \Psi \in \{(1,3,1)\}; \nonumber\\
		(ii)\quad \psi &=& q_L \in (3,2,\frac{1}{6}),\hspace{0.75cm} \Psi \in \{(\bar{3},3,-\frac{2}{3}),\,(\bar{3},3,\frac{1}{3})\}; \nonumber\\
		(iii)\quad \psi &=& e_R \in (1,1,-1),\hspace{0.5cm} \Psi \in \{(1,2,\frac{1}{2}),\,(1,2,\frac{3}{2})\}; \nonumber\\
		(iv)\quad \psi &=& u_R \in (3,1,\frac{2}{3}),\hspace{0.75cm} \Psi \in \{(\bar{3},2,-\frac{7}{6}),\,(\bar{3},2,-\frac{1}{6})\}; \nonumber\\
		(v)\quad \psi &=& d_R \in (3,1,-\frac{1}{3}),\hspace{0.5cm} \Psi \in \{(\bar{3},2,-\frac{1}{6}),\,(\bar{3},2,\frac{5}{6})\}.
	\end{eqnarray}

	While the two diagrams together encompass all the combinations listed in Table~\ref{table:V4-quantum-numbers}, such nuances are vital for establishing concrete relations between the effective operators and the heavy fields.
\end{itemize}

\hrule

\subsubsection*{\centering$ \boxed{\,\,\,\,\,\text{\texttt{Heavy-heavy mixing}}\,\,\,\,\,}$}

\begin{figure}[!htb]
		\centering
		\renewcommand{\thesubfigure}{\roman{subfigure}}
		\hspace{-0.6cm}
		\begin{subfigure}[t]{3.7cm}
			\centering
			\includegraphics[scale=0.37]{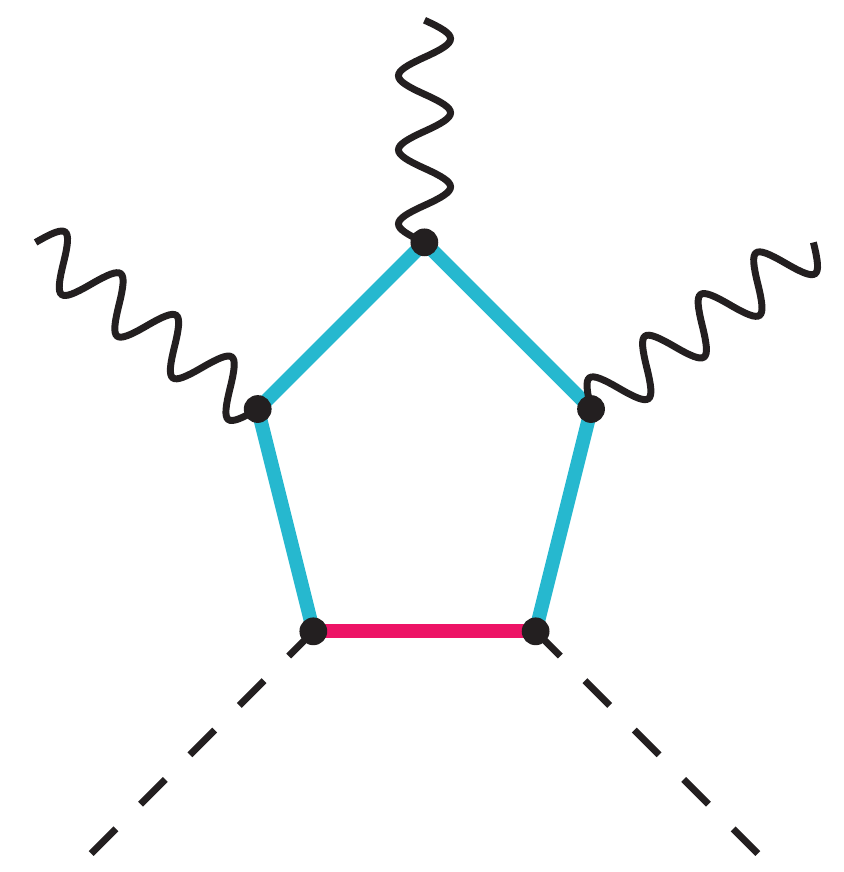}
			\caption{\textsf{$\phi^2X^3$}}
			\label{subfig:x3p2-3}
		\end{subfigure}
		\begin{subfigure}[t]{4cm}
			\centering
			\includegraphics[scale=0.3]{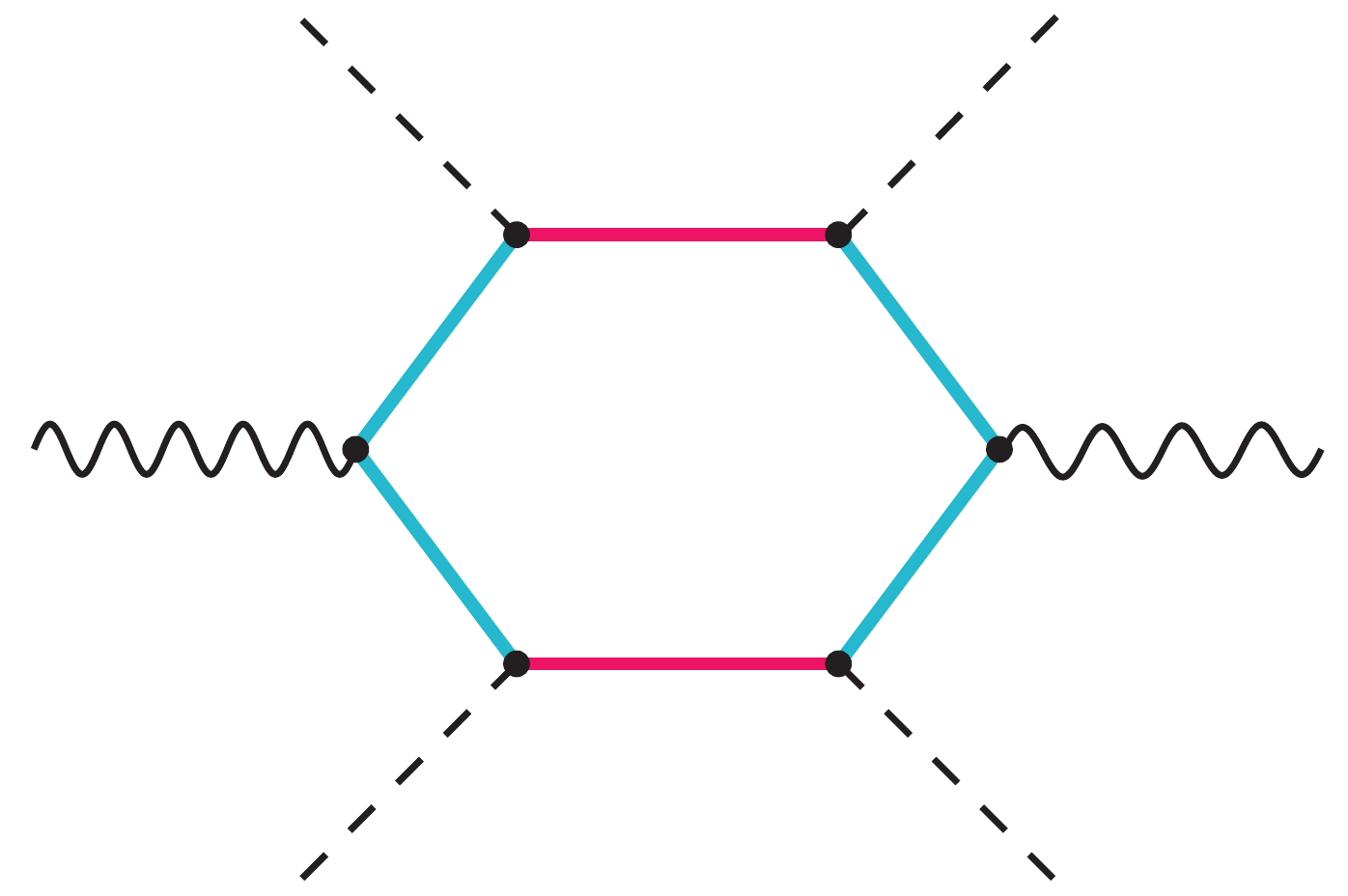}
			\caption{$\phi^4X^2$}
			\label{subfig:x2p4-3}
		\end{subfigure}
		\begin{subfigure}[t]{3.7cm}
			\centering
			\includegraphics[scale=0.35]{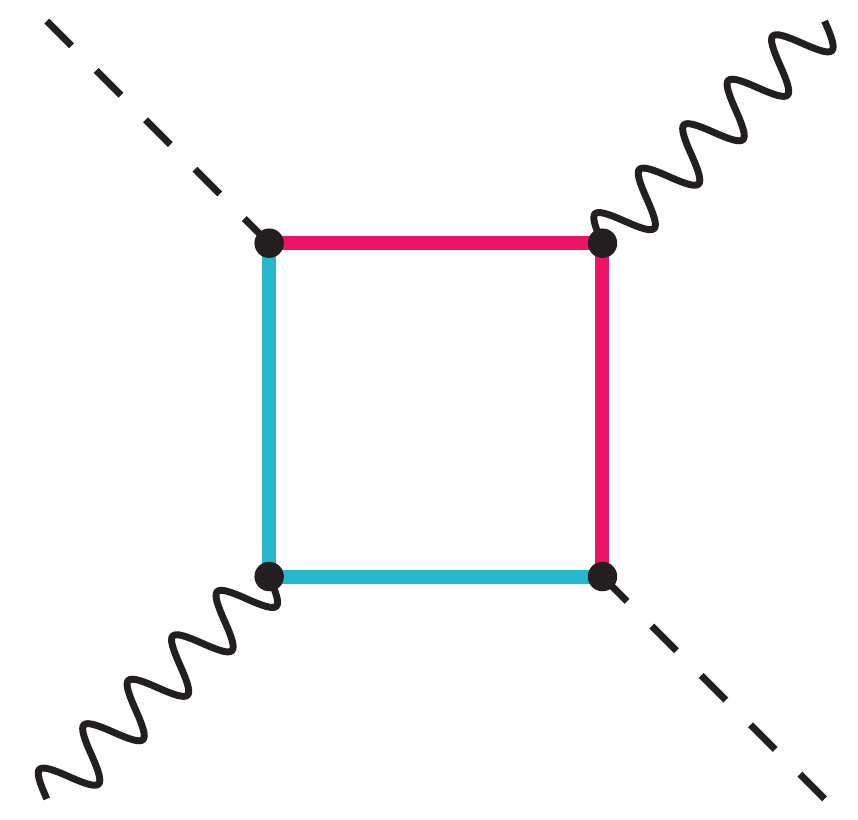}
			\label{subfig:x2phi2d2-3}
			\caption{$\phi^2X^2\mathcal{D}^2$}
		\end{subfigure}
		\begin{subfigure}[t]{3.7cm}
			\centering
			\includegraphics[scale=0.37]{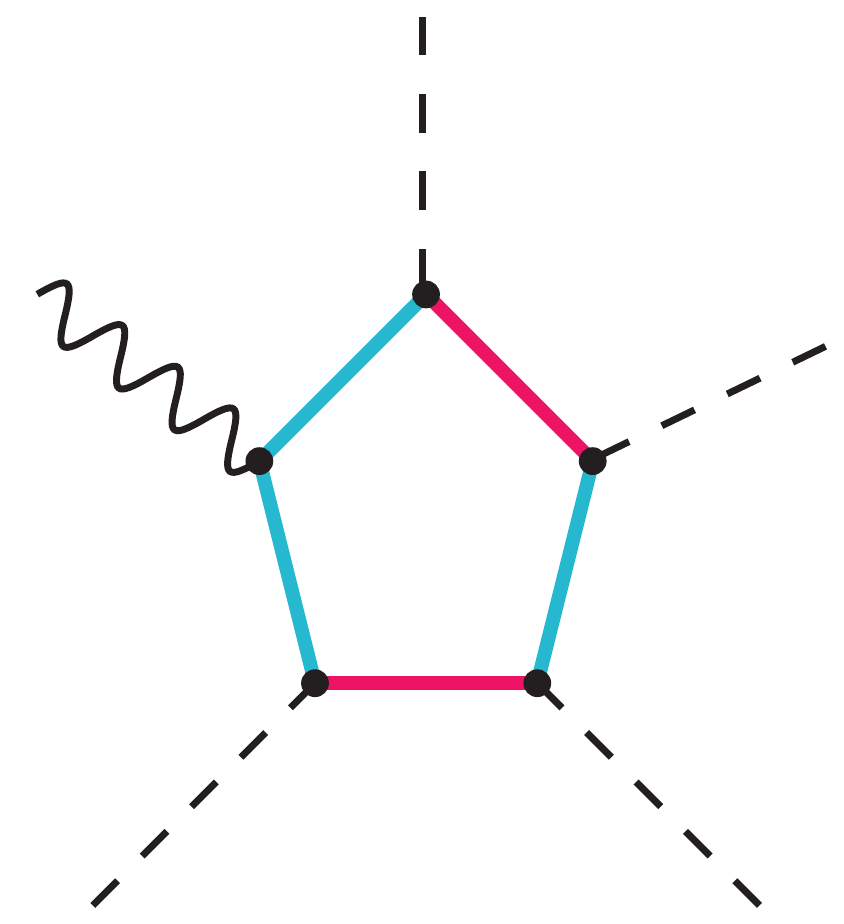}
			\caption{$\phi^4X\mathcal{D}^2$}
			\label{subfig:xphi4d2-3}
		\end{subfigure}
		\caption{\sf Schematic diagrams revealing heavy-heavy fermion mixing incorporated within (i) $ \phi^2X^3 $, (ii) $ \phi^4X^2 $, (iii) $ \phi^2X^2\mathcal{D}^2 $, and (iv) $ \phi^4X\mathcal{D}^2 $ classes of SMEFT operators.}
		\label{fig:phi-X-heavy-heavy}
\end{figure}

Fig.\,\ref{fig:phi-X-heavy-heavy} contains one-loop schematic diagrams corresponding to the $\phi^2X^3$, $\phi^4X^2$,  $\phi^2X^2\mathcal{D}^2$ and  $\phi^4X\mathcal{D}^2$ operator classes, exhibiting mixing between two heavy fermions in the loop. The constituent vertices for each case are \{\hyperlink{vertex-5}{$V{}_5$},  \hyperlink{vertex-7}{$V{}_7$}\}. This case permits a plethora of heavy field quantum numbers but each of those must satisfy Eq.~\eqref{eq:vertex-5-rel} as well as Eq.~\eqref{eq:x4-vertex-constraint} at the appropriate vertices. These diagrams are necessary to account for the CP-violating operators and the heavy fermions have the characteristics of vector-like fermions.

\subsection{Validation of the results}
 Below we provide a discussion on how the results catalogued so far can be validated against the \textit{top-down} EFT methodology. We will elucidate how D8 operators made up of the SM Higgs, its conjugate and their derivatives originate after \textbf{heavy scalar} fields are integrated out. The starting point for this approach is the following schematic Lagrangian:
\begin{eqnarray}
	\mathcal{L}[\chi,\phi]= B(\phi)^\dagger\chi +h.c. +\frac{1}{2}\chi^T \big(P^2-M^2-U(\phi)\big) \chi + \mathcal{O}(\Phi^3),
\end{eqnarray}
where $ \chi $ denotes a generic heavy scalar field, $\phi$ denotes the  light fields collectively but for simplicity we shall infer it as the SM Higgs boson. $ P_\mu\equiv i\mathcal{D}_\mu $ and $ B,\,U $ are functions of the light fields.

\subsubsection*{\underline{Tree-level}}
 Effective operators at tree-level are generated by substituting the classical solution of the heavy field equation of motion\footnote{The classical solution $\chi_c$ solves the Euler-Lagrange equation: $\cfrac{\delta \mathcal{L}}{\delta \chi}\,\,\,\bigg|_{\chi = \chi_c} = 0.$  } ($ \chi_c $) back into the Lagrangian. After the covariant derivative expansion the effective Lagrangian at tree-level assumes the form \cite{Henning:2014wua}:
 
{\small\begin{eqnarray}\label{eq:tree-level-eff-lag}
	\mathcal{L}^{\text{eff}}_{\text{tree}} = B^\dagger\frac{1}{M^2}B + B^\dagger\frac{1}{M^2}(P^2-U)\frac{1}{M^2}B+ B^\dagger\frac{1}{M^2}(P^2-U)\frac{1}{M^2}(P^2-U)\frac{1}{M^2}B+\cdot\cdot
\end{eqnarray}} \label{eq:tree-eff}
The term containing $ B $ in the Lagrangian refers to interactions which are linear with respect to the heavy field but contain multiple light fields. In terms of the terminology established in this article, these correspond to the vertices \hyperlink{vertex-1}{$V{}_1$} and \hyperlink{vertex-2}{$V{}_2$} shown in Fig.\;\ref{fig:vertices}. Only a finite number of heavy scalar extensions of the SM permit such a term in the Lagrangian \cite{Henning:2014wua} and these are:
\begin{eqnarray}
	\Phi \in \{ (1,1,0),\,(1,3,0),\,(1,3,1),\,(1,2,\frac{1}{2}),\,(1,4,\frac{1}{2}),\,(1,2,\frac{3}{2}),\,(1,4,\frac{3}{2})\}.
\end{eqnarray}

\begin{table}[h]
	\centering
	\renewcommand{\arraystretch}{2.8}
	{\small\begin{tabular}{||c|c|c|c||}
		\hline
		\hline
		$\Phi$&
		\textsf{Interaction terms}&
		\textsf{Effective operators}&
		\textsf{Op. class}\\
		\hline
		\hline
		
		\multirow{2}{*}{$(1,1,0)$}&
		\multirow{2}{*}{$\underbrace{c_1\,(H^{\dagger}H)}_{B}\,\Phi$,}&
		\hspace{-1cm}$\cfrac{1}{M^6}\,(B^\dagger\,U\,U\,B)\, \rightarrow \, \cfrac{c_1^2}{M^6}\,(H^{\dagger}H)^4$&
		$\phi^8$\\
		
		&
		\multirow{2}{*}{$\underbrace{(H^{\dagger}H)}_{U}\,\Phi^2$}&
		$\cfrac{1}{M^6}\,(B^\dagger\,P^2\,U\,B)\, \rightarrow \, \cfrac{c_1^2}{M^6}(H^{\dagger}H)\square(H^{\dagger}H)^2$&
		$\phi^6\mathcal{D}^2$\\
		
		&
		&
		$\cfrac{1}{M^6}\,(B^\dagger\,P^2\,P^2\,B)\, \rightarrow \, \cfrac{c_1^2}{M^6}(H^{\dagger}H)[\square(H^{\dagger}H)]^2$&
		$\phi^4\mathcal{D}^4$\\
		
		\hline
		
		\multirow{2}{*}{$(1,4,\frac{3}{2})$}&
		$\underbrace{(H^iH^jH^k)}_{B}\,\Phi^{\dagger}_{ijk}$,&
		\hspace{-1cm}$\cfrac{1}{M^4}\,(B^\dagger\,U\,B)\, \rightarrow \, \cfrac{1}{M^4}\,(H^{\dagger}H)^4$&
		$\phi^8$\\
		
		&
		$\underbrace{(H^{\dagger}_iH^i)}_{U}\,\Phi^\dagger_{jkl}\Phi^{jkl}$ + \texttt{perm.}&
		$\cfrac{1}{M^4}\,(B^\dagger\,P^2\,B)\, \rightarrow \, \cfrac{1}{M^4}\,(H^{\dagger}H)\square(H^{\dagger}H)^2$&
		$\phi^6\mathcal{D}^2$\\
		\hline
	\end{tabular}}
	\caption{\sf Origin of D8 SMEFT operators through tree-level integrating out from UV models containing (i) a real singlet scalar $\Phi \in (1,1,0)$ and (ii) an $SU(2)_L$ quadruplet $\Phi \in (1,4,\frac{3}{2})$. Here, $i,j,k,l$ refer to $SU(2)$ indices and in the last row \texttt{"perm."} refers to the possible permutations of these indices.}
	\label{table:integ-out-B-U}
\end{table}

The term proportional to $U$ in the Lagrangian, on the other hand describes an interaction which is quadratic in both light and heavy fields. This is akin the vertex \hyperlink{vertex-3}{$V{}_3$} and is ubiquitous across all single heavy scalar extensions irrespective to their quantum numbers. In fact, $U(\phi) \equiv U(H, H^{\dagger}) = H^{\dagger}\,H$. Based on the various combinations of $B$ and $U$, the origin of SMEFT operators through the different terms of Eq.~\eqref{eq:tree-level-eff-lag} can be understood. A couple of illustrative examples demonstrating the origin of D8 SMEFT operators within this formalism have been summarized in Table~\ref{table:integ-out-B-U}. These relations between heavy field quantum numbers and SMEFT operator classes exactly match the ones established through tree-level diagrammatic unfolding for the $\phi^8$, $\phi^6\mathcal{D}^2$ and $\phi^4\mathcal{D}^4$ operator classes as part of our analysis. 

\subsubsection*{\underline{One-loop-level}}
To discuss the integration out of heavy fields at one-loop-level we have adopted the covariant diagram approach of ref.~\cite{Zhang:2016pja} and we have focussed only on the $ \phi^6 \mathcal{D}^2 $ operator class. Reiterating the fact that $P_\mu \equiv i\mathcal{D}_{\mu}$ and $U = (H^{\dagger}H)$ across single heavy scalar extensions of SM with arbitrary $SU(3)_C\times SU(2)_L \times  U(1)_Y$ quantum numbers, i.e., $\Phi \in (R_C, R_L, Y)$, we can refer to operators of the $ \phi^6 \mathcal{D}^2 $ class as $ P^2 U^3 $ in the language of ref.~\cite{Zhang:2016pja}.

The covariant diagram consisting of only scalar heavy particles in the loop along with two $P_{\mu}$ insertions and three $U$ insertions is shown below:

\begin{tabular}{c c}
	\multirow{3}{*}{\includegraphics[scale=0.4]{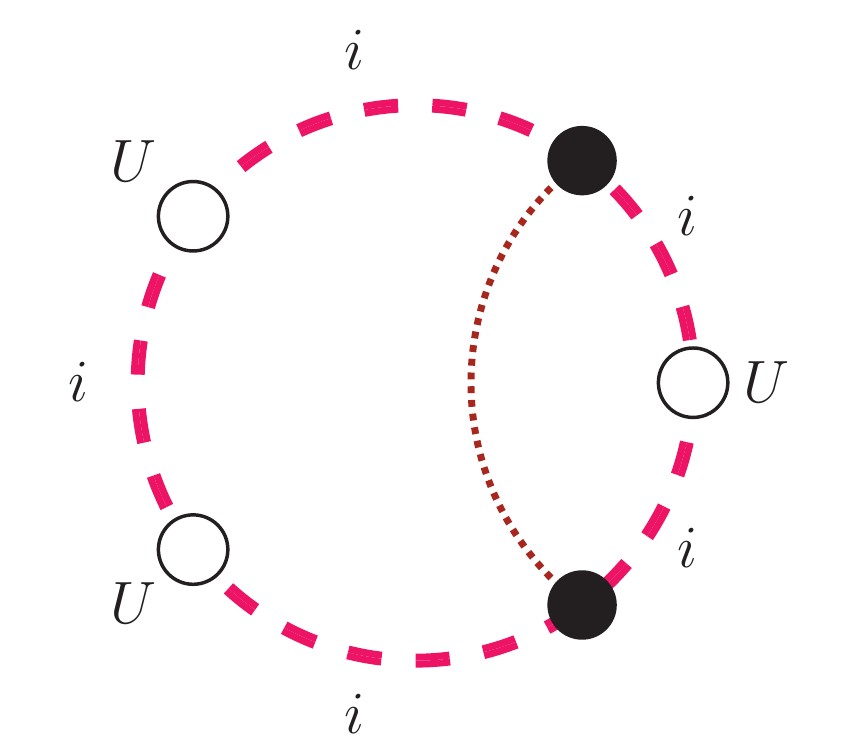}} &\\
	&\multirow{3}{*}{$ = -i\, c_s\, (2)^2\, I[q^2]_{i}^{5}\,\,\textsf{tr}\left( U\,[P_\mu \,, U]\,[P^\mu\,, U]\right)$.}\\
	&\\
	&\\
	&\\
	&\\
	&\\
\end{tabular}\label{eq:p6d2-op}

Adjacent to the covariant diagram, we have also noted the corresponding effective operator using the conventions of ref.~\cite{Zhang:2016pja}. This can be re-written as  $\mathcal{C}^{(\phi^6\mathcal{D}^2)}\,\textsf{tr}\left( U\,[P_\mu \,, U]\,[P^\mu\,, U]\right) $, and $\mathcal{C}^{(\phi^6\mathcal{D}^2)} = -i\, c_s\, (2)^2\, I[q^2]_{i}^{5}$ contributes to the Wilson coefficient. Here, $ c_s $ depends on the type of the heavy field in the loop (e.g., $ c_s=1/2 $ for real scalar fields.), and 
\begin{eqnarray}
	I[q^2]_i^5 = \int \frac{d^d q}{(2 \pi)^d}\frac{q^2}{(q^2-M_i^2)^5} = \frac{1}{12\,M_i^4}.	
\end{eqnarray}

The $i$ in the subscript on the left hand side marks the variety of propagators in the covariant diagram. The loop factor $I[q^2]_i^5$ can be evaluated using tools such as \texttt{PackageX} \cite{Patel:2015tea}. For a comprehensive discussion, we refer the reader to \cite{Zhang:2016pja}.

The noteworthy point is that such covariant diagrams can be drawn for any heavy scalar, hence leading to the same family of SMEFT operators. Thus, our results for the unfolding of operators consisting of the SM Higgs, it's conjugate and their derivatives into one-loop diagrams constituted of a single heavy scalar can be corroborated.

\section{Common origin of subsets of D6 and D8 operators}\label{sec:dim6-dim8-common}

The search for clues of new physics, within D8 SMEFT operators, is slowly becoming a focal point of contemporary phenomenological as well as experimental research. As we take the next stride in probing beyond the Standard Model, we must be mindful of the subtleties as well as overarching patterns evident across operators of different mass dimensions. Even when restricting ourselves to the purely bosonic operators, a comparison with the results of \cite{DasBakshi:2021xbl} reveals clear connections between the UV origin of D6 and D8 SMEFT operators. In what follows, we have underlined these connections as well as the subtleties involved when studying operators of dimensions-6 and -8 together.

\begin{itemize}
 \item Direct comparison between the results for D6 operator classes catalogued in \cite{DasBakshi:2021xbl} and section \ref{sec:cpv} of this work and those of D8, as discussed in section \ref{sec:dim-8-bosonic} that differ only by $\phi^2 \equiv (H^{\dagger}H)$ reveals common UV origin for pairs of D6, D8 classes, e.g.,
 \begin{eqnarray}\label{eq:d6-d8-op-class-pairs}
 	(\phi^6,\,\phi^8);\,\,\,\,\,\,
 	 (\phi^4\mathcal{D}^2,\,\phi^6\mathcal{D}^2);\,\,\,\,\,\,(\phi^2X^2,\,\phi^4X^2).
 \end{eqnarray}
 
  This is not surprising as the additional $\phi^2$ piece is an overall singlet. Therefore, the pairs of operator classes in Eq.~\eqref{eq:d6-d8-op-class-pairs} exhibit successive orders of perturbation theory and provide corrections to the same SM processes, most notably the electroweak precision observables. The corrections are weighted by powers of $(v/\Lambda)^{2n}$, with $n = 1$ and $2$ for dimensions-6 and -8 respectively.
  
  \item On the other hand, operator classes that differ by a single $X_{\mu\nu}$, which also increases the mass dimension by 2, do not necessarily share the same origin. For instance, the $X^3$ class contains only two types of operators, one trilinear in $W^I_{\mu\nu}$ and the other in $G^A_{\mu\nu}$. They enforce the constraint of non-triviality of the $SU(2)_L$ and $SU(3)_C$ representations of the heavy field respectively. The $X^4$ class not only contains operators describing self-interactions of $B_{\mu\nu}$, $W^I_{\mu\nu}$ and $G^A_{\mu\nu}$ but also various mixed cases. Operators of the latter sub-category impose additional constraints on the heavy field quantum numbers. This underlines the significance of D8 operators for discriminating between UV models that furnish similar results at D6. 
  \end{itemize}

These associations between D6 and D8 operators and their shared UV origin also serve to streamline any phenomenological analysis involving them. This is because the Wilson coefficients of all the operators generated after integrating out a particular heavy field are functions of the same finite number of parameters, irrespective of whether they belong to D6 or D8 classes. This way, \textit{top-down} analyses, informed by the results of this work, can be extended to account for D8 operators without worrying about an unmanageable number of free parameters. Such a luxury cannot be ensured in \textit{bottom-up} analyses, where free parameters start to proliferate as the mass-dimension increases. This has been the main source of scepticism regarding the incorporation of D8 operators for statistical analyses of experimental data. A few areas where D8 operators have gained a lot of prominence have been described below:

 \begin{itemize}
  \item Operator classes such as, $\phi^4\mathcal{D}^4$, $\phi^2X^2\mathcal{D}^2$, $X^4$, etc.\;encapsulate vertices relevant for vector boson scattering (VBS) processes \cite{Khachatryan:2015cms,ATLAS:2014atlas,Sirunyan:2018cms,Sirunyan:2017cms,Sirunyan:2019cms,Aaboud:2019atlas,ATLAS:2019atlas}. Interactions involving only the charged electroweak gauge bosons ($W^+, W^-$) and those involving charged as well as neutral gauge bosons ($Z,\,\gamma$) manifest within the SM Lagrangian itself. But, trilinear and quartic interactions within the neutral sector are scarce. The relevant vertices are first encountered within D8 operators. A hint of such processes will certainly widen the scope for new physics. 
 
Also, owing to their common UV origin, D8 operators, (e.g., $\phi^4\mathcal{D}^4$) enveloping such rare processes will always be accompanied by D6 operators ($\phi^4\mathcal{D}^2$ in this case) that contribute to observables such as Higgs signal strength (HSS). Therefore, studying the rare variety of VBS in tandem with HSS could provide better exclusion limits on the parameter space for the BSM proposal. 

 \item  Constraints can be enforced on BSM physics based on electroweak precision data (EWPD) in terms of the oblique ($ S, T, U $) parameters. The first two receive contributions from D6 operators. The $ U $ parameter, on the other hand, receives a vanishing contribution at D6. It obtains non-zero contributions only at mass dimension-8  \cite{Grinstein:1991cd,Murphy:2020rsh}. The operator classes that affect these parameters have been highlighted below:
 \begin{eqnarray}\label{eq:s-t-u}
 	S \rightarrow \underbrace{\phi^2X^2}_{D6},\,\,\underbrace{\phi^4X^2}_{D8}; \qquad T \rightarrow \underbrace{\phi^4\mathcal{D}^2}_{D6},\,\,\underbrace{\phi^6\mathcal{D}^2}_{D8}; \qquad U \rightarrow \underbrace{\phi^4X^2}_{D8}.
 \end{eqnarray}
Naively, one may assume that the $\phi^4X^2$ class only serves as the next-order contributor to observables affected by the $\phi^2X^2$ operators as suggested by the discourse around Eq.~\eqref{eq:d6-d8-op-class-pairs}. But Eq.~\eqref{eq:s-t-u} elucidates how the D8 classes can be leading order contributors for certain observables and thus supply novel means of distinguishing between BSM proposals.
 
 \item The unfolding of the CP-violating subsets of operators requires pairs of heavy vector-like fermions. The signature of the violation is found in Yukawa-like interactions with a $\gamma_5$ matrix present at the vertices. This pattern occurs across both D6 and D8 operators as demonstrated in this work. 
 For instance, the VLF pair possessing the following quantum numbers:
 \begin{eqnarray}
 	\Psi^{(1)}_{L, R} \in (1,3,1)  \hspace{0.5cm}\text{and}\hspace{0.5cm} \Psi^{(2)}_{L, R} \in (1,2,\frac{1}{2}),
 \end{eqnarray}
 generate the CP-violating subsets of the $\phi^2X^2$ and $X^3$ operator classes at D6. Simultaneously, these also generate CP-violating operators belonging to the $\phi^4X^2$, $\phi^2X^3$, $\phi^2X^2\mathcal{D}^2$ and $X^4$ classes at D8. A scrupulous investigation into the phenomena of CP-violation would necessitate the inclusion of dimension-8 operators.  
\end{itemize}

Thus, operators at mass dimension-8 are significant not just as the next order contributors to SM processes or to bolster the predictions of dimension-6 SMEFT operators but also as purveyors of novel and esoteric phenomena.

\section{Conclusion}

In this work, we have assembled a catalogue of admissible quantum numbers for heavy scalars and fermions that can generate specific subsets of SMEFT operators. We have subjected purely bosonic, CP-violating operators of mass dimension-6 and all purely bosonic operators of dimension-8 to an elaborate unfolding procedure. It commences with the identification of Lorentz invariant vertices describing interactions between heavy and light fields. These vertices are then employed to build tree-level as well as loop-level diagrams with the same external legs as the SMEFT operator classes. The operation culminates with the identification of one or more heavy propagators within those diagrams. We have adhered to a notion of conciseness in our discussion by restricting to descriptions at the level of operator class in most cases and we have delved into detailed examples only for a few pertinent scenarios. We have vividly highlighted recurrent patterns across different operator classes as well as the nuances present within operators with the similar constitution.

We have validated our findings through direct comparison with the results of \textit{top-down} analyses that generate CP-violating SMEFT operators from models containing vector-like fermions. We have also surveyed how heavy scalar loops can engender SMEFT operators with bosonic legs within the covariant diagram approach.

By studying the parallels as well as the disparity between the ways in which operators of dimensions-6 and -8 originate from UV models, we have accentuated the significance of the latter for current as well as future phenomenological analyses dedicated to the search for new physics. We have shed light on the firm ties between CP-violation in SMEFT and extensions of the SM through the inclusion of vector-like fermions. We have also elucidated the role of dimension-8 operators as the radix of atypical phenomena such as the scattering of neutral electroweak gauge bosons through trilinear and quartic vertices. We have underlined the significance of conducting investigations into these rare phenomena in conjunction with the study of high precision observables for pinpointing the most appropriate beyond the Standard Model scenario.
	
\section*{Acknowledgement}
	The authors would like to acknowledge the discussions with Joydeep Chakrabortty. SR is supported by the MHRD, Government of India.
	SP is supported by the MHRD, Government of India, under the Prime Minister's Research Fellows (PMRF) Scheme, 2020. 
\appendix

\section{The Standard Model field content and Lagrangian}\label{sec:app}

The Standard Model degrees of freedom along with their representations under the $SU(3)_C$ and $SU(2)_L$ groups, their respective $U(1)_Y$ hypercharges, baryon and lepton numbers and their spins have been collected in Table~\ref{table:SM-fields}. 
\begin{table}[h]
	\centering
	\renewcommand{\arraystretch}{2}
	\begin{tabular}{|c|c|c|c|c|c|c|}
			\hline
			\textsf{Field} & \textbf{$SU(3)_C$} & \textbf{$SU(2)_{L}$}&\textbf{$U(1)_{Y}$}&\textsf{Baryon No.}&\textsf{Lepton No.}&\textsf{Spin}\\
			\hline
			$H$    &1&2&1/2&0&0&0\\
			$q^p_L$       &3&2&1/6&1/3&0&1/2\\
			$u^p_R$     &3&1&2/3&1/3&0&1/2\\
			$d^p_R$     &3&1&-1/3&1/3&0&1/2\\
			$l^p_L$       &1&2&-1/2&0&-1&1/2\\
			$e^p_R$     &1&1&-1&0&-1&1/2\\
			\hline
			$G^A_{\mu}$ &8&1&0&0&0&1\\
			$W^I_{\mu}$ &1&3&0&0&0&1\\
			$B_{\mu}$   &1&1&0&0&0&1\\
			\hline
	\end{tabular}
	\caption{\sf Standard Model: Gauge and global quantum numbers and spins of the fields. Here, $A = 1,2,\cdots,8$; $I = 1,2,3;\,\, p = 1,2,3$ and $\mu = 0,1,2,3$ refer to the $SU(3), SU(2)$, flavour and Lorentz indices respectively.} 
	\label{table:SM-fields}
\end{table}

\section{Pure bosonic dimension-8 operators}\label{sec:app1}
 The complete list of purely bosonic SMEFT operators of mass dimension-8 have been presented in Tables~\ref{tab:smeft8class_1_2_3}-\ref{tab:smeft8class_5_6_7_8}. These were first constructed in \cite{Murphy:2020rsh,Li:2020gnx}.
\begin{table}[h]
	\begin{center}
		\begin{adjustbox}{width=1.0\textwidth,center}
			\small
			\begin{minipage}[t]{2.4cm}
				\renewcommand{\arraystretch}{1.8}
				\begin{tabular}[t]{c|c}
					\multicolumn{2}{c}{\boldmath$1:\phi^8$} \\
					\hline
					$Q_{H^8}$  &  $(H^\dag H)^4$ 
				\end{tabular}
			\end{minipage}
			\hspace{1cm}
			\begin{minipage}[t]{5.6cm}
				\renewcommand{\arraystretch}{1.8}
				\begin{tabular}[t]{c|c}
					\multicolumn{2}{c}{\boldmath$2:\phi^6\mathcal{D}^2$} \\
					\hline
					$Q_{H^6}^{(1)}$  & $(H^{\dag} H)^2 (\mathcal{D}_{\mu} H^{\dag} \mathcal{D}^{\mu} H)$ \\
					$Q_{H^6}^{(2)}$  & $(H^{\dag} H) (H^{\dag} \tau^I H) (\mathcal{D}_{\mu} H^{\dag} \tau^I \mathcal{D}^{\mu} H)$
				\end{tabular}
			\end{minipage}
			\hspace{1cm}
			\begin{minipage}[t]{4.9cm}
				\renewcommand{\arraystretch}{1.8}
				\begin{tabular}[t]{c|c}
					\multicolumn{2}{c}{\boldmath$3:\phi^4\mathcal{D}^4$} \\
					\hline
					$Q_{H^4}^{(1)}$  &  $(\mathcal{D}_{\mu} H^{\dag} \mathcal{D}_{\nu} H) (\mathcal{D}^{\nu} H^{\dag} \mathcal{D}^{\mu} H)$ \\ 
					$Q_{H^4}^{(2)}$  &  $(\mathcal{D}_{\mu} H^{\dag} \mathcal{D}_{\nu} H) (\mathcal{D}^{\mu} H^{\dag} \mathcal{D}^{\nu} H)$ \\ 
					$Q_{H^4}^{(3)}$  &  $(\mathcal{D}^{\mu} H^{\dag} \mathcal{D}_{\mu} H) (\mathcal{D}^{\nu} H^{\dag} \mathcal{D}_{\nu} H)$
				\end{tabular}
			\end{minipage}
		\end{adjustbox}
	\end{center}
	\caption{\sf The dimension-8 SMEFT operators with only the SM scalar, its conjugate and their derivatives as the building blocks.}
	\label{tab:smeft8class_1_2_3}
\end{table}
\begin{table}[h]
	\begin{center}
		\begin{adjustbox}{width=0.8\textwidth,center}
			\small
			\begin{minipage}[t]{6cm}
				\renewcommand{\arraystretch}{2.2}
				\begin{tabular}[t]{c|c}
					\multicolumn{2}{c}{\boldmath$4:X^4,\, X^3 X^{\prime}$} \\
					\hline
					$Q_{G^4}^{(1)}$  &  $(G_{\mu\nu}^A G^{A\mu\nu}) (G_{\rho\sigma}^B G^{B\rho\sigma})$ \\
					$Q_{G^4}^{(2)}$  &  $(G_{\mu\nu}^A \widetilde{G}^{A\mu\nu}) (G_{\rho\sigma}^B \widetilde{G}^{B\rho\sigma})$ \\
					$Q_{G^4}^{(3)}$  &  $(G_{\mu\nu}^A G^{B\mu\nu}) (G_{\rho\sigma}^A G^{B\rho\sigma})$ \\
					$Q_{G^4}^{(4)}$  &  $(G_{\mu\nu}^A \widetilde{G}^{B\mu\nu}) (G_{\rho\sigma}^A \widetilde{G}^{B\rho\sigma})$ \\
					$Q_{G^4}^{(5)}$  &  $(G_{\mu\nu}^A G^{A\mu\nu}) (G_{\rho\sigma}^B \widetilde{G}^{B\rho\sigma})$ \\
					$Q_{G^4}^{(6)}$  &  $(G_{\mu\nu}^A G^{B\mu\nu}) (G_{\rho\sigma}^A \widetilde{G}^{B\rho\sigma})$ \\
					$Q_{G^4}^{(7)}$  &  $d^{ABE} d^{CDE} (G_{\mu\nu}^A G^{B\mu\nu}) (G_{\rho\sigma}^C G^{D\rho\sigma})$ \\
					$Q_{G^4}^{(8)}$  &  $d^{ABE} d^{CDE} (G_{\mu\nu}^A \widetilde{G}^{B\mu\nu}) (G_{\rho\sigma}^C \widetilde{G}^{D\rho\sigma})$ \\
					$Q_{G^4}^{(9)}$  &  $d^{ABE} d^{CDE} (G_{\mu\nu}^A G^{B\mu\nu}) (G_{\rho\sigma}^C \widetilde{G}^{D\rho\sigma})$ \\
					$Q_{W^4}^{(1)}$  &  $(W_{\mu\nu}^I W^{I\mu\nu}) (W_{\rho\sigma}^J W^{J\rho\sigma})$ \\
					$Q_{W^4}^{(2)}$  &  $(W_{\mu\nu}^I \widetilde{W}^{I\mu\nu}) (W_{\rho\sigma}^J \widetilde{W}^{J\rho\sigma})$ \\
					$Q_{W^4}^{(3)}$  &  $(W_{\mu\nu}^I W^{J\mu\nu}) (W_{\rho\sigma}^I W^{J\rho\sigma})$ \\
					$Q_{W^4}^{(4)}$  &  $(W_{\mu\nu}^I \widetilde{W}^{J\mu\nu}) (W_{\rho\sigma}^I \widetilde{W}^{J\rho\sigma})$ \\
					$Q_{W^4}^{(5)}$  &  $(W_{\mu\nu}^I W^{I\mu\nu}) (W_{\rho\sigma}^J \widetilde{W}^{J\rho\sigma})$ \\
					$Q_{W^4}^{(6)}$  &  $(W_{\mu\nu}^I W^{J\mu\nu}) (W_{\rho\sigma}^I \widetilde{W}^{J\rho\sigma})$ \\
					$Q_{B^4}^{(1)}$  &  $(B_{\mu\nu} B^{\mu\nu}) (B_{\rho\sigma} B^{\rho\sigma})$ \\
					$Q_{B^4}^{(2)}$  &  $(B_{\mu\nu} \widetilde{B}^{\mu\nu}) (B_{\rho\sigma} \widetilde{B}^{\rho\sigma})$ \\
					$Q_{B^4}^{(3)}$  &  $(B_{\mu\nu} B^{\mu\nu}) (B_{\rho\sigma} \widetilde{B}^{\rho\sigma})$ \\
					$Q_{G^3B}^{(1)}$  &  $d^{ABC} (B_{\mu\nu} G^{A\mu\nu}) (G_{\rho\sigma}^B G^{C\rho\sigma})$ \\
					$Q_{G^3B}^{(2)}$  &  $d^{ABC} (B_{\mu\nu} \widetilde{G}^{A\mu\nu}) (G_{\rho\sigma}^B \widetilde{G}^{C\rho\sigma})$ \\
					$Q_{G^3B}^{(3)}$  &  $d^{ABC} (B_{\mu\nu} \widetilde{G}^{A\mu\nu}) (G_{\rho\sigma}^B G^{C\rho\sigma})$ \\
					$Q_{G^3B}^{(4)}$  &  $d^{ABC} (B_{\mu\nu} G^{A\mu\nu}) (G_{\rho\sigma}^B \widetilde{G}^{C\rho\sigma})$
				\end{tabular}
			\end{minipage}
			\hspace{1cm}
			\begin{minipage}[t]{5cm}
				\renewcommand{\arraystretch}{2.2}
				\begin{tabular}[t]{c|c}
					\multicolumn{2}{c}{\boldmath$4:X^2 X^{\prime 2}$} \\
					\hline
					$Q_{G^2W^2}^{(1)}$  &  $(W_{\mu\nu}^I W^{I\mu\nu}) (G_{\rho\sigma}^A G^{A\rho\sigma})$ \\
					$Q_{G^2W^2}^{(2)}$  &  $(W_{\mu\nu}^I \widetilde{W}^{I\mu\nu}) (G_{\rho\sigma}^A \widetilde{G}^{A\rho\sigma})$ \\
					$Q_{G^2W^2}^{(3)}$  &  $(W_{\mu\nu}^I G^{A\mu\nu}) (W_{\rho\sigma}^I G^{A\rho\sigma})$ \\
					$Q_{G^2W^2}^{(4)}$  &  $(W_{\mu\nu}^I \widetilde{G}^{A\mu\nu}) (W_{\rho\sigma}^I \widetilde{G}^{A\rho\sigma})$ \\
					$Q_{G^2W^2}^{(5)}$  &  $(W_{\mu\nu}^I \widetilde{W}^{I\mu\nu}) (G_{\rho\sigma}^A G^{A\rho\sigma})$ \\
					$Q_{G^2W^2}^{(6)}$  &  $(W_{\mu\nu}^I W^{I\mu\nu}) (G_{\rho\sigma}^A \widetilde{G}^{A\rho\sigma})$ \\
					$Q_{G^2W^2}^{(7)}$  &  $(W_{\mu\nu}^I G^{A\mu\nu}) (W_{\rho\sigma}^I \widetilde{G}^{A\rho\sigma})$ \\
					$Q_{G^2B^2}^{(1)}$  &  $(B_{\mu\nu} B^{\mu\nu}) (G_{\rho\sigma}^A G^{A\rho\sigma})$ \\
					$Q_{G^2B^2}^{(2)}$  &  $(B_{\mu\nu} \widetilde{B}^{\mu\nu}) (G_{\rho\sigma}^A \widetilde{G}^{A\rho\sigma})$ \\
					$Q_{G^2B^2}^{(3)}$  &  $(B_{\mu\nu} G^{A\mu\nu}) (B_{\rho\sigma} G^{A\rho\sigma})$ \\
					$Q_{G^2B^2}^{(4)}$  &  $(B_{\mu\nu} \widetilde{G}^{A\mu\nu}) (B_{\rho\sigma} \widetilde{G}^{A\rho\sigma})$ \\
					$Q_{G^2B^2}^{(5)}$  &  $(B_{\mu\nu} \widetilde{B}^{\mu\nu}) (G_{\rho\sigma}^A G^{A\rho\sigma})$ \\
					$Q_{G^2B^2}^{(6)}$  &  $(B_{\mu\nu} B^{\mu\nu}) (G_{\rho\sigma}^A \widetilde{G}^{A\rho\sigma})$ \\
					$Q_{G^2B^2}^{(7)}$  &  $(B_{\mu\nu} G^{A\mu\nu}) (B_{\rho\sigma} \widetilde{G}^{A\rho\sigma})$ \\
					$Q_{W^2B^2}^{(1)}$  &  $(B_{\mu\nu} B^{\mu\nu}) (W_{\rho\sigma}^I W^{I\rho\sigma})$ \\
					$Q_{W^2B^2}^{(2)}$  &  $(B_{\mu\nu} \widetilde{B}^{\mu\nu}) (W_{\rho\sigma}^I \widetilde{W}^{I\rho\sigma})$ \\
					$Q_{W^2B^2}^{(3)}$  &  $(B_{\mu\nu} W^{I\mu\nu}) (B_{\rho\sigma} W^{I\rho\sigma})$ \\
					$Q_{W^2B^2}^{(4)}$  &  $(B_{\mu\nu} \widetilde{W}^{I\mu\nu}) (B_{\rho\sigma} \widetilde{W}^{I\rho\sigma})$ \\
					$Q_{W^2B^2}^{(5)}$  &  $(B_{\mu\nu} \widetilde{B}^{\mu\nu}) (W_{\rho\sigma}^I W^{I\rho\sigma})$ \\
					$Q_{W^2B^2}^{(6)}$  &  $(B_{\mu\nu} B^{\mu\nu}) (W_{\rho\sigma}^I \widetilde{W}^{I\rho\sigma})$ \\
					$Q_{W^2B^2}^{(7)}$  &  $(B_{\mu\nu} W^{I\mu\nu}) (B_{\rho\sigma} \widetilde{W}^{I\rho\sigma})$
				\end{tabular}
			\end{minipage}
		\end{adjustbox}
	\end{center}
	\caption{\sf The dimension-8 SMEFT operators constituted only of field strength tensors.}
	\label{tab:smeft8class_4}
\end{table}

\begin{table}[h]
	\begin{center}
		\begin{adjustbox}{width=1.0\textwidth,center}
			\small
			\begin{minipage}[t]{8cm}
				\renewcommand{\arraystretch}{1.8}
				\begin{tabular}[t]{c|c}
					\multicolumn{2}{c}{\boldmath$5:\phi^2X^3$} \\
					\hline
					$Q_{G^3H^2}^{(1)}$  &  $f^{ABC} (H^\dag H) G_{\mu}^{A\nu} G_{\nu}^{B\rho} G_{\rho}^{C\mu}$ \\
					$Q_{G^3H^2}^{(2)}$  &  $f^{ABC} (H^\dag H) G_{\mu}^{A\nu} G_{\nu}^{B\rho} \widetilde{G}_{\rho}^{C\mu}$ \\
					$Q_{W^3H^2}^{(1)}$  &  $\epsilon^{IJK} (H^\dag H) W_{\mu}^{I\nu} W_{\nu}^{J\rho} W_{\rho}^{K\mu}$ \\
					$Q_{W^3H^2}^{(2)}$  &  $\epsilon^{IJK} (H^\dag H) W_{\mu}^{I\nu} W_{\nu}^{J\rho} \widetilde{W}_{\rho}^{K\mu}$ \\
					$Q_{W^2BH^2}^{(1)}$  &  $\epsilon^{IJK} (H^\dag \tau^I H) B_{\mu}^{\,\nu} W_{\nu}^{J\rho} W_{\rho}^{K\mu}$ \\
					$Q_{W^2BH^2}^{(2)}$  &  $\epsilon^{IJK} (H^\dag \tau^I H) (\widetilde{B}^{\mu\nu} W_{\nu\rho}^J W_{\mu}^{K\rho} + B^{\mu\nu} W_{\nu\rho}^J \widetilde{W}_{\mu}^{K\rho})$
				\end{tabular}
			\end{minipage}
			\hspace{1cm}
			\begin{minipage}[t]{6cm}
				\renewcommand{\arraystretch}{1.8}
				\begin{tabular}[t]{c|c}
					\multicolumn{2}{c}{\boldmath$6:\phi^4X^2$} \\
					\hline
					$Q_{G^2H^4}^{(1)}$  & $(H^\dag H)^2 G^A_{\mu\nu} G^{A\mu\nu}$ \\
					$Q_{G^2H^4}^{(2)}$  & $(H^\dag H)^2 \widetilde G^A_{\mu\nu} G^{A\mu\nu}$ \\
					$Q_{W^2H^4}^{(1)}$  & $(H^\dag H)^2 W^I_{\mu\nu} W^{I\mu\nu}$ \\
					$Q_{W^2H^4}^{(2)}$  & $(H^\dag H)^2 \widetilde W^I_{\mu\nu} W^{I\mu\nu}$ \\
					$Q_{W^2H^4}^{(3)}$  & $(H^\dag \tau^I H) (H^\dag \tau^J H) W^I_{\mu\nu} W^{J\mu\nu}$ \\
					$Q_{W^2H^4}^{(4)}$  & $(H^\dag \tau^I H) (H^\dag \tau^J H) \widetilde W^I_{\mu\nu} W^{J\mu\nu}$ \\
					$Q_{WBH^4}^{(1)}$  & $ (H^\dag H) (H^\dag \tau^I H) W^I_{\mu\nu} B^{\mu\nu}$ \\
					$Q_{WBH^4}^{(2)}$  & $(H^\dag H) (H^\dag \tau^I H) \widetilde W^I_{\mu\nu} B^{\mu\nu}$ \\
					$Q_{B^2H^4}^{(1)}$  & $ (H^\dag H)^2 B_{\mu\nu} B^{\mu\nu}$ \\
					$Q_{B^2H^4}^{(2)}$  & $(H^\dag H)^2 \widetilde B_{\mu\nu} B^{\mu\nu}$ \\
				\end{tabular}
			\end{minipage}
		\end{adjustbox}
		\begin{adjustbox}{width=1.0\textwidth,center}
			\small
			\begin{minipage}[t]{7.8cm}
				\renewcommand{\arraystretch}{1.8}
				\begin{tabular}[t]{c|c}
					\multicolumn{2}{c}{\boldmath$7:\phi^2X^2\mathcal{D}^2$} \\
					\hline
					$Q_{G^2H^2\mathcal{D}^2}^{(1)}$  &  $(\mathcal{D}^{\mu} H^{\dag} \mathcal{D}^{\nu} H) G_{\mu\rho}^A G_{\nu}^{A \rho}$ \\
					$Q_{G^2H^2\mathcal{D}^2}^{(2)}$  &  $(\mathcal{D}^{\mu} H^{\dag} \mathcal{D}_{\mu} H) G_{\nu\rho}^A G^{A \nu\rho}$ \\
					$Q_{G^2H^2\mathcal{D}^2}^{(3)}$  &  $(\mathcal{D}^{\mu} H^{\dag} \mathcal{D}_{\mu} H) G_{\nu\rho}^A \widetilde{G}^{A \nu\rho}$ \\
					$Q_{W^2H^2\mathcal{D}^2}^{(1)}$  &  $(\mathcal{D}^{\mu} H^{\dag} \mathcal{D}^{\nu} H) W_{\mu\rho}^I W_{\nu}^{I \rho}$ \\
					$Q_{W^2H^2\mathcal{D}^2}^{(2)}$  &  $(\mathcal{D}^{\mu} H^{\dag} \mathcal{D}_{\mu} H) W_{\nu\rho}^I W^{I \nu\rho}$ \\
					$Q_{W^2H^2\mathcal{D}^2}^{(3)}$  &  $(\mathcal{D}^{\mu} H^{\dag} \mathcal{D}_{\mu} H) W_{\nu\rho}^I \widetilde{W}^{I \nu\rho}$ \\
					$Q_{W^2H^2\mathcal{D}^2}^{(4)}$  &  $i \epsilon^{IJK} (\mathcal{D}^{\mu} H^{\dag} \tau^I \mathcal{D}^{\nu} H) W_{\mu\rho}^J W_{\nu}^{K \rho}$ \\
					$Q_{W^2H^2\mathcal{D}^2}^{(5)}$  &  $\epsilon^{IJK} (\mathcal{D}^{\mu} H^{\dag} \tau^I \mathcal{D}^{\nu} H) (W_{\mu\rho}^J \widetilde{W}_{\nu}^{K \rho} - \widetilde{W}_{\mu\rho}^J W_{\nu}^{K \rho})$ \\
					$Q_{W^2H^2\mathcal{D}^2}^{(6)}$  &  $i \epsilon^{IJK} (\mathcal{D}^{\mu} H^{\dag} \tau^I \mathcal{D}^{\nu} H) (W_{\mu\rho}^J \widetilde{W}_{\nu}^{K \rho} + \widetilde{W}_{\mu\rho}^J W_{\nu}^{K \rho})$ \\
					$Q_{WBH^2\mathcal{D}^2}^{(1)}$  &  $(\mathcal{D}^{\mu} H^{\dag} \tau^I \mathcal{D}_{\mu} H) B_{\nu\rho} W^{I \nu\rho}$ \\
					$Q_{WBH^2\mathcal{D}^2}^{(2)}$  &  $(\mathcal{D}^{\mu} H^{\dag} \tau^I \mathcal{D}_{\mu} H) B_{\nu\rho} \widetilde{W}^{I \nu\rho}$ \\
					$Q_{WBH^2\mathcal{D}^2}^{(3)}$  &  $i (\mathcal{D}^{\mu} H^{\dag} \tau^I \mathcal{D}^{\nu} H) (B_{\mu\rho} W_{\nu}^{I \rho} - B_{\nu\rho} W_{\mu}^{I\rho})$ \\
					$Q_{WBH^2\mathcal{D}^2}^{(4)}$  &  $(\mathcal{D}^{\mu} H^{\dag} \tau^I \mathcal{D}^{\nu} H) (B_{\mu\rho} W_{\nu}^{I \rho} + B_{\nu\rho} W_{\mu}^{I\rho})$ \\
					$Q_{WBH^2\mathcal{D}^2}^{(5)}$  &  $i (\mathcal{D}^{\mu} H^{\dag} \tau^I \mathcal{D}^{\nu} H) (B_{\mu\rho} \widetilde{W}_\nu^{^I \rho} - B_{\nu\rho} \widetilde{W}_\mu^{^I \rho})$ \\
					$Q_{WBH^2\mathcal{D}^2}^{(6)}$  &  $(\mathcal{D}^{\mu} H^{\dag} \tau^I \mathcal{D}^{\nu} H) (B_{\mu\rho} \widetilde{W}_\nu^{^I \rho} + B_{\nu\rho} \widetilde{W}_\mu^{^I \rho})$ \\
					$Q_{B^2H^2\mathcal{D}^2}^{(1)}$  &  $(\mathcal{D}^{\mu} H^{\dag} \mathcal{D}^{\nu} H) B_{\mu\rho} B_{\nu}^{\,\,\,\rho}$ \\
					$Q_{B^2H^2\mathcal{D}^2}^{(2)}$  &  $(\mathcal{D}^{\mu} H^{\dag} \mathcal{D}_{\mu} H) B_{\nu\rho} B^{\nu\rho}$ \\
					$Q_{B^2H^2\mathcal{D}^2}^{(3)}$  &  $(\mathcal{D}^{\mu} H^{\dag} \mathcal{D}_{\mu} H) B_{\nu\rho} \widetilde{B}^{\nu\rho}$
				\end{tabular}
			\end{minipage}
			\hspace{0.8cm}
			\begin{minipage}[t]{6.4cm}
				\renewcommand{\arraystretch}{1.8}
				\begin{tabular}[t]{c|c}
					\multicolumn{2}{c}{\boldmath$8:\phi^4X\mathcal{D}^2$} \\
					\hline
					$Q_{WH^4\mathcal{D}^2}^{(1)}$  & $(H^{\dag} H) (\mathcal{D}^{\mu} H^{\dag} \tau^I \mathcal{D}^{\nu} H) W_{\mu\nu}^I$ \\
					$Q_{WH^4\mathcal{D}^2}^{(2)}$  & $(H^{\dag} H) (\mathcal{D}^{\mu} H^{\dag} \tau^I \mathcal{D}^{\nu} H) \widetilde{W}_{\mu\nu}^I$ \\
					$Q_{WH^4\mathcal{D}^2}^{(3)}$  & $\epsilon^{IJK} (H^{\dag} \tau^I H) (\mathcal{D}^{\mu} H^{\dag} \tau^J \mathcal{D}^{\nu} H) W_{\mu\nu}^K$ \\
					$Q_{WH^4\mathcal{D}^2}^{(4)}$  & $\epsilon^{IJK} (H^{\dag} \tau^I H) (\mathcal{D}^{\mu} H^{\dag} \tau^J \mathcal{D}^{\nu} H) \widetilde{W}_{\mu\nu}^K$ \\
					$Q_{BH^4\mathcal{D}^2}^{(1)}$  & $(H^{\dag} H) (\mathcal{D}^{\mu} H^{\dag} \mathcal{D}^{\nu} H) B_{\mu\nu}$ \\
					$Q_{BH^4\mathcal{D}^2}^{(2)}$  & $(H^{\dag} H) (\mathcal{D}^{\mu} H^{\dag} \mathcal{D}^{\nu} H) \widetilde{B}_{\mu\nu}$
				\end{tabular}
			\end{minipage}
		\end{adjustbox}
	\end{center}
	\caption{\sf Bosonic dimension-8 operators in the SMEFT containing both field strength tensors and Higgs boson fields.}
	\label{tab:smeft8class_5_6_7_8}
\end{table}

\clearpage

\section{Additional vertices describing interactions of SM and BSM fields}\label{sec:extra-vertices}

\begin{figure}[h]
	\centering
	\renewcommand{\thesubfigure}{\roman{subfigure}}
	\begin{subfigure}[h]{4.5cm}
		\centering
		\includegraphics[scale=0.4]{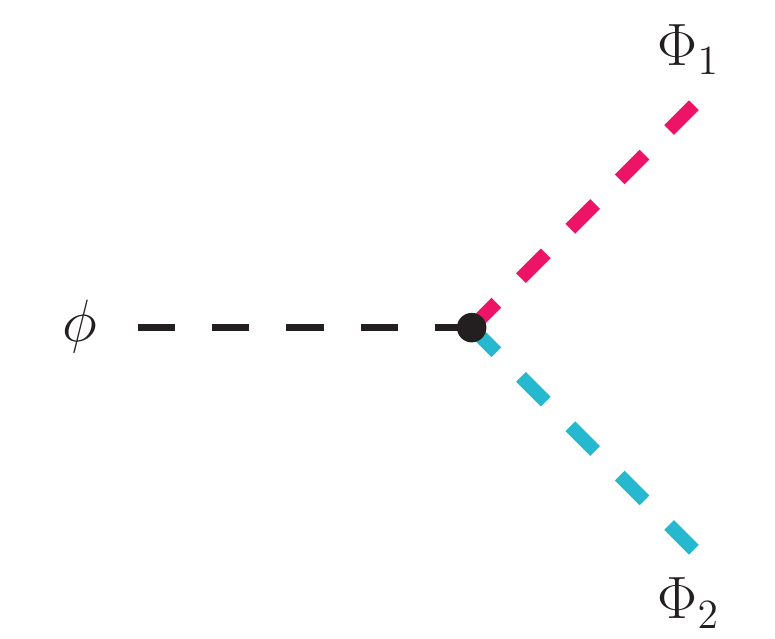}
		\caption{\hypertarget{vertex-8}{$V{}_8$}}\label{vertex:V2-extra}
	\end{subfigure}
	\begin{subfigure}[h]{4.5cm}
		\centering
		\includegraphics[scale=0.4]{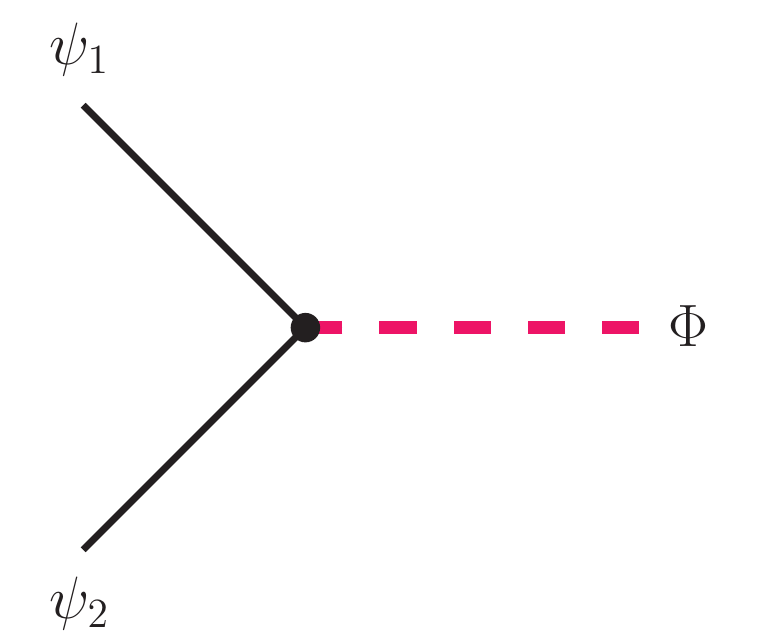}
		\caption{\hypertarget{vertex-9}{$V{}_9$}}\label{vertex:V5-extra}
	\end{subfigure}
	\begin{subfigure}[h]{4.5cm}
		\centering
		\includegraphics[scale=0.4]{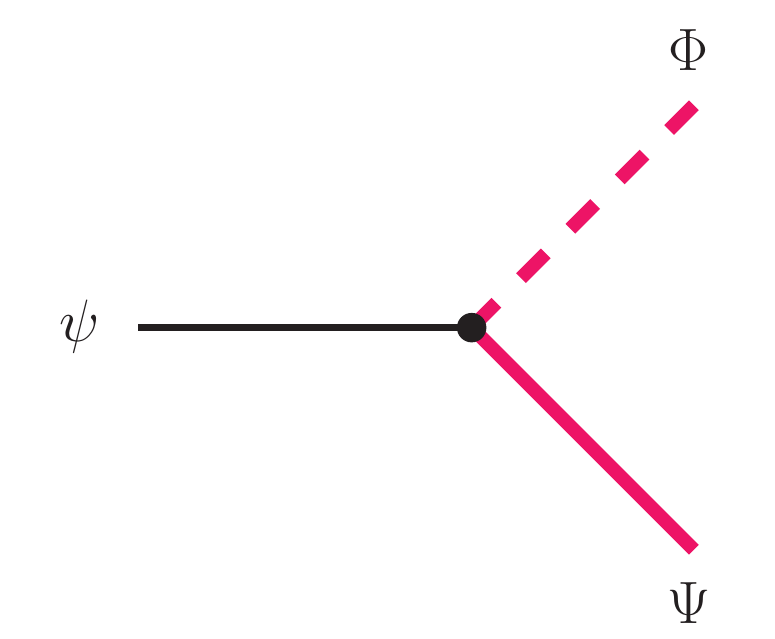}
		\caption{\hypertarget{vertex-10}{$V{}_{10}$}}\label{vertex:V7-extra}
	\end{subfigure}
	\caption{\sf Vertices describing interactions between (i) the SM scalar and two distinct heavy scalars (ii) two SM fermions and a heavy scalar and (iii) an SM fermion and a heavy scalar as well as a heavy fermion.}
	\label{fig:vertices-extra}
\end{figure}

In addition to the fundamental vertices elucidated in Fig.~\ref{fig:vertices} that form the building blocks of the diagrams illustrated in this article, there are a few more ways in which the SM fields can interact with heavy scalars and (or) fermions. These scenarios have been depicted in Fig.~\ref{fig:vertices-extra}. The reasons behind the exclusion of each of these interactions, as well as diagrams from our discussion, can be summarized as follows:

\begin{enumerate}
	\item $V{}_{8}$: Such a vertex does not uniquely fix the quantum numbers of the two heavy scalars $\Phi_{1,2} \in (R_{C_{1,2}}, R_{L_{1,2}}, Y_{1,2})$. As a matter of fact, we can only impose the following constraints on their quantum numbers:
	\begin{eqnarray}
		R_{C_1} \otimes R_{C_2} = 1, \hspace{1cm} R_{L_1} \otimes R_{L_2} = 2,  \hspace{1cm} Y_1 = Y_2  \pm \frac{1}{2}
	\end{eqnarray}
	with $+$ or $-$ in the last relation appearing based on whether $\phi = H^{\dagger}$ or $H$ is present at the vertex. Among the operators considered in this work, $H$ and $H^{\dagger}$ appear together. If the diagrams are unfolded into one-loop diagrams using such vertices, they then essentially predict a two scalar extension of the SM to describe the origin of the particular operator. On the other hand, by working with diagrams containing the vertex $V{}_3$, we limit ourselves to the more minimal case of single-particle extensions of the SM to describe the source of the same operator. 
	
	\item $V{}_{9}$: Since our focus is on SMEFT operators, the external states are always described by SM degrees of freedom. This vertex could only be accommodated if we had taken into account operators composed of SM fermions, but when we restrict ourselves to the purely bosonic sector, such a vertex offers no contributions to any of the diagrams.
	
	\item $V{}_{10}$: Similar to the case of $V{}_{9}$, the exclusion of this vertex from the main discussion is explained by the absence of SM fermions as external states of the operators considered in this work.
\end{enumerate}

\clearpage

	\providecommand{\href}[2]{#2}
	\addcontentsline*{toc}{section}{}
	\bibliographystyle{JHEP}
	\bibliography{CPV_VBS}
	
\end{document}